\documentclass[prd,preprint,superscriptaddress,preprintnumbers,eqsecnum,showpacs,nolongbibliography,nofootinbib,nobibnotes,noeprint]{revtex4-2}
\usepackage[linkcolor=blue,citecolor=blue,urlcolor=blue,colorlinks=true,breaklinks]{hyperref} 

\usepackage{amsfonts,bm}
\usepackage{amsfonts,amssymb,amsmath}

\usepackage[export]{adjustbox}
\usepackage{enumitem}
\usepackage[english]{babel}
\usepackage{slashed}
\usepackage[usenames]{xcolor}
\usepackage{mathtools}
\usepackage[makeroom]{cancel}
\usepackage{color}

\def\srm#1{{\rm{\scriptscriptstyle #1}}}

\newcommand{\be}{\begin{equation}}
\newcommand{\bea}{\begin{eqnarray}}
\newcommand{\ee}{\end{equation}}
\newcommand{\eea}{\end{eqnarray}}

\def\1eq#1{Eq.~(\ref{#1})}

\def\2eqs#1#2{Eqs.~(\ref{#1}) and~(\ref{#2})}
\def\3eqs#1#2#3{Eqs.~(\ref{#1}),~(\ref{#2}) and~(\ref{#3})}

\def\fig#1{Fig.~\ref{#1}}

\def\ie{{\it i.e.}, }
\def\eg{{\it e.g.}, }

\newcommand{\Ls}{ \mathit{L}_{{sg}}}   
\def\g{\Gamma}

\def\s#1{{\scriptscriptstyle #1}}

\newcommand{\fatg}{{\rm{I}}\!\Gamma}

\newcommand{\gb}{\overline{\Gamma}}

\newcommand{\Bfat}{\mathbb{B}}

\newcommand{\Cfat}{{\mathbb C}}

\def\sat{h}

\newcommand{\Itilde}{ {\omega} }

\newcommand{\Deltaeff}{\Delta_{\rm eff}}

\newcommand{\Khat}{{\widehat K}}
\newcommand{\Ko}{K_{\rm oge}}
\newcommand{\Ro}{{\cal R}_{\rm oge}}
\newcommand{\kin}{K_{\rm{kin}}}

\begin{document}

\title{Nonlinear Schwinger mechanism 
in QCD, \\and Fredholm alternatives theorem}

\author{M.~N.~Ferreira}
\affiliation{\mbox{School of Physics, Nanjing University},
Nanjing, Jiangsu 210093, China}

\affiliation{\mbox{Institute for Nonperturbative Physics, Nanjing University},
Nanjing, Jiangsu 210093, China}

\author{J.~Papavassiliou}
\affiliation{\mbox{Department of Theoretical Physics and IFIC, 
University of Valencia and CSIC},
E-46100, Valencia, Spain}
\affiliation{\mbox{ExtreMe Matter Institute EMMI}, GSI,
Planckstrasse 1, 64291, Darmstadt, Germany}

\begin{abstract}

We present a novel implementation 
of the Schwinger mechanism in QCD, which fixes uniquely the scale of the effective
gluon mass and streamlines considerably   
the procedure of multiplicative renormalization. 
The key advantage of this method stems from   
the nonlinear nature of the dynamical equation that generates    
massless poles in the longitudinal sector of the 
three-gluon vertex. 
An exceptional feature of this approach is  
an extensive cancellation 
involving 
the components of 
the integral expression that 
determines the gluon mass;
it is triggered 
once the 
Schwinger-Dyson equation of the pole-free part of the three-gluon vertex 
has been appropriately exploited. It turns out that 
this cancellation is 
driven by the so-called 
Fredholm alternatives theorem,
which operates among the 
set of integral equations
describing this system.  Quite remarkably, in the 
linearized approximation  
this theorem 
enforces the exact masslessness of the gluon. Instead, 
the nonlinearity   
induced by the full treatment of
the relevant kernel 
evades this 
theorem, allowing for the 
emergence of 
a nonvanishing mass. 
The numerical results 
obtained from 
the resulting equations
are compatible with the 
lattice findings, 
and may be further refined 
through the inclusion of 
the remaining fundamental 
vertices of the theory. 

\end{abstract}

\maketitle

\newpage 

\section{Introduction}\label{sec:intro}

As is well-known,  
the gauge symmetry 
of the classical Yang-Mills action~\cite{Yang:1954ek,Gross:1973id,Politzer:1973fx,Marciano:1977su}  
prohibits the inclusion of a 
mass term $m^2 A^a_\mu A^{a\mu} $ for the gauge field $A_\mu^a$, and this prohibition 
persists after the 
gauge-fixing through the Faddeev-Popov procedure~\cite{Faddeev:1967fc} has been carried out, due to the residual 
Becchi-Rouet-Stora-Tyutin (BRST)
symmetry~\cite{Becchi:1974md,Tyutin:1975qk,Becchi:1975nq}. Moreover,  
symmetry-preserving regularization schemes, 
such as dimensional regularization~\cite{tHooft:1972tcz,Bollini:1972ui}, prevent
the generation of a mass term at any finite 
order in perturbation theory. 
Nevertheless, as has been argued over four decades 
ago~\cite{Cornwall:1981zr}, a non-perturbative effective gluon mass may be
generated through the non-Abelian realization  
of the Schwinger mechanism (SM)~\cite{Schwinger:1962tn,Schwinger:1962tp}. 
This dynamical scenario has received particular attention 
over the years, 
becoming an integral part 
of the intense activity associated  with the emergence of a mass gap in the gauge
sector of QCD~\cite{Eichten:1974et,Smit:1974je,Parisi:1980jy,Bernard:1981pg,Bernard:1982my,Donoghue:1983fy,Mandula:1987rh,Cornwall:1989gv,Lavelle:1991ve,Halzen:1992vd,Wilson:1994fk,Mihara:2000wf,Kondo:2001nq,Philipsen:2001ip,Aguilar:2002tc,Pawlowski:2003hq,Aguilar:2004sw,Aguilar:2006gr,Epple:2007ut,Aguilar:2008xm,Braun:2007bx,
Tissier:2010ts,Campagnari:2010wc,Tissier:2011ey,Serreau:2012cg,Fagundes:2011zx,Binosi:2012sj,Maas:2011se,Siringo:2015wtx,Cyrol:2016tym,Glazek:2017rwe,Cyrol:2018xeq,Gracey:2019xom,Roberts:2020hiw,Pelaez:2021tpq,Eichmann:2021zuv,Horak:2022aqx,Papavassiliou:2022wrb,Ding:2022ows,Ferreira:2023fva}. 

In its original formulation,
the SM is based on the 
fundamental 
observation that
a mass for 
a vector meson may be generated, without 
clashing with the gauge symmetry of the action,  
if the corresponding vacuum polarization develops a pole with positive residue
($m^2$)
at zero-momentum transfer~\cite{Schwinger:1962tn,Schwinger:1962tp}.
In that case, the gluon propagator, $\Delta(q)$,
saturates to a fixed value 
at the origin, namely  
$\Delta^{-1}(0) = m^2$, 
such that the residue of the pole plays the role of an effective mass.

The implementation of this basic idea 
in the context of Yang-Mills theories, 
in general, and in the gauge sector of QCD 
in particular, is technically rather complicated. 
Specifically, the required pole 
is provided by the propagator $ \delta^{ab}/q^2$ of 
a massless color-carrying   
scalar bound state, formed out of the fusion of a pair of gluons~\cite{Eichten:1974et,Smit:1974je,Poggio:1974qs,Cornwall:1981zr,Aguilar:2011xe,Ibanez:2012zk,Aguilar:2015bud,Aguilar:2017dco,Aguilar:2021uwa}. 
The amplitude 
for the formation of such a pole,
denoted by $\Bfat(r)$, 
is governed by a homogeneous Bethe-Salpeter equation 
(BSE). The existence of a nontrivial solution for $\Bfat(r)$
endows the three-gluon vertex 
with a longitudinally coupled 
pole~\cite{Aguilar:2011xe,Ibanez:2012zk,Aguilar:2015bud,Aguilar:2017dco,Aguilar:2021uwa},  
which is transmitted to the Schwinger-Dyson equation (SDE)~\cite{Roberts:1994dr,Alkofer:2000wg,Maris:2003vk,Fischer:2006ub,Roberts:2007ji,Fischer:2008uz,Binosi:2009qm,Bashir:2012fs,Fister:2013bh,Cloet:2013jya,Binosi:2014aea,Aguilar:2015bud,Binosi:2016rxz,Binosi:2016nme,Huber:2018ned,Huber:2020keu,Papavassiliou:2022wrb,Ferreira:2023fva} 
that governs the momentum evolution of 
the gluon propagator,
triggering the SM. 
In the  
limit $q \to 0$, this SDE expresses 
the effective gluon mass as an integral 
involving 
$\Bfat(r)$. In addition,  
the Ward identity (WI) satisfied 
by the three-gluon vertex is 
modified by an amount 
controlled by the so-called 
{\it ``displacement function''}~\cite{Aguilar:2016vin,Aguilar:2021uwa,Aguilar:2022thg}, denoted by 
$\Cfat(r)$, which is intimately connected to $\Bfat (r)$.

Even though considerable 
advances have been made in 
our understanding of the dynamical scenario   
outlined above,
the treatments presented 
in the literature leave 
certain key  
questions unanswered. First, in the linearized version of the BSE 
studied thus far, 
the scale of the solutions for $\Bfat (r)$
remains undetermined~\cite{Aguilar:2011xe,Ibanez:2012zk,Aguilar:2015bud,Aguilar:2017dco,Aguilar:2021uwa}; thus,  
the value of the gluon mass 
is adjusted from the saturation point of the gluon propagator 
obtained in lattice simulations~\cite{Alexandrou:2001fh,Cucchieri:2007md,Bogolubsky:2007ud,Bowman:2007du,Kamleh:2007ud,Cucchieri:2007rg,Bogolubsky:2009dc,Oliveira:2009eh,Cucchieri:2009kk,Cucchieri:2009zt,Cucchieri:2011pp,Oliveira:2010xc,Ayala:2012pb,Sternbeck:2012mf,Bicudo:2015rma,Duarte:2016iko,Dudal:2018cli,Aguilar:2019uob}. 
Second, 
the integral expression 
furnishing the mass 
is multiplied 
by $Z_3$, the renormalization constant of the three-gluon vertex. As a result, the full 
treatment of the 
gluon mass equation requires 
the implementation 
of multiplicative 
renormalization at a nonperturbative level; 
however, the numerical 
evaluation of this integral 
has been carried out 
by simply 
setting $Z_3=1$~\cite{Aguilar:2011xe,Binosi:2012sj,Ibanez:2012zk,Aguilar:2015bud,Aguilar:2021uwa,Aguilar:2023mam}.

In the present work we report significant progress towards the self-consistent resolution of the open questions mentioned above. The main points 
of this analysis may 
be summarized as follows.

({\it i})
The BSE governing $\Bfat(r)$ 
is made nonlinear, by including a term cubic in $\Bfat$, 
not considered in previous studies~\cite{Aguilar:2011xe,Ibanez:2012zk,Aguilar:2015bud,Aguilar:2017dco,Aguilar:2021uwa,Aguilar:2023mam}. Quite interestingly, the  
quadratic part of this dependence on 
$\Bfat$ is absorbed into a 
constant denoted by $\omega$; as a result, 
the total contribution 
of this term 
to the BSE is a factor of the form 
$\omega \,\Bfat(r)$. Consequently,  
the resulting integral equation 
may still be treated as an
eigenvalue problem, exactly 
as was done 
in the past with the linear BSE. 
The presence of this novel term determines 
the scale of the solutions 
up to an overall sign;  
since the  
basic quantities of interest, such as the gluon mass and the displacement function, depend quadratically on 
$\Bfat$, this residual 
sign ambiguity is immaterial. 

({\it ii})
The multiplicative renormalization is 
carried out by introducing the 
SDE for the three-gluon vertex, 
which depends on the renormalization 
constant $Z_3$. It turns out that, 
by virtue of a massive cancellation
triggered by the form of the 
BSE in ({\it i}), the multiplicative 
renormalization of the mass equation may be carried out exactly. 
The resulting manifestly finite expression 
is proportional to the parameter 
$\omega$. Thus, rather remarkably, 
in the absence of the 
non-cubic term in the BSE, the finite gluon mass vanishes, even in the presence of a non-vanishing $\Bfat (r)$. 
Instead, when $\omega$ assumes its natural value dictated by the BSE, the resulting 
gluon mass is non-vanishing; its value 
can be made 
compatible with the $\Delta^{-1}(0)$
obtained from lattice simulations by adjusting the kernel of the BSE.

({\it iii})
It turns out that
the extensive cancellations 
mentioned in 
({\it ii}) 
can be attributed to 
a concrete 
mathematical reason, 
namely the 
Fredholm alternatives theorem~\cite{vladimirov1971equations,polyanin2008handbook}, 
which is in full operation when 
$\omega =0$. In that case, the 
kernels of the homogeneous BSE for $\Bfat (r)$
and of the inhomogeneous SDE for the three-gluon vertex 
are identical. Then, according to the 
theorem, no simultaneous solutions  
exist to these two integral equations 
unless a certain integral expression vanishes; and, quite notably, this expression is none other than the mass equation 
derived from the gluon SDE. Instead, the presence of a non-vanishing $\omega$ introduces a difference between the two kernels, thus evading the main assumption of the theorem
and leading to the emergence of a gluon mass. 

The article is organized as follows. 
In Sec.~\ref{sec:SM} we review the theoretical framework 
employed in this work, focusing on the set of 
special relations pertaining to the SM. Then, certain key results are derived 
in Sec.~\ref{sec:key},
which are extensively used
in the ensuing analysis. 
In Sec.~\ref{sec:rengen} 
the renormalization of the 
relevant quantities is outlined, placing 
particular emphasis on the 
treatment of the components emerging from the activation of the SM.
In Sec.~\ref{sec:BSE}
we derive in detail the BSE that 
controls the formation of the bound state pole, paying special attention to the novel term that causes 
the non-linearity of the resulting integral equation.  
In Sec.~\ref{sec:scalefix} we explain in detail how 
the BSE of the previous section leads to the fixing of the scale of the solutions obtained.  
Then, in Sec.~\ref{sec:trick} we carry out the 
multiplicative renormalization of the 
gluon mass, which proceeds through 
the implementation of a crucial cancellation. 
In Sec.~\ref{sec:Fredholm}
we state the Fredholm alternatives theorem, and apply it to the set of integral equations considered in the previous section. 
In Sec.~\ref{sec:res} we perform a detailed numerical analysis of the above equations, and study the dependence of the gluon mass and the displacement function
on the details of the BSE kernel. 
In Sec.~\ref{sec:conc} we discuss our results 
and summarize our conclusions. Finally, 
in Appendix \ref{sec:toy} we illustrate the central mathematical   
construction of Sec.~\ref{sec:trick} by means of  
a concrete example. 

\section{General theoretical framework}\label{sec:SM} 

In this section we introduce the necessary notation, and 
present a brief overview of the main notions associated 
with the implementation of the SM in the context of 
Yang-Mills theories; for recent reviews on the subject, see~\cite{Papavassiliou:2022wrb,Ferreira:2023fva}.

Let us consider the (Landau-gauge) gluon propagator 
\be
\Delta^{ab}_{\mu\nu}(q) = -i\delta^{ab} {P}_{\mu\nu}(q)
\Delta(q)\,,
\qquad 
{P}_{\mu\nu}(q) := g_{\mu\nu} - q_\mu q_\nu/{q^2}\,,  \label{gluon_def}
\ee 
and the associated dimensionless vacuum polarization 
${\bf \Pi}(q)$,  
\be 
\Delta^{-1}({q})=q^2 [1 + i {\bf \Pi}(q)]\,.
\label{vacpol}
\ee
In addition, 
the three-gluon vertex, 
${\cal G}^{abc}_{\alpha \mu \nu}(q,r,p)$,
with all momenta 
incoming and $q+r+p =0$,
is defined as 
\be
{\cal G}^{abc}_{\alpha \mu \nu}(q,r,p) \ = \  \langle 0 \vert T[\widetilde{A}^a_\alpha(q) \widetilde{A}^b_\mu(r) \widetilde{A}^c_\nu(p)] \vert 0\rangle  \,,  \label{eq:Green3g}
\ee 
where 
$\widetilde{A}^a_\alpha$ denotes the  Fourier 
transformed SU(3) 
gauge field, and $T$ the standard time-ordering operation. 
It is convenient for the analysis that follows to 
explicitly factor out 
of ${\cal G}^{abc}_{\alpha \mu \nu}(q,r,p)$
the gauge coupling $g$, 
thus defining the vertex $\fatg^{abc}_{\alpha\mu\nu}(q,r,p)$,
according to 
\bea
{\cal G}^{abc}_{\alpha \mu \nu}(q,r,p) &=& g \fatg^{abc}_{\alpha\mu\nu}(q,r,p) \,,
\nonumber\\
\fatg^{abc}_{\alpha\mu\nu}(q,r,p) &=& f^{abc}\fatg_{\alpha\mu\nu}(q,r,p) \,,
\label{GgG}
\eea
where $f^{abc}$ denotes the structure constants of the 
group SU(3). 

\begin{figure}[!ht]
  \includegraphics[width=0.45\textwidth]{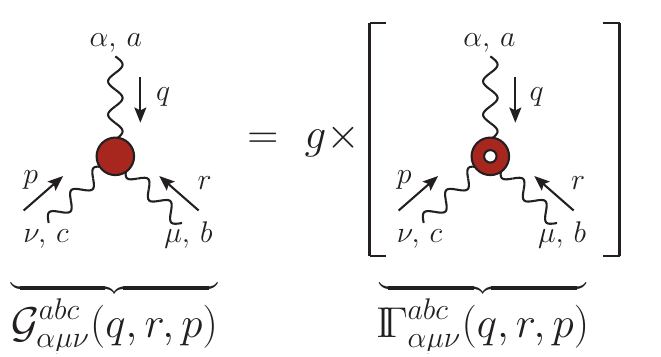}\hfill
  \includegraphics[width=0.45\textwidth]{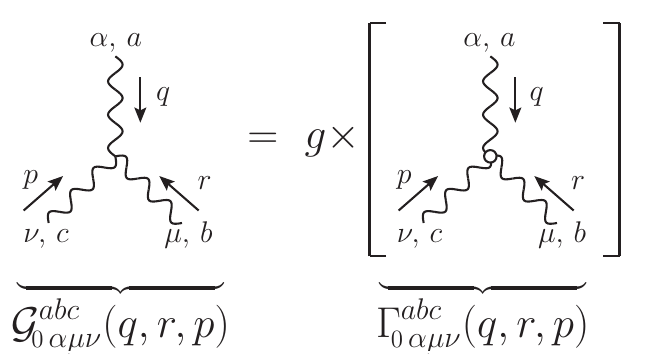}
  \caption{ Diagrammatic conventions for the fully dressed three-gluon vertex (left) and its tree-level counterpart (right).}\label{fig:3g_def}
\end{figure}

Note that both ${\cal G}^{abc}_{\alpha \mu \nu}(q,r,p)$ 
and $\fatg^{abc}_{\alpha\mu\nu}(q,r,p)$
will be employed throughout this work; therefore, a clear diagrammatic distinction between them has been introduced in \fig{fig:3g_def}. 
At tree level, 
\be 
\fatg_{\!0}^{\alpha\mu\nu}(q,r,p) = (q - r)^\nu g^{\alpha\mu} + (r - p)^\alpha g^{\mu\nu} + (p - q)^\mu g^{\nu\alpha} \,.
\label{bare3g}
\ee 

The cornerstone of the SM is 
 the observation that  
if the function ${\bf \Pi}(q)$ develops  
a pole at $q^2=0$ ({\it ``massless pole''}), 
then, in that kinematic limit, the propagator of the gauge boson 
saturates at a  non-zero value~\cite{Schwinger:1962tn,Schwinger:1962tp}, according to the sequence (Euclidean space)
\be
\lim_{q^2 \to 0} {\bf \Pi}(q) = m^2/q^2 \,\,\Longrightarrow \,\,\lim_{q^2 \to 0} \,\Delta^{-1}(q) = \lim_{q^2 \to 0} \,(q^2 + m^2) \,\,\Longrightarrow \,\,\Delta^{-1}(0) = m^2\,.
\label{schmech}
\ee
This effect is interpreted as the 
dynamical generation of a mass, which is  
identified with the positive residue  
of the pole. 

In the absence of elementary scalar fields, massless poles may arise due to a variety of reasons~\cite{Zumino:1965rka,Jackiw:1973tr,Jackiw:1973ha,Cornwall:1973ts}. 
In Yang-Mills theories, the 
pole formation proceeds through 
the fusion of two gluons 
or of a ghost-antighost pair 
into a {\it color-carrying} scalar bound state 
of vanishing mass~\cite{Eichten:1974et,Smit:1974je,Poggio:1974qs,Cornwall:1981zr,Aguilar:2011xe,Ibanez:2012zk,Aguilar:2015bud,Aguilar:2017dco,Eichmann:2021zuv,Aguilar:2021uwa}, to be denoted by 
$\Phi^{a}$. In the case of the gluon fusion that we will consider throughout this work, the formation of this bound state is controlled  by a special BSE; details of the solutions obtained 
are given in the following sections. 

\begin{figure}[!ht]
  \centering
  \includegraphics[width=\textwidth]{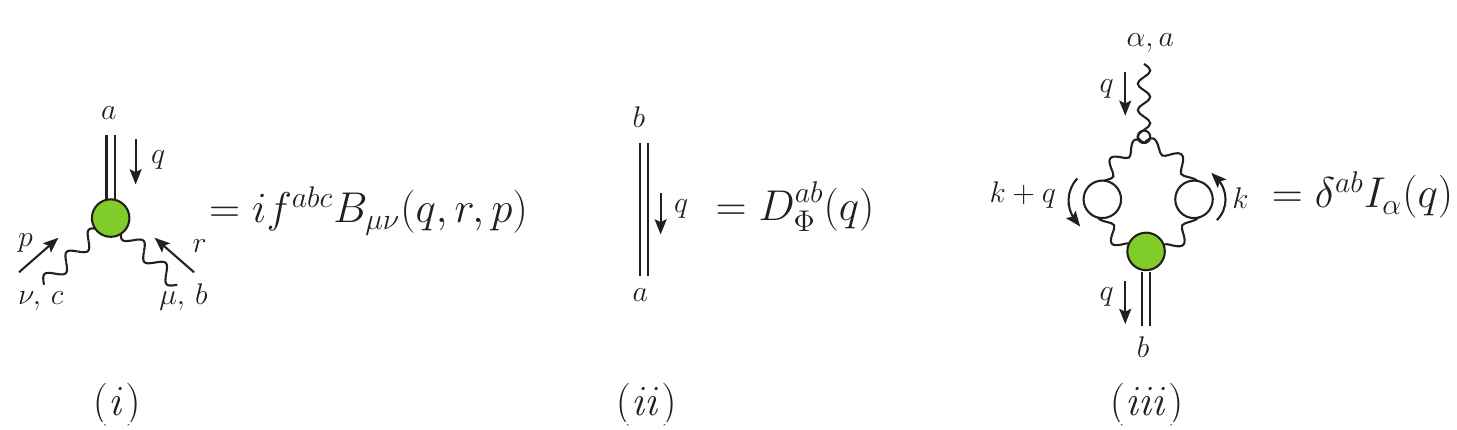}
  \caption{Left: The effective vertex $B^{abc}_{\mu\nu}(q,r,p)$, 
  with Lorentz, color, and momentum conventions indicated. Center: The propagator of the massless composite scalar, $D_{\Phi}^{ab}(q)$. Right: Gluon-scalar transition amplitude, $I^{ab}_{\alpha}(q)$.}\label{fig:B_def}
\end{figure}

There are three main structures 
associated with this nonperturbative process, which are diagrammatically depicted 
in \fig{fig:B_def}: 

({\it i}) The effective vertex describing the 
interaction between $\Phi^{a}$ and two gluons, 
denoted by 
\be
B^{abc}_{\mu\nu} (q,r,p) = i f^{abc} B_{\mu\nu} (q,r,p) \,.
\label{eq:Bdef}
\ee

({\it ii}) The propagator of the 
massless composite scalar, denoted  
by 
\be
D^{ab}_{\Phi}(q) = 
\frac{i \delta^{ab}}{q^2} \,. 
\label{eq:scpr}
\ee

({\it iii}) The transition amplitude, 
$I^{ab}_{\alpha}(q) = \delta^{ab} I_{\alpha}(q)$, 
connecting a gluon 
$A^{a}_{\alpha}$ with a scalar $\Phi^{b}$; 
Lorentz invariance imposes that 
\be
I_{\alpha}(q) = q_{\alpha} I(q) \,,
\label{eq:theI}
\ee
where $I(q)$ is a scalar form factor. 

\begin{figure}[h!]
  \includegraphics[width=0.7\textwidth]{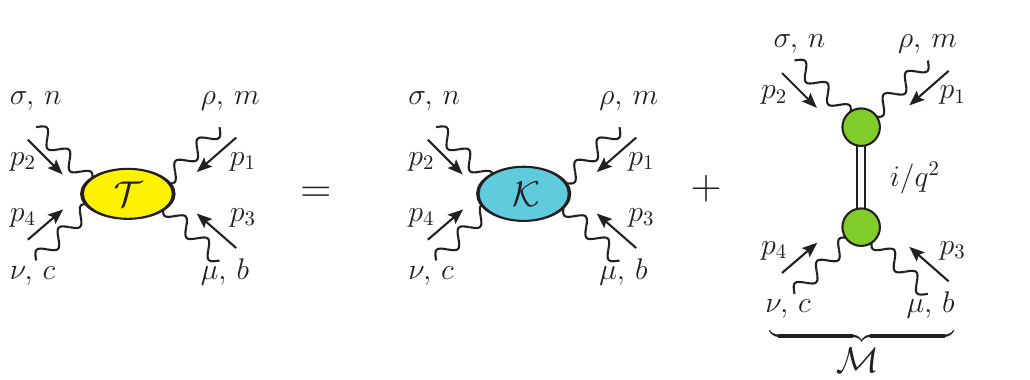}
  \caption{Decomposition of the four-gluon scattering kernel, ${\cal T}^{mnbc}_{\rho\sigma\mu\nu}(p_1,p_2,p_3,p_4)$, into a regular part, ${\cal K}^{mnbc}_{\rho\sigma\mu\nu}(p_1,p_2,p_3,p_4)$, and a massless pole part, ${\cal M}^{mnbc}_{\rho\sigma\mu\nu}(p_1,p_2,p_3,p_4)$, according to \1eq{kern_decomp}. }\label{fig:Kern_pole}
\end{figure}

\begin{figure}[t]
  \includegraphics[width=0.8\textwidth]{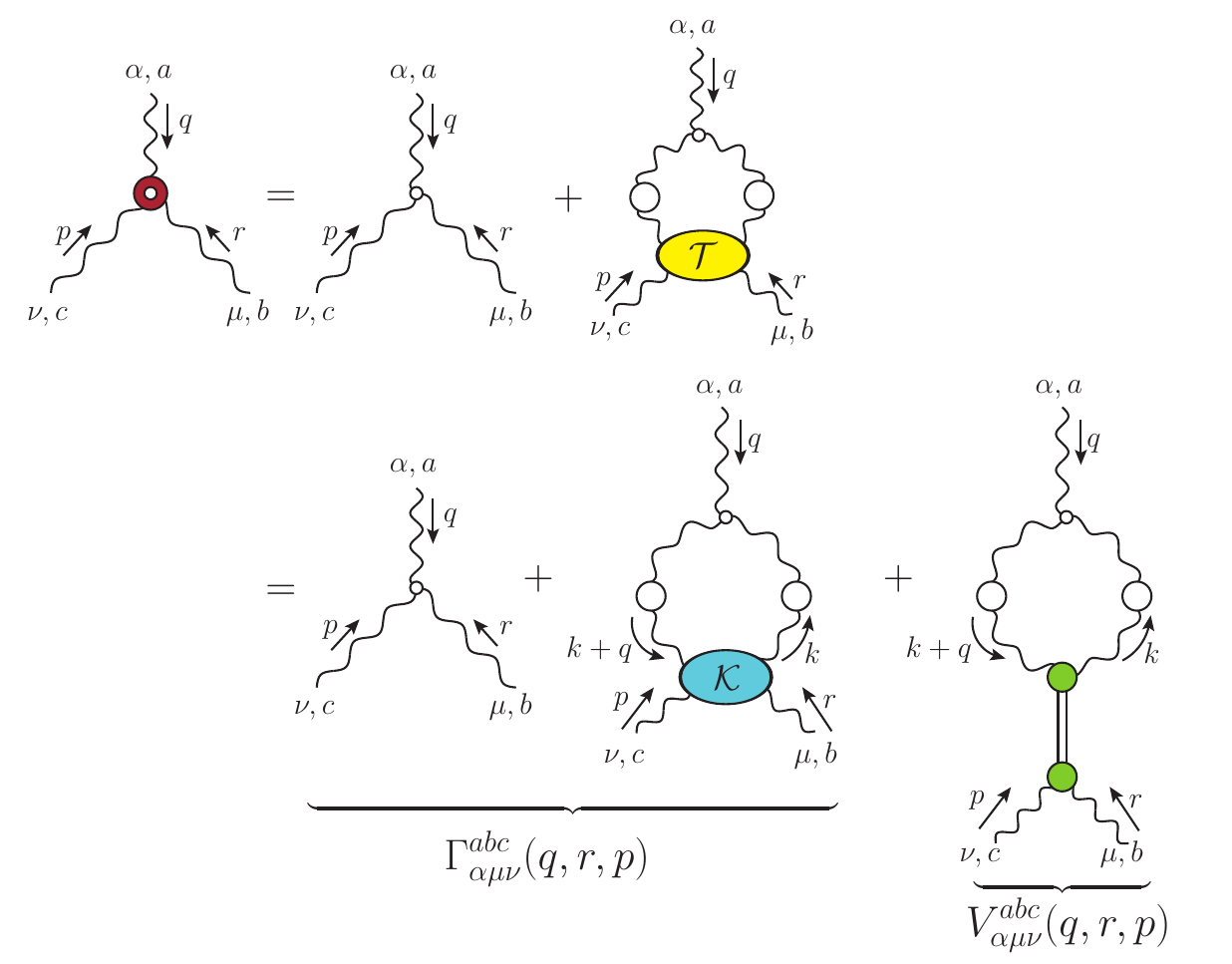}
\caption{{\it First line}: SDE for the three-gluon vertex.
{\it Second line}: The pole induced to the three-gluon vertex due to the ${\cal M}$
component of ${\cal T}$ in \1eq{kern_decomp} (see also \fig{fig:Kern_pole}).} 
\label{fig:BKBSE}
\end{figure}

The emergence of 
items ({\it i})-({\it iii})
modifies 
profoundly the structure 
of the four-gluon kernel, 
${\cal T}^{mnbc}_{\rho\sigma\mu\nu}(p_1,p_2,p_3,p_4)$,
entering in the 
skeleton expansion of the three-gluon vertex SDE, see \fig{fig:BKBSE}. In particular, 
as shown diagrammatically in \fig{fig:Kern_pole}, we have that 
\be
{\cal T}^{mnbc}_{\rho\sigma\mu\nu}(p_1,p_2,p_3,p_4)
= {\mathcal K}^{mnbc}_{\rho\sigma\mu\nu}(p_1,p_2,p_3,p_4) + {\mathcal M}^{mnbc}_{\rho\sigma\mu\nu}(p_1,p_2,p_3,p_4) \,,
\label{kern_decomp}
\ee
where ${\mathcal K}$
denotes the regular, pole-free term, while ${\mathcal M}^{mnbc}$
is given by 
\be
{\mathcal M}^{mnbc}_{\rho\sigma\mu\nu}(p_1,p_2,p_3,p_4) 
= B^{mnx}_{\rho\sigma} (p_1,p_2,q) D^{xe}_{\Phi}(q) B^{ebc}_{\mu\nu} (p_3,p_4,-q) \,,
\label{kernM}
\ee
with \mbox{$ q= p_1+p_2  =-p_3-p_4$}.

We emphasize that, despite appearances, the 
kernel ${\cal T}$
is one-particle irreducible, because the propagator $D_{\Phi}(q)$, 
present in the term ${\mathcal M}$,
does not correspond to a fundamental field, but rather 
to a composite 
excitation,  
representing 
a multitude of 
gluon interactions. 

The way how the pole in  ${\mathcal M}$ causes  
the saturation of the 
gluon propagator is through 
the three-gluon vertex. 
In particular, consider the 
SDE for the three-gluon vertex,
whose diagrammatic representation 
is given in \fig{fig:BKBSE}; note 
that only the 
diagrams relevant to our purposes 
are shown (first row).
The insertion of 
${\mathcal M}^{mnbc}$
in this SDE
generates a new term for 
the three-gluon vertex
(second row), to be denoted by 
$V_{\alpha\mu\nu}(q,r,p)$. 
Specifically, $\fatg_{\alpha\mu\nu}(q,r,p)$
assumes the general 
form\footnote{Due to the Bose symmetry of $\fatg_{\alpha\mu\nu}(q,r,p)$,
poles appear also in the 
channels carrying momenta $r$ or $p$,
and are 
proportional to 
$I_{\mu}(r)$ or 
$I_{\nu}(p)$.
However, when these legs  
are internal to Feynman diagrams,  
they get annihilated due to the contraction with the corresponding 
Landau-gauge propagators.}
\be
\fatg_{\alpha\mu\nu}(q,r,p) = \g_{\alpha\mu\nu}(q,r,p) + V_{\alpha\mu\nu}(q,r,p)\,,
\label{3g_split}
\ee
where $\g_{\alpha\mu\nu}(q,r,p)$ represents the pole-free 
component, while 
$V_{\alpha\mu\nu}(q,r,p)$ 
is given by
\be
V_{\alpha\mu\nu}(q,r,p) = I_{\alpha}(q) \left(\frac{i}{q^2}\right) iB_{\mu\nu}(q,r,p) \,,
\label{eq:Valt}
\ee
as may be read off from \fig{fig:BKBSE}. 
Then, due to \1eq{eq:theI}, the term  
$V_{\alpha\mu\nu}(q,r,p)$ becomes 
\be
V_{\alpha\mu\nu}(q,r,p) = - \left(\frac{q_{\alpha}}{q^2}\right) I(q) B_{\mu\nu}(q,r,p) \,.
\label{eq:Valt2}
\ee
Evidently, $V_{\alpha\mu\nu}(q,r,p)$ satisfies 
$
P^{\alpha'\alpha}(q) V_{\alpha\mu\nu}(q,r,p) =0$; in that sense, the poles 
are said to be 
{\it ``longitudinally coupled''}.

\begin{figure}[!ht]
  \includegraphics[scale=0.6]{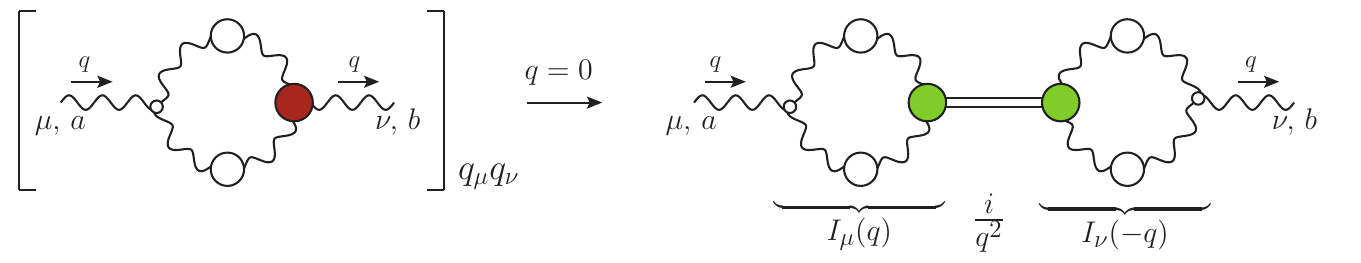}
  \caption{ Relation between the gluon mass and the transition amplitude, $I_{\alpha}(q)$. }\label{fig:squared_diag}
\end{figure}

Now, since 
the SDE controlling the momentum evolution of the 
gluon propagator depends on the vertex $\fatg_{\alpha\mu\nu}(q,r,p)$, the component  
$V_{\alpha\mu\nu}(q,r,p)$
provides the massless pole 
required for the activation of the SM.
Specifically, focusing on the 
part of the gluon propagator proportional to $q_{\mu}q_{\nu}$, 
we obtain the characteristic diagram shown in \fig{fig:squared_diag}, 
composed by the ``square'' of 
the transition amplitude 
$I_{\alpha}(q)$. In particular, 
in the limit $q \to 0$, we have 
the central result~\cite{Aguilar:2011xe,Ibanez:2012zk,Aguilar:2015bud}
\be
m^2 = g^2 I^2 \,,
\label{glmf}
\ee
where we have defined the 
short-hand notation $I := I(0)$.
%

Let us finally  
emphasize that, in order to elucidate the novel aspects of the method,  
in Figs.~\ref{fig:BKBSE} and \ref{fig:squared_diag} 
we have omitted all diagrams 
containing ghosts as internal lines or 
four-gluon vertices. This omission poses 
no problem for the following two reasons. 
First, the omitted graphs do not interfere  
in any way with the results derived 
in what follows; for a brief discussion, see Sec.~\ref{sec:conc}. 
Second, we do not rely on the 
vertex SDE for an accurate 
description of the 
form factor $\Ls$ [see \1eq{PPGamma}], appearing  
in some of the central formulas;
instead, we use the available lattice results for this quantity~\cite{Athenodorou:2016oyh,Boucaud:2017obn,Sternbeck:2017ntv,Aguilar:2021lke,Aguilar:2021okw,Maas:2020zjp,Catumba:2021hng,Catumba:2021yly,Pinto-Gomez:2022brg}.

\section{Setting the stage: Key results and main properties}\label{sec:key}

In this section we present 
a collection of 
relations and properties 
that serve as prerequisites for 
deriving the main results of this work.  
Note that, while several of these items have already 
appeared in the literature, 
here we focus on 
aspects that are especially pertinent  
for the task at hand.

\subsection{An important limit}\label{subsec:q0lim}

The kinematic limit relevant for 
the gluon mass generation is 
that of vanishing momentum transfer, or, in terms of the propagator diagram in 
\fig{fig:squared_diag}, the limit 
$q\to 0$. It is therefore important to 
determine the behavior of the 
effective vertex $B_{\mu\nu}(q,r,p)$
in that limit. 
 
The general tensorial 
decomposition of this vertex is given by 
\be 
B_{\mu\nu}(q,r,p) = B_1 \, g_{\mu\nu}  + B_2\, r_\mu r_\nu  + B_3 \, p_\mu p_\nu  +  B_4 \, r_\mu p_\nu  + B_5 \,  p_\mu r_\nu  \,,
\label{eq:Bdec}
\ee
and $B_i := B_i(q,r,p)$.
The Bose symmetry of the full vertex 
$B^{abc}_{\mu\nu} (q,r,p)$ 
under the exchange of the two incoming gluons imposes that 
$B^{abc}_{\mu\nu} (q,r,p) = 
B^{acb}_{\nu\mu}(q,p,r)$, 
from which follows that  
\mbox{$B_{\mu\nu}(q,r,p) =- 
B_{\nu\mu}(q,p,r)$}. Then, setting  in this last relation 
$q=0$ \, ($p=-r$), we obtain 
\mbox{$B_{\mu\nu}(0,r,-r) =- 
B_{\nu\mu}(0,r,-r)$}. 
Given that 
\be 
B_{\mu\nu}(0,r,-r) = B_1(0,r,-r) \, g_{\mu\nu} +  C_1(0,r,-r) r_\mu r_\nu  \,,
\label{B0rr}
\ee 
where $C_1:=B_2+B_3-B_4 -B_5$,
we conclude immediately that 
\be
B_1(0,r,-r) = 0 = C_1(0,r,-r)\,\,\,
\Longrightarrow \,\,
B_{\mu\nu}(0,r,-r) =0 \,.
\label{Bat0}
\ee

\begin{figure}[ht]
\includegraphics[width=0.45\textwidth]{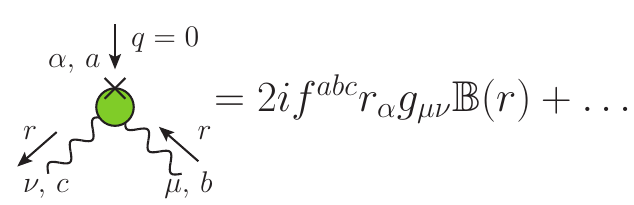}
\caption{Diagrammatic representation of \1eq{theBB}, corresponding to the first nonvanishing term in the Taylor expansion of $B_{\mu\nu}^{abc}(q,r,p)$ around $q = 0$.}
\label{fig:B_derivative}
\end{figure}

Let us then 
consider the Taylor expansion 
of $B_{\mu\nu}(q,r,p)$ around $q=0$,  
\be 
B_{\mu\nu}(0,r,-r) = q^{\alpha} {\cal B}_{\alpha\mu\nu}(0,r,-r) + \ldots \,,
\label{BandB}
\ee
where we have 
introduced 
the shorthand notation
\be 
{\cal B}_{\alpha\mu\nu}(0,r,-r) :=
\left[\frac{\partial}{\partial q^\alpha} B^{\mu\nu}(q,r,\, -r-q) \right]_{q = 0} \,,
\label{Bcal_def} 
\ee
accompanied by the diagrammatic representation 
shown in \fig{fig:B_derivative}, and 
the ellipsis denotes 
terms of higher order in $q$.
When taking the limit indicated in \1eq{Bcal_def} we consider  
$r$ to be independent of $q$, in which case, 
due to conservation of four-momentum, $p$ depends on $q$, 
since \mbox{$p=-q-r$};
therefore,  
$\partial p^{\mu}/\partial q^\alpha = -g^{\mu}_\alpha$. 

The next step is to contract
${\cal B}_{\alpha\mu\nu}(0,r,-r)$ by 
$P_{\mu'}^{\mu}(r) P_{\nu'}^{\nu}(r)$; 
this contraction appears naturally 
in the 
diagrams where 
$B_{\mu\nu}(q,r,p)$ is inserted, because 
the two gluon propagators attached to it are in the Landau gauge.
Then, since 
the tensorial decomposition of 
${\cal B}_{\alpha\mu\nu}(0,r,-r)$ is given by 
\be 
{\cal B}_{\alpha\mu\nu}(0,r,-r) = 
{\cal B}_1 (r) \,r_{\alpha} g_{\mu\nu} 
+
{\cal B}_2 (r)
\left[r_{\alpha} g_{\mu\nu} + r_{\mu} g_{\alpha\nu} + r_{\nu} g_{\mu\alpha}\right] 
+ {\cal B}_3 (r) \, r_{\alpha} r_{\mu} r_{\nu} \,,
\label{Bcaltens} 
\ee
only the first term survives this contraction,
\ie 
\be 
{\cal B}_{\alpha\mu\nu}(0,r,-r)
P_{\mu'}^{\mu}(r) P_{\nu'}^{\nu}(r)
= {\cal B}_1 (r) \,r_{\alpha} 
P_{\mu'\nu'}(r) \,.
\label{Bsurv} 
\ee

It is now straightforward to establish that 
\be 
r_{\alpha} {\cal B}_1 (r)  
= \left[\frac{\partial}{\partial q^\alpha} B_1(q,r,\, -r-q) \right]_{q = 0} = 2 r_{\alpha}
\underbrace{\left[\frac{\partial {B}_1(q,r,p)}{\partial p^2} \right]_{q = 0}}_{{\mathbb B}(r)} \,,
\label{cb2}
\ee 
or
\be
{\cal B}_1 (r) = 2 {\mathbb B}(r) \,,
\label{morebees}
\ee
from which follows that 
\be
{\cal B}_{\alpha\mu\nu}(0,r,-r) 
= 2 \,\Bfat(r) \, r_\alpha g_{\mu\nu} 
+ \ldots \,,
\label{theBB} 
\ee
and then, from \1eq{BandB}, 
\be 
B_{\mu\nu}(0,r,-r) = 2 \, (q\cdot r )\,
\Bfat(r) \, g_{\mu\nu}
+ \ldots \,,
\label{BqB}
\ee
where the ellipses denote 
terms that get annihilated upon the 
aforementioned contraction. 
The diagrammatic representation of 
\1eq{theBB}
is shown in
\fig{fig:B_derivative}, and will be used in the analysis that 
follows.

\subsection{The gluon-scalar transition amplitude}\label{subsec:theI}

Given the central relation of 
\1eq{glmf}, we next 
derive a convenient expression 
that allows us to compute the form factor $I$ in terms of 
the basic quantities entering in its diagrammatic definition, see  
\fig{fig:B_def}, item ({\it iii}). 

For general momentum $q$, the 
transition amplitude
${I}^{\alpha}(q)$ is given by 
\be
I^{\alpha}(q) =  -\frac{iC_{\rm A}}{2} 
\int_k \Gamma_{\!0}^{\alpha\beta\lambda}(q,k,-k-q) \Delta_{\beta\mu}(k) \Delta_{\lambda\nu}(k+q)  B^{\mu\nu}(-q,-k,k+q)\,,
\label{I1}
\ee
where \1eq{eq:Bdef} was employed,
the symmetry factor 
$\frac{1}{2}$ 
has been supplied, and $C_{\rm A}$ is the Casimir eigenvalue of the adjoint representation [$N$ for SU($N$)]. In addition, we use the shorthand notation
\be 
\int_{k} := \frac{1}{(2\pi)^4} \int \!\!{\rm d}^4 k \,, \label{msr}
\ee
where 
the use of a symmetry-preserving regularization scheme is implicitly assumed.

Then, by virtue of \1eq{eq:theI}, 
it is elementary to establish that
\be 
I = \frac{1}{4}\left[ \frac{\partial I^\alpha(q)}{\partial q^\alpha} \right]_{q = 0} \,.
\label{Ider} 
\ee
So, from \1eq{I1} we obtain 
\begin{align} 
4I =&\, -\frac{iC_{\rm A}}{2} \int_k \left[ \frac{\partial}{\partial q^\alpha}\Gamma_{\!0}^{\alpha\beta\lambda}(q,k,-k-q) \Delta_{\beta\mu}(k) \Delta_{\lambda\nu}(k+q) \right]_{q = 0} \!\!\!\! \!\! B^{\mu\nu}(0,-k,k) \nonumber\\
&\, -\frac{iC_{\rm A}}{2}  \int_k \Gamma_{\!0}^{\alpha\beta\lambda}(0,k,-k)  \Delta_{\beta}^{\mu}(k) \Delta_{\lambda}^{\nu}(k) 
\, 
{\cal B}_{\alpha\mu\nu}(0,-k,k) \,,\label{I_scalar_step1}
\end{align}
and since  $B_{\mu\nu}(0,-k,k) =0$ 
[see \1eq{Bat0}],  
we have 
\be
4I = -\frac{iC_{\rm A}}{2}  \int_k 
\Gamma_{\!0}^{\alpha\beta\lambda}(0,k,-k)  \Delta_{\beta}^{\mu}(k) \Delta_{\lambda}^{\nu}(k) 
\, 
{\cal B}_{\alpha\mu\nu}(0,-k,k) \,.
\label{I_scalar_step2}
\ee
To further evaluate \1eq{I_scalar_step1},
use \1eq{bare3g} to get  
\be 
\Gamma_{\!0}^{\alpha\beta\lambda}(0,k,-k) = 2 k^\alpha g^{\beta\lambda} - k^\lambda g^{\alpha\beta} - k^\beta g^{\alpha\lambda} \,,
\label{Gtree0}
\ee
and then \1eq{theBB} (with $r \to -k$), 
to finally obtain
\be 
I = -\frac{3iC_{\rm A}}{2} \int_k k^2 \Delta^2(k) \Bfat(k) \,.
\label{I_scalar_step3}
\ee

\begin{figure}[ht]
\includegraphics[width=0.45\textwidth]{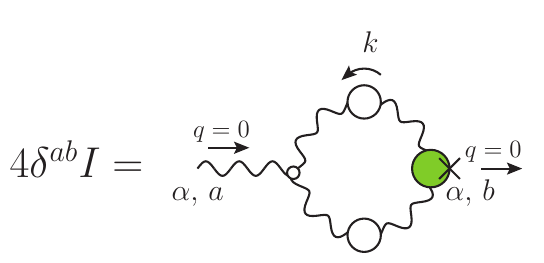}
\caption{Diagrammatic definition of the scalar form factor, $I$, of the transition amplitude.}
\label{fig:I_from_B_fat}
\end{figure}

\subsection{SDE of the three-gluon vertex and its soft gluon limit}\label{subsec:theL}

\begin{figure}[!ht]
\includegraphics[scale=0.7]{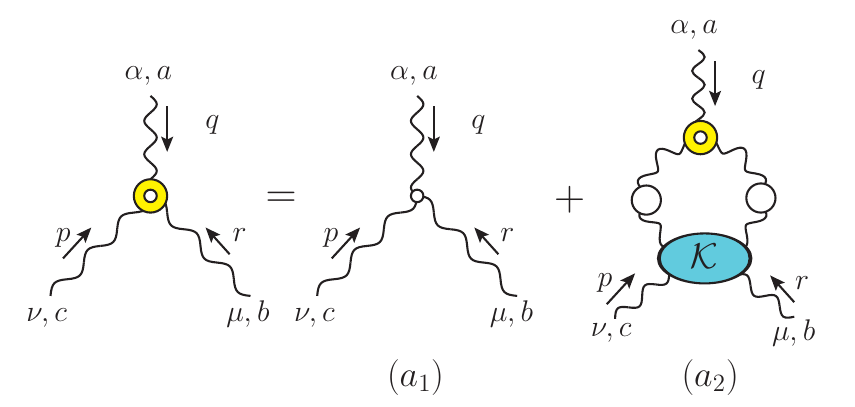}
\caption{SDE for the regular part of the three-gluon vertex, $\g^{abc}_{\alpha\mu\nu}(q,r,p)$, in BS form.}
\label{fig:L_BSE}
\end{figure}

The soft-gluon limit of the SDE for  
$\Gamma_{\alpha\mu\nu}^{abc}(q,r,p)$ is of central importance for the ensuing analysis; 
indeed, as will be shown in detail 
in Sec.~\ref{sec:trick}, 
its proper use 
leads to considerable simplifications 
at the level of the mass equation. 
In order to elucidate this important aspect,  
we employ a simplified version of the SDE, 
which contains precisely the structure required 
for certain crucial relations to be triggered. 
Specifically, we consider the SDE
represented in \fig{fig:L_BSE}, 
composed only by the tree-level contribution 
$(a_1)$ and the diagram ($a_2$), 
while 
diagrams where the 
gluon with momentum $q$ couples to a 
ghost-gluon or a four-gluon vertex are omitted\footnote{The role of these terms will be discussed in section Sec.~\ref{sec:conc}.}. 
The reason why diagram ($a_2$) is singled out  
is because its  
kernel ${\cal K}$  appears also in 
the BSE for $\Bfat(r)$, shown in 
\fig{fig:B_BSE}.
This, in turn, will be 
instrumental for the implementation of the 
renormalization procedure outlined in 
Sec.~\ref{sec:trick}

\begin{figure}[!ht]
  \centering
  \includegraphics[width=\textwidth]{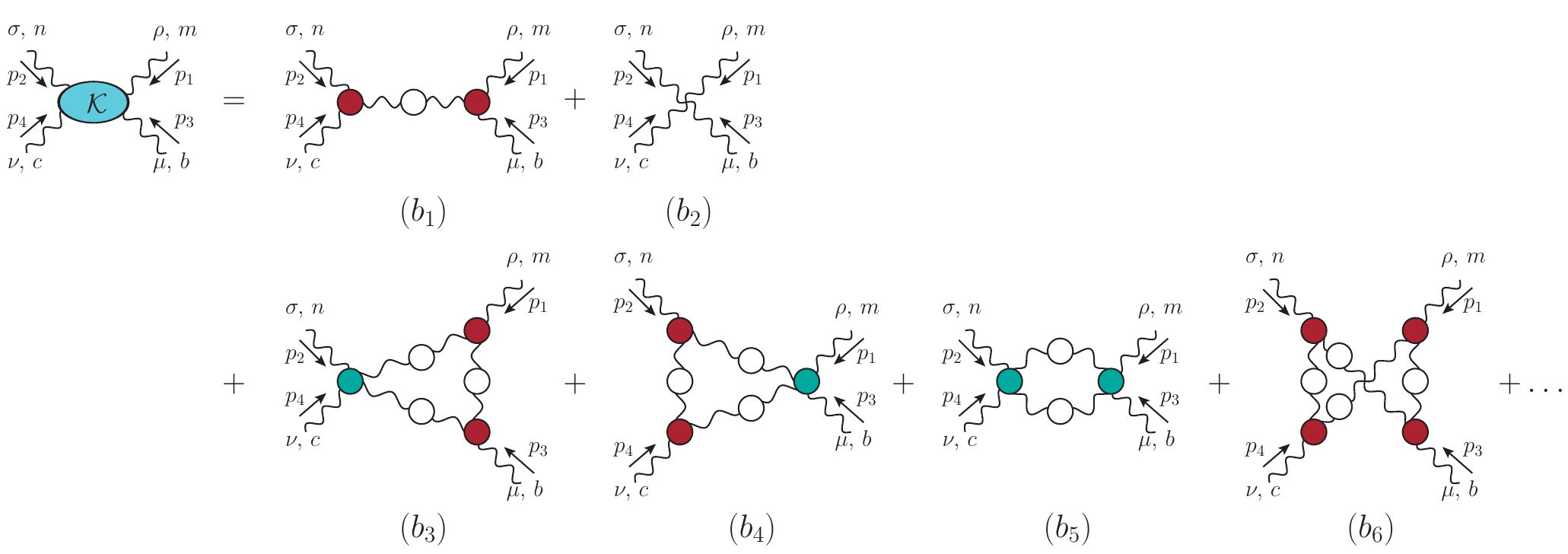}
\caption{Skeleton expansion of the scattering kernel, ${\cal K}^{mnbc}_{\rho\sigma\mu\nu}(p_1,p_2,p_3,p_4)$. The one-gluon exchange is highlighted and the ellipsis denotes contributions with two or more loops. }\label{fig:Kern_diags}
\end{figure}

Note that, in the standard version of 
this SDE,  
the three-gluon 
vertex 
in diagram $(a_2)$
is kept at tree level.
The form employed here,
with the three-gluon vertex 
fully-dressed, 
corresponds to the BSE 
analogue of this 
equation~\cite{Berges:2004pu,Carrington:2010qq,York:2012ib,Mueller:2015fka,Williams:2015cvx,Aguilar:2023qqd}. 
Consequently, the 
skeleton expansion of the 
kernel $\cal K$, given in 
\fig{fig:Kern_diags},
does not contain certain classes of diagrams
(\eg ladder graphs) in 
order to avoid overcounting. In fact, ghost loops aside,  the diagrams of  \fig{fig:Kern_diags} comprise precisely the kernel of the standard glueball BSE~\cite{Huber:2020ngt,Huber:2021yfy}. 
The main advantage of this 
version of the SDE 
is that the additional fully-dressed vertex 
absorbs the vertex renormalization 
$Z_3$, defined in \1eq{renconst},
which otherwise would be 
multiplying the tree-level vertex; 
as a result,
renormalization may be carried out subtractively 
rather than multiplicatively.
In particular, the 
validity of the 
special formula in 
\1eq{BSE_Lren} hinges precisely on 
graph ($a_2$) having a fully-dressed three-gluon vertex.

We next factor out the color structure $f^{abc}$ by setting  
$\Gamma_{\alpha\mu\nu}^{abc}(q,r,p) = f^{abc}\Gamma_{\alpha\mu\nu}(q,r,p)$, and focus on the soft gluon limit,  
$\Gamma_{\alpha\mu\nu}(0,r,-r)$.  
The reason for this choice 
is that, eventually, 
this vertex will be used in 
the graph of \1eq{fig:squared_diag}, 
which is evaluated precisely in this 
kinematic limit. 
Setting $q=0$ into 
\fig{fig:L_BSE}, 
we get
\be 
\g_{\alpha\mu\nu}^{abc}(0,r,-r) = \g_{\!0\,\alpha\mu\nu}^{abc}(0,r,-r) \,+\, \int_k \Gamma^{axe}_{\alpha\gamma\delta}(0,k,-k)
\Delta_{xm}^{\gamma\rho}(k)
\Delta_{en}^{\delta\sigma}(k)
{\cal K}^{mnbc}_{\rho\sigma\mu\nu}(-k,k,r,-r) \,. \label{L_BSE_step1}
\ee
%

To proceed further, 
consider the 
contraction of $\Gamma_{\alpha\mu\nu}(0,r,-r)$
by $P_{\mu'}^{\mu}(r) P_{\nu'}^{\nu}(r)$; it is 
elementary to establish that this vertex projection is described by a single form factor, 
denoted by $\Ls(r)$, namely~\cite{Aguilar:2021okw,Aguilar:2021uwa}
\be
P_{\mu'}^{\mu}(r) P_{\nu'}^{\nu}(r) 
\Gamma_{\alpha\mu\nu}(0,r,-r) 
= 2 \Ls(r) r_{\alpha} P_{\mu'\nu'}(r) \,.
\label{PPGamma}
\ee
Note that, 
in the Landau gauge that we employ, the $\Gamma^{axe}_{\alpha\gamma\delta}(0,k,-k)$ on the r.h.s. of \1eq{L_BSE_step1} is automatically contracted by two transverse projectors, thus triggering \1eq{PPGamma} therein. Then, 
contracting both sides of \1eq{L_BSE_step1} 
by $P^\mu_{\mu'}(r)P^\nu_{\nu'}(r)$, 
we obtain 
\begin{align} 
f^{abc}\Ls(r)r_\alpha P_{\mu'\nu'}(r) =&\, f^{abc} r_\alpha P_{\mu'\nu'}(r) \label{L_BSE_step2} \\
- &\, f^{amn} \int_k k_{\alpha} \Ls(k)
\Delta^2(k)
P^{\rho\sigma}(k){\cal K}^{mnbc}_{\rho\sigma\mu\nu}(-k,k,r,-r)P^\mu_{\mu'}(r)P^\mu_{\mu'}(r) \,. \nonumber
\end{align}

Lastly, we contract \1eq{L_BSE_step2} with $f^{abc}r^\alpha g^{\mu'\nu'}$ to eliminate the color and Lorentz indices. Using that $P^\mu_\mu(r) = 3$, and that  for SU($N$),
$f^{abc}f^{abc} = C_{\rm A}(N^2 - 1)$, we obtain
\be
\Ls(r) =\, 1 + \alpha_s\! \int_k  k^2 \Delta^2(k) \,{K}(r,k) \,\Ls(k) \,, \label{BSE_L}
\ee
with the kernel $K(r,k)$ defined as
\be
\alpha_s K(r,k) :=
- \frac{(r\cdot k)}{c \,r^2k^2} f^{abc}f^{amn}P^{\mu\nu}(r)P^{\rho\sigma}(k)\,
{\cal K}^{mnbc}_{\rho\sigma\mu\nu}(-k,k,r,-r) \,,
\label{K_def}
\ee
where we explicitly factor out the strong charge $\alpha_s = g^2/4\pi$. The numerical factor 
\mbox{$c:= 3 C_{\rm A}( N^2 - 1 )$} 
arises from the projections of Lorentz and color indices; 
for $N = 3$, $c = 72$.
%

\subsection{Displacement of the Ward identity of the three-gluon vertex}\label{subsec:disp}

In addition to triggering the SM,  the massless poles  
are crucial for preserving gauge invariance: 
their inclusion in the off-shell interaction vertices guarantees that the Slavnov-Taylor identities (STIs)~\cite{Taylor:1971ff,Slavnov:1972fg} of the theory retain the exact same form before and after mass generation~\cite{Aguilar:2023mdv}.
In the case of the three-gluon 
vertex, 
the presence 
of these poles 
gives rise to a characteristic effect, 
namely 
the displacement of the 
WI satisfied by the soft-gluon form factor 
$\Ls (r)$~\cite{Aguilar:2021uwa,Aguilar:2022thg}, introduced in 
\1eq{PPGamma}.

In particular, let us denote by 
$\Ls (r)$ and 
$\Ls^{\star} (r)$ the corresponding form factor when 
the SM is turned on and off, respectively. When the SM is off, 
the three-gluon vertex is simply given by $\Gamma_{\alpha \mu \nu}(q,r,p)$, and the STI it satisfies 
is given by~\cite{Marciano:1977su,Ball:1980ax,Davydychev:1996pb,Gracey:2019mix} 
\be
q^\alpha \Gamma_{\alpha \mu \nu}(q,r,p)
\,{\cal T}_{\mu'\nu'}^{\mu\nu}(r,p)
\!=\! F(q^2)\,[ 
 \Delta^{-1}(p^2) H_{\nu\mu}(p,q,r) - \Delta^{-1}(r^2) H_{\mu\nu}(r,q,p)]
\,{\cal T}_{\mu'\nu'}^{\mu\nu}(r,p)\,,
\label{STI} 
\ee
where ${\cal T}_{\mu'\nu'}^{\mu\nu}(r,p) :=
P_{\mu'}^{\mu}(r)P_{\nu'}^{\nu}(p)$,   
$F(q)$ denotes the ghost dressing function, defined from the ghost propagator, $D^{ab}(q)$, by $D^{ab}(q) = i\delta^{ab}F(q)/q^2$, and $H_{\nu\mu}(p,q,r)$ is the 
ghost-gluon kernel. 

Taking the limit $q\to 0$ of both sides of \1eq{STI}, also known as the ``soft-gluon limit'', 
one obtains the 
corresponding WI, which simply states that 
\be 
\Ls^{\!\star}(r^2) = 
[{\rm r.h.s. \,\, of \, \,STI \,\,in \,\, \1eq{STI}}]_{q\to 0} \, \,. 
\label{Lstar}
\ee 
When the SM is activated, 
the vertex that satisfies 
the STI of \1eq{STI} 
is the {\it full} $\fatg_{\alpha \mu \nu}(q,r,k)$, given by
\1eq{3g_split}.  In particular,  
with the aid of 
\1eq{eq:Valt2},
the l.h.s. 
of \1eq{STI} becomes 
\be
q^\alpha \fatg_{\alpha \mu \nu}(q,r,p)
\,{\cal T}_{\mu'\nu'}^{\mu\nu}(r,p)=
[q^\alpha \g_{\alpha\mu\nu}(q,r,p) - 
I(q) B_{\mu\nu}(q,r,p)]
{\cal T}_{\mu'\nu'}^{\mu\nu}(r,p) \,,
\ee
while the r.h.s. remains
exactly the same. 
Then, as has been shown in 
~\cite{Aguilar:2021uwa,Aguilar:2022thg},
the limit $q\to 0$ of the STI yields
\be
\Ls(r^2) =
\Ls^{\!\star}(r^2) + \Cfat(r^2) \,,
\label{3gldis}
\ee
where 
\be 
\Cfat(r) := - I \, \Bfat(r) \,,
\label{C_from_B}
\ee
is denominated the 
``{\it displacement function}''. 

The importance of this result 
is that one may determine  
$\Cfat(r)$ as the difference 
between the $\Ls(r^2)$, 
extracted directly from the lattice,
and the expression for 
$\Ls^{\!\star}(r^2)$ on the r.h.s. 
of \1eq{Lstar}, 
which is evaluated using lattice inputs for its various components~\cite{Aguilar:2021uwa,Aguilar:2022thg}.
The result for $\Cfat(r)$ 
that emerges through this procedure 
will serve as benchmark  
for the numerical analysis 
presented in Sec.~\ref{sec:res}.

\section{Renormalization: general considerations}\label{sec:rengen}

In this section we discuss in detail the renormalization of the key dynamical 
components that emerge from the activation of the SM. In particular, we will focus on the renormalization properties of the central 
quantities $I$ and $B_{\mu\nu}$.  

We focus exclusively on the gluon propagator and the 
three-gluon vertex, which are the two basic ingredients 
comprising the structures considered in this work.
Denoting by the index 
``R'' the renormalized 
quantities, we have 
\be
\Delta_{\s R}(q^2) = Z^{-1}_{A} \Delta(q^2) \,, \quad\quad\quad 
\fatg^{\alpha\mu\nu}_{\!\!\s R}(q,r,p) =  Z_3 \fatg^{\alpha\mu\nu}(q,r,p)\,, \quad\quad\quad
g_{\s R} = Z_g^{-1} g\,,
\label{renconst}
\ee
where $Z^{1/2}_{A}$ is the wave function 
renormalization constants of the gluon  
field, $Z_3$ the 
renormalization constant of the three-gluon vertex, 
and $Z_g$ the coupling renormalization constant,
given by 
\be
Z_g =  Z_3 Z_A^{-3/2} \,.
\label{eq:sti_renorm}
\ee 
Furthermore, we will introduce two additional 
renormalization constants, 
to be denoted by $Z_{I}$ and $Z_{B}$, 
which renormalize the quantities
$I$ and $B^{\mu\nu}(q,r,p)$, 
respectively, 
according to~\cite{Aguilar:2014tka} 
\be
I_{\s R} = Z^{-1}_{I} I \,,
\quad\quad
B_{\s R}^{\mu\nu}(q,r,p) =
Z^{-1}_{B} B^{\mu\nu}(q,r,p) \,.
\label{IBren}
\ee
The partial derivatives of 
$B^{\mu\nu}(q,r,p)$ also renormalize 
in the same way, namely
\be
{\cal B}_{\s R}^{\alpha\mu\nu}(0,r,-r) 
= Z^{-1}_{B} 
{\cal B}_{\s R}^{\alpha\mu\nu}(0,r,-r) \,,
\quad\quad
\Bfat_{\s R}(r) = Z^{-1}_{B} \Bfat(r) \,.
\label{derBren}
\ee
Given that both the $I$ and 
$B^{\mu\nu}$ are 
composed by the fundamental fields 
and vertices, it is natural to expect
that the $Z_{I}$ and $Z_{B}$
should be expressed in terms of the 
$Z_{A}$ and $Z_{3}$ introduced in 
\1eq{renconst}; indeed, we find that~\cite{Aguilar:2014tka} 
\be
Z_{I} = Z^{-1}_{3} Z_{A}  \,,
\quad\quad\quad
Z_{B} = Z^{-1}_{A}  \,.
\label{ZIZB}
\ee
To see how the relations in \1eq{ZIZB} arise, we 
first turn to \1eq{glmf}. 
Since $m^2 := \Delta^{-1}(0)$,
from the first relation in \1eq{renconst} we have that~\cite{Aguilar:2014tka,Aguilar:2015bud} 
\be
m^2 = Z^{-1}_A m^2_{\s R} \,.
\label{mren}
\ee
On the other hand, combining 
\1eq{glmf}, the first relation in \1eq{IBren}, 
and \1eq{eq:sti_renorm}, we have 
\be
m^2 =g^2 I^2 = Z^2_g \, Z^2_{I} \, 
\underbrace{g^2_{\s R} I^2_{\s R}}_{m^2_{\s R}} 
= Z^2_3 \, Z^{-3}_A \, Z^2_{I} \,m^2_{\s R} \,.
\label{mren2}
\ee
Then, the direct comparison of \2eqs{mren}{mren2} 
leads immediately to the first relation of 
\1eq{ZIZB}.

The second relation in \1eq{ZIZB} may be obtained 
from \1eq{C_from_B}, by noticing 
that, 
since $V_{\alpha\mu\nu}(q,r,p)$ is a component of the three-gluon vertex 
$\fatg^{\alpha\mu\nu}(q,r,p)$, 
it is renormalized 
as in \1eq{renconst}, namely
\be
V^{\alpha\mu\nu}(q,r,p)
=  Z^{-1}_3 \,
V^{\alpha\mu\nu}_{\!\!\s R}(q,r,p)
\,,
\label{V_ren}
\ee
and, therefore,
\be
\Cfat(k) = Z^{-1}_3 \,\Cfat_{\!\s R}(k) \,.
\label{Cfatren}
\ee
On the other hand, from \1eq{C_from_B}, using \1eq{IBren} and the first relation of 
\1eq{ZIZB}, we have that 
\be
\Cfat(k) = - Z_{I} Z_{B} \underbrace{I_{\s R}\Bfat_{\s R}(k)}_{-\Cfat_{\!\s R}(k)} =  
Z^{-1}_{3} Z_{A} Z_{B} \,\Cfat_{\!\s R}(k) \,.
\label{renCB}
\ee
Then, the comparison between 
\2eqs{Cfatren}{renCB} yields 
the second relation of \1eq{ZIZB}.

We next turn to the renormalization of the kernel
${\cal K}^{mnbc}_{\rho\sigma\mu\nu}
(q,r,p,t)$, appearing in 
\1eq{L_BSE_step1}, and later on in
\1eq{d1d2}.
To that end, notice that  
${\cal K}$ is a part of the 
four-gluon amplitude 
${\cal G}^{mnbc}_{\rho\sigma\mu\nu}
(q,r,p,t) = 
\langle 0 \vert T[\widetilde{A}^m_\rho(q) \widetilde{A}^n_\sigma(r) \widetilde{A}^b_\mu(p)\widetilde{A}^c_\nu(t)] \vert 0\rangle$, with the external 
legs amputated, \ie 
\be
{\cal G}(q,r,p,t) = \Delta(q)\Delta(r)\Delta(p)\Delta(t) {\cal K}(q,r,p,t)
+ \cdots 
\label{GKD}
\ee
where the ellipsis denotes 
the diagrams excluded when switching from the SDE to the BSE kernel,
as discussed in Sec.~\ref{subsec:theL}. 
Since, 
$A^{a \mu}_{\s R} = Z^{1/2}_{A}
A^{a \mu}$, 
the definition 
of ${\cal G}$ in terms of gauge fields implies that 
${\cal G}_{\s R}= Z^{-2}_{A} \, {\cal G}$.
Therefore, 
from \2eqs{GKD}{renconst}, we get 
\be
{\cal K}_{\s R}(q,r,p,t) = Z^{2}_{A} \,{\cal K}(q,r,p,t) \,,
\label{Kcalren}
\ee
and, since 
${\cal K} \sim \alpha_{s} K$, we have 
\be
{K}_{\s R}(q,r,p,t) = Z^{2}_{A}Z^{2}_{g} \,{K}(q,r,p,t) \,. 
\label{Kren}
\ee
From the above relations it is 
immediate to establish that 
the combinations 
$\Delta{\Bfat}$, 
$\Delta^2 {\cal K}$,
and $\alpha_{s} \Delta^2  K$
are
renormalization-group invariant (RGI), \ie
\be
\Delta \Bfat = 
\Delta_{\s R} \Bfat_{\s R}\,,
\qquad\qquad
\Delta^2 {\cal K} =\Delta^2_{\s R} \,{\cal K}_{\s R} \,,
\qquad\qquad 
\alpha_{s} \Delta^2  K
=
\alpha_{s}^{\s R}\Delta^2_{\s R} K_{\s R} \,.
\label{Krgi}
\ee
Note that, by virtue of
the renormalization rule
$B_{\s R}^{\mu\nu} =
Z_{A} B^{\mu\nu}$, 
given by 
\2eqs{IBren}{ZIZB},  
the kernel ${\mathcal M}$ 
defined in 
\1eq{kernM} 
renormalizes as 
${\cal M}_{\s R}= Z^{2}_{A} \,{\cal M}$,
\ie, exactly as 
the kernel ${\mathcal K}$ in 
\1eq{Kcalren}; this is 
consistent with the fact that both ${\mathcal K}$ and 
${\mathcal M}$ are parts 
of the same four-gluon kernel, see  
\1eq{kern_decomp} and \fig{fig:BKBSE}.

Armed with the above results, 
we may now derive a useful 
expression for the $I_{\s R}$,
using \1eq{I_scalar_step2} as our 
point of departure. 
Substituting the bare quantities
comprising \1eq{I_scalar_step2} 
by renormalized ones,  we have
\be
4 Z_{I} I_{\s R} =   
-\frac{iC_{\rm A}}{2}  Z^2_{A} Z_{B} \int_k 
\Gamma_{\!0}^{\alpha\beta\lambda}(0,k,-k)  \Delta_{{\s R}\,\beta}^{\mu}(k) \Delta_{{\s R}\,\lambda}^{\nu}(k) 
\, 
{\cal B}_{\!{\s R} \,\alpha\mu\nu}(0,-k,k) \,,
\label{I_scalar_ren1}
\ee
and, after 
employing \1eq{ZIZB},
\be
4I_{\s R} =   
-\frac{iC_{\rm A}}{2} Z_3 \int_k 
\Gamma_{\!0}^{\alpha\beta\lambda}(0,k,-k)  \Delta_{{\s R}\,\beta}^{\mu}(k) \Delta_{{\s R}\,\lambda}^{\nu}(k) 
\, 
{\cal B}_{\!{\s R} \,\alpha\mu\nu}(0,-k,k)
\,,
\label{I_scalar_ren2}
\ee
or, from 
\1eq{I_scalar_step3}, 
\be 
I_{\s R} = -\frac{3iC_{\rm A}}{2} Z_3 \int_k k^2 \Delta_{\s R}^2(k) \Bfat_{\s R}(k) \,.
\label{I_scalar_ren3}
\ee

We finally turn to the renormalization of \1eq{BSE_L}. 
By virtue of 
\1eq{PPGamma}, 
it is clear that 
$\Ls^{\!\s R}(r) = Z_3 \Ls(r)$.
Then, using 
\1eq{Krgi},
it is 
straightforward to show that 
the renormalized version of \1eq{BSE_L} is given by 
\be
\Ls^{\!\s R}(r) =\, Z_3 +   \alpha_{s}^{\s R}\!\! \int_k  k^2 \Delta^2_{\s R}(k) \,{K}_{\s R}(r,k) \,\Ls^{\!\s R}(k) \,. \label{BSE_Lren}
\ee
As announced in 
Sec.~\ref{subsec:theL}, 
the  renormalization required for \1eq{BSE_Lren} 
is subtractive. 

In what follows 
we will suppress the index ``R'' in order to avoid notational clutter.

\section{Non-linear Bethe-Salpeter equation for pole formation}\label{sec:BSE}

In this section we derive the BSE satisfied by the amplitude 
$\Bfat(r)$. This type of equation has been 
derived in the literature before~\cite{Aguilar:2011xe,Ibanez:2012zk,Aguilar:2015bud,Aguilar:2017dco,Aguilar:2021uwa}; however,
in contradistinction to these earlier 
versions, the present BSE is not linearized, having 
its cubic nature fully retained.

\begin{figure}[!ht]
  \centering
  \includegraphics[width=0.9
\textwidth]{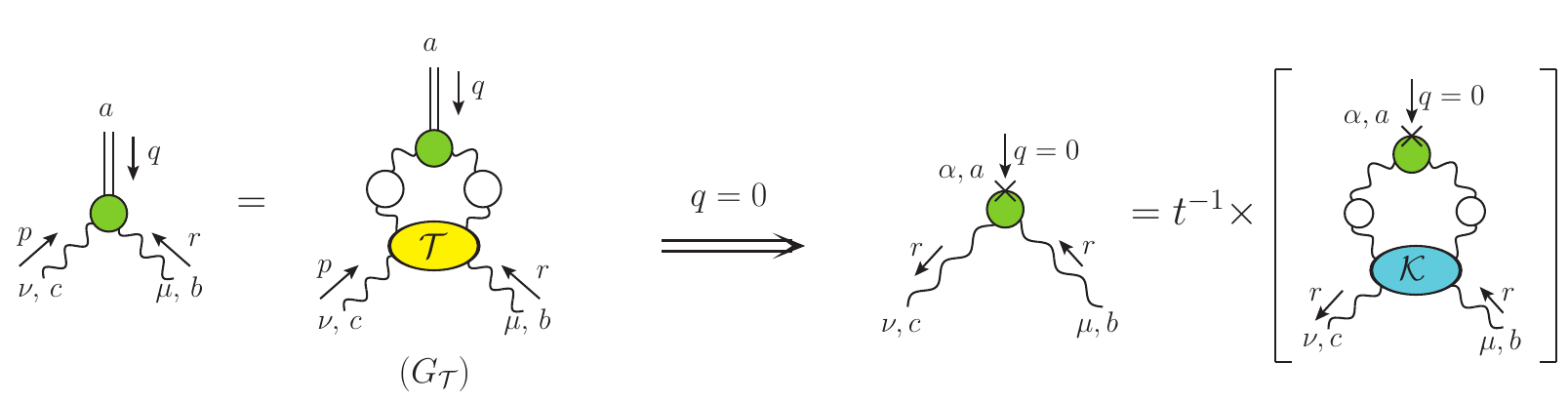}
\caption{BSE for the effective vertex $B^{abc}_{\mu\nu}(q,r,p)$ (left) and its $q = 0$ limit (right).}
\label{fig:B_BSE}
\end{figure}

The diagrammatic form of the BSE 
for general $q$
is shown in the left part of \fig{fig:B_BSE}, 
where the four-gluon kernel 
${\cal T}$ is depicted in \fig{fig:Kern_pole}. 
Specifically, from the 
first equality of \fig{fig:B_BSE}
we have
\be
B^{abc}_{\mu\nu}(q,r,p) 
= (G_{\cal T})^{abc}_{\mu\nu}(q,r,p)
\,,\label{bse1} 
\ee
where
\be
(G_{\cal T})^{abc}_{\mu\nu}(q,r,p) = \int_k B^{axe}_{\alpha\beta}(q,k,-k-q)
\Delta_{xm}^{\alpha\rho}(k)
\Delta_{en}^{\beta\sigma}(k+q)
{\cal T}^{mnbc}_{\rho\sigma\mu\nu}(-k,k+q,r,p) \,.
\label{GT}
\ee
Then, 
after using \2eqs{kern_decomp}{kernM}, two 
distinct terms arise, namely 
\be
(G_{\cal T})^{abc}_{\mu\nu}(q,r,p) = 
(G_{\cal K})^{abc}_{\mu\nu}(q,r,p) +(G_{\!{\cal M}})^{abc}_{\mu\nu}(q,r,p) \,,
\label{G_K_M}
\ee
with 
\bea
(G_{\cal K})^{abc}_{\mu\nu}(q,r,p) &=& 
\int_k B^{axe}_{\alpha\beta}(q,k,-k-q)
\Delta_{xm}^{\alpha\rho}(k)
\Delta_{en}^{\beta\sigma}(k+q)
{\cal K}^{mnbc}_{\rho\sigma\mu\nu}(-k,k+q,r,p) \,,
\nonumber\\
&{}&
\nonumber\\
(G_{\!{\cal M}}) ^{abc}_{\mu\nu}(q,r,p)  &=& 
\Omega^{ad}(q)
D^{ds}_{\Phi}(q) 
B^{sbc}_{\mu\nu}(q,r,p) \,,
\label{d1d2}
\eea
where (symmetry factor $\frac{1}{2}$ included)
\be
\Omega^{ad}(q) = \frac{1}{2}
\int_k B^{axe}_{\alpha\beta}(q,k,-k-q)
\Delta_{xm}^{\alpha\rho}(k)
\Delta_{en}^{\beta\sigma}(k+q) B^{dnm}_{\sigma\rho}(-q,k+q, -k) \,.
\label{omega}
\ee
Note that, while the term $(G_{\cal K})$ is linear in 
$B^{abc}_{\mu\nu}$, the term $(G_{\!{\cal M}})$ is cubic.  
As a result, the scale ambiguity of the solutions, present 
when only the term $(G_{\cal K})$ is considered, 
is now eliminated (see next section).

Next, it is important to 
establish that 
the product $\Omega^{ad}(q)
D^{ds}_{\Phi}(q)$ appearing in $(G_{\!{\cal M}})$ 
is finite as 
$q \to 0$; this is so, because,  in that limit,  
$\Omega^{ad}(q) \sim q^2 \delta^{ad}$, thus cancelling exactly the massless pole in  
$D^{ds}_{\Phi}(q)$. 

To derive this 
key result from \1eq{omega}, 
set first  
$\Omega^{ad}(q) = \delta^{ad} \Omega(q)$ 
and carry out the color algebra 
to obtain   
\be
\Omega(q) = \frac{C_{\rm A}}{2} \int_k 
B_{\alpha\beta}(q,k,-k-q)
P^{\alpha\rho}(k) 
P^{\beta\sigma}(k+q) 
B_{\rho\sigma}(-q,-k, k+q)
\Delta(k)\Delta(k+q) \,.
\label{omsc}
\ee
Then, taking the limit $q\to 0$
of \1eq{omsc} using \1eq{BqB},
we find 
\bea
\lim_{q \to 0} \Omega(q) &=& 
\frac{C_{\rm A}}{2} \int_k 
{B}_{\alpha\beta}(0,k,-k)
P^{\alpha\rho}(k) 
P^{\beta\sigma}(k) 
{B}_{\sigma\rho}(0,k,-k)
\Delta^2(k)
\nonumber\\
&=&
6 C_{\rm A} \int_k 
(q\cdot k)^2 \Delta^2(k) {\mathbb B}^2(k) \,,
\label{omsc3}
\eea
and thus 
\be
\lim_{q \to 0} \Omega(q) = q^2 {\widetilde \omega} \,,
\qquad\qquad
{\widetilde \omega} := 
\frac{3C_{\rm A}}{2} \int_k 
k^2 \Delta^2(k) {\mathbb B}^2(k) \,.
\label{omsc2}
\ee
With this result in hand, it is 
immediate to establish that 
\be
(G_{\!{\cal M}})^{abc}_{\mu\nu}(0,r,-r)
= \omega \,
{B}_{\mu\nu}^{abc}(0,r,-r) \,, \qquad
\omega := i \, {\widetilde \omega} \,;
\label{d20}
\ee
therefore, this term 
combines directly 
with the term 
$B^{abc}_{\mu\nu}(q,r,p)$ on the 
l.h.s of 
\1eq{bse1} in the same limit.
In particular, 
after defining the 
new variable,
\be 
t := 1 - \omega \,,
\label{thet}
\ee
\1eq{bse1} becomes 
\be
t \,
{B}^{abc}_{\mu\nu}(0,r,-r)
= \int_k B^{axe}_{\alpha\beta}(0,k,-k)
\Delta_{xm}^{\alpha\rho}(k)
\Delta_{en}^{\beta\sigma}(k)
{\cal K}^{mnbc}_{\rho\sigma\mu\nu}(-k,k,r,-r) \,. \label{bse3}
\ee
%

Given \1eq{Bat0}, it is clear that,
in the limit $q=0$,
\1eq{bse3}
yields to lowest order 
a trivial result ($0=0$).  So, in order 
to extract  nontrivial 
information from it, one must equate the terms linear in $q$ on both of its sides. 
Specifically, 
using \1eq{BandB} for the $B_{\mu\nu}(0,r-r)$ and 
$B_{\alpha\beta}(0,k-k)$
appearing in \1eq{bse3}, one obtains 
the equation\footnote{We assume that $t\neq 0$; if $t=0$, the l.h.s. of  \1eq{bse3} vanishes, and the resulting equation
is no longer of the 
BSE type.}
\be 
{\cal B}^{abc}_{\lambda\mu\nu}(0,r,-r) = t^{-1}\int_k {\cal B}^{axe}_{\lambda\alpha\beta}(0,k,-k)
\Delta_{xm}^{\alpha\rho}(k)
\Delta_{en}^{\beta\sigma}(k)
{\cal K}^{mnbc}_{\rho\sigma\mu\nu}(-k,k,r,-r) \,, \label{BSE_calB}
\ee
which admits the diagrammatic representation given in the second line of \fig{fig:B_BSE}.

Then, contracting both sides by 
$P_{\mu'}^{\mu}(r) P_{\nu'}^{\nu}(r)$
and employing \1eq{Bsurv}, we get 
\be
f^{abc} \,\Bfat(r) \, r_\lambda 
P_{\mu'\nu'}(r) \,
= - t^{-1} f^{amn} 
\int_k
\Bfat(k)\, \Delta^2(k) \, k_\lambda P^{\sigma\rho}(k) \,
{\cal K}^{mnbc}_{\rho\sigma\mu\nu}(-k,k,r,-r)
P_{\mu'}^{\mu}(r) P_{\nu'}^{\nu}(r) \,,
\label{BSEalm}
\ee
and, after contracting both sides by 
$f^{abc} \, r^{\lambda} g^{\mu'\nu'}$, and using that 
$P_{\mu}^{\mu}(r) =3$, 
we arrive at the following 
equation for the amplitude $\Bfat(r)$, 
\be 
\Bfat(r) = t^{-1} 
\alpha_s
\int_k  k^2 \Delta^2(k) {K}(r,k) \Bfat(k) \,, 
\label{BSEhom}
\ee
where $K(r,k)$ 
is {\it exactly} the kernel 
defined in \1eq{K_def}.

We end this section by pointing out that 
\1eq{BSEhom} 
is RGI.
This follows directly by observing that,
due to the first relation in \1eq{Krgi}, 
${\widetilde \omega}$ itself, defined through \1eq{omsc2}, 
is RGI, and so is the $t$ in 
\1eq{thet} \ie $\omega= \omega_{\s R}$ 
and $t= t_{\s R}$. 
Then, the full BSE in \1eq{BSEhom}
is RGI due to the third relation in \1eq{Krgi}, 
and the fact that the 
$Z_B$ introduced to renormalize 
the $\Bfat$
cancels on both sides. As a result, one may 
substitute into \1eq{BSEhom} directly bare for renormalized quantities, without 
any renormalization constants 
appearing in the final expression, \ie
\be 
\Bfat_{\s R}(r) = t_{\s R}^{-1} 
\alpha_s^{\s R}
\int_k  k^2 \Delta_{\s R}^2(k) {K}_{\s R}(r,k) \Bfat_{\s R}(k) \,.
\label{BSEhomren}
\ee
Again, the index ``R'' will be suppressed in what follows.

\section{Dynamical Scale-fixing}\label{sec:scalefix}

In order to proceed further with our analysis, it is necessary to 
pass certain key equations 
to Euclidean space, following standard transformation rules. Specifically, 
we consider all physical momenta to 
be space-like; in particular,   
$r^2 \to - r_{\srm E}^2$, where $r_{\srm E}^2 > 0$. In addition, we use the 
conversion relations 
\begin{align}
\int_k =&\, i\int_{k_{\srm E}} \,, \qquad &\Delta_{\srm E}(r_{\srm E}^2) =&\, - \Delta(-r_{\srm E}^2) \,, \qquad &\Bfat_{\srm E}(r_{\srm E}^2) =&\, -\Bfat(-r^2_{\srm E}) \,, \nonumber\\
\Cfat_{\srm E}(r_{\srm E}^2) =&\, -\Cfat(-r^2_{\srm E}) \,, \qquad &\Ls^{\!{\srm E}}(r_{\srm E}^2) =&\, \Ls(-r_{\srm E}^2) \,. \label{rule_euc}
\end{align}
Applying the above relations to Eqs.~\eqref{I_scalar_ren3}, \eqref{omsc2}, and \eqref{d20}, the expressions for $\omega_{\srm E}$ and $I_{\srm E}$ read
\bea 
\omega_{\srm E} &=& \, \frac{3 C_{\rm A}}{2} \int_{k_{\srm E}} k_{\srm E}^2 \Delta^2_{\srm E}(k_{\srm E}^2) \Bfat_{\srm E}^2(k_{\srm E}^2) \,, 
\label{omegaEu}
\\
I_{\srm E} &=& \, \frac{3 C_{\rm A} Z_3}{2} \int_{k_{\srm E}} k_{\srm E}^2 \Delta^2_{\srm E}(k_{\srm E}^2) \Bfat_{\srm E}(k_{\srm E}^2) \,. 
\label{IEu}
\eea

The Euclidean versions of 
\2eqs{BSE_Lren}{BSEhomren} may be derived 
in a similar way. In doing so, 
note that the kernel $K$
must be cast in the form\footnote{The origin of the factor ``$i$'' 
may be easily understood at the level of the one-gluon exchange diagram
of \fig{fig:Kern_diags}; it is included in the definition of the gluon propagator, see \1eq{gluon_def}.
}
\be
K(r,k) = i K_{\srm E}(r_{\srm E},k_{\srm E}) \,,
\ee
where
\be 
i \, \alpha_s K_{\srm E}(r_{\srm E},k_{\srm E}) :=
\frac{(r_{\srm E}\cdot k_{\srm E})}{c \,r^2_{\srm E}k^2_{\srm E}} f^{abc}f^{amn}\left[ P^{\mu\nu}(r)P^{\rho\sigma}(k)\,
{\cal K}^{mnbc}_{\rho\sigma\mu\nu}(-k,k,r,-r) \right]_{\srm E}\,,
\label{K_def_euc}
\ee
and $\left[ \ldots \right]_{\srm E}$ denotes the application of \1eq{rule_euc} to the expressions in square brackets.

Then, with $t_{\srm E} := 1 - \omega_{\srm E}$,
\be 
\Bfat_{\srm E}(r_{\srm E}^2) = t_{\srm E}^{-1} 
\alpha_s
\int_{k_{\srm E}}  k_{\srm E}^2 \Delta_{\srm E}^2(k_{\srm E}^2) {K}_{\srm E}(r_{\srm E},k_{\srm E}) \Bfat_{\srm E}(k_{\srm E}^2) \,, 
\label{BSEhom_euc}
\ee
whereas the equation for $\Ls^{\srm E}(r)$ reads
\be
\Ls^{\!{\srm E}}(r_{\srm E}^2) =\, Z_3 +   \alpha_{s}\! \int_{k_{\srm E}}  k_{\srm E}^2 \Delta_{\srm E}^2(k_{\srm E}^2) \,{K}_{\srm E}(r_{\srm E},k_{\srm E}) \,\Ls^{\!{\srm E}}(k_{\srm E}^2) \,. 
\label{BSE_Lren_euc}
\ee

Finally, the expressions for the Euclidean displacement function and mass, retain the same form as in \2eqs{glmf}{C_from_B}, but now carry indices ``E'', \ie
\be 
\Cfat_{\srm E}(r_{\srm E}^2) := - I_{\srm E} \, \Bfat_{\srm E}(r_{\srm E}^2) \,, \qquad\quad m^2 = g^2I^2_{\srm E} \,.
\label{C_from_B_euc}
\ee
In order to simplify the 
notation, in what follows the index ``E''  
will be omitted.

We next focus on 
the BSE of \1eq{BSEhom_euc}. 
Strictly speaking, 
 \1eq{BSEhom_euc}
is a non-linear 
integral equation, due to the quadratic 
dependence of $t$ on the 
unknown function $\Bfat$.
However, the fact that 
$t$ is a constant, independent of the variable $r$,
allows one to  
solve \1eq{BSEhom_euc} 
as an eigenvalue problem,
typical of a linear homogeneous 
equation, with one crucial difference: 
the solutions do not suffer  
from the 
scale indeterminacy intrinsic  
to such a linear equation,
because the presence of $t$ 
fixes the scale, up to an overall sign. 

To appreciate the role played by the parameter $t$,
we temporarily set $t=1$. Then, 
let us suppose that 
the kernel $K$ is such that 
the eigenvalue that yields
a nontrivial 
solution for $\Bfat(k)$ requires 
that $\alpha_s \to \alpha_\star$, where 
$\alpha_\star$ differs from the value predicted for 
$\alpha_s$ within the renormalization scheme employed. 

Then, restoring $t$ at the level of \1eq{BSEhomren}, 
it becomes clear that it has to compensate precisely 
for the difference between $\alpha_s$ and $\alpha_\star$.
In particular, we must have 
\be
t = \frac{\alpha_s}{\alpha_\star} \,,
\label{thet2}
\ee
which brings \1eq{BSEhomren} into the required form  
\be 
\Bfat(r) = \alpha_\star \int_k  k^2 \Delta^2(k) {K}(r,k) \Bfat(k) \,. 
\label{BSE_hom2}
\ee

To understand how the presence of the parameter 
$t$ in \1eq{BSEhomren}
leads to the determination 
of the  scale of $\Bfat(k)$, 
note that, by combining \2eqs{thet}{thet2}
we obtain the condition 
\be 
\Itilde =
1 - \alpha_s /\alpha_\star \,.
\label{cond}
\ee
Since the value of the l.h.s. 
of \1eq{cond} is fixed, 
$\Itilde$ 
is completely determined,
and, therefore, 
the size of the $\Bfat(k)$ 
entering in the definition of 
\1eq{omegaEu} is constrained.

To see this, let $\Bfat_0(k)$ be a solution of \1eq{BSEhomren} with a scale set arbitrarily, \eg by imposing that the global maximum of $\Bfat_0(k)$ is at $1$, 
and denote by $\omega_0$ 
the value of $\omega$ when $\Bfat(k) \to \Bfat_0(k)$ is substituted in \1eq{omegaEu}. 
The 
$\Bfat$ that is compatible with \1eq{cond}
is related to $\Bfat_0$ by a multiplicative constant, $\sigma$, \ie $\Bfat(k) = \sigma\,\Bfat_0(k)$, 
whose value is determined from the equation
\be 
\sigma^2 \omega_0 = 1 - \alpha_s /\alpha_\star \,,
\label{sigma}
\ee
or, equivalently, 
\be 
\sigma = \pm \sqrt{\frac{ 1 - \alpha_s/\alpha_\star }{ {\Itilde}_0 } } \,.
\label{scale_setting}
\ee
Note that, since ${\Itilde }$ is quadratic in $\Bfat(k)$, the sign of $\sigma$, and hence the sign of $\Bfat$ itself, is left undetermined. However, as we can see from \2eqs{IEu}{C_from_B_euc}, 
$\Cfat(k)$ is quadratic in $\Bfat(k)$, and therefore 
does not get affected by the sign ambiguity of \1eq{scale_setting}. In fact, the overall sign of $\Cfat(k)$
turns out to be negative for the entire range of Euclidean momenta, in agreement with the sign found in the lattice extraction of $\Cfat(k)$
presented in~\cite{Aguilar:2021uwa,Aguilar:2022thg}.

\section{Renormalization of the gluon mass equation}\label{sec:trick}

The compact relation in  
\1eq{glmf} expresses 
the gluon mass in terms of  
the gluon-scalar transition 
amplitude $I$, 
which depends, in turn, on 
the BS amplitude $\Bfat(r)$.
After the renormalization 
and the rotation 
to the Euclidean space, 
$I$ is given by \1eq{IEu}.
The obvious difficulty associated with the use of the latter equation 
is the renormalization constant $Z_3$ multiplying the relevant integral. 

It turns out 
that a subtle sequence of relations 
allows one to implement the 
multiplicative renormalization implicit 
in \1eq{IEu}
without having to carry out any integrations. 
The upshot of these considerations is that 
the renormalization
of $I$
finally amounts to 
the {\it effective} 
replacement $Z_3 \to \omega \Ls(k)$ 
at the level of 
\1eq{IEu}, \ie 
\be 
I = \frac{3C_{\rm A}}{2} \omega \int_k k^2 \Delta^2(k) \Ls(k)\Bfat(k) \,.
\label{Iomega}
\ee
In what follows we will present a detailed 
derivation of the above 
key
result.

The first observation is that, 
by virtue of 
\1eq{BSE_Lren}, the $Z_3$ in \1eq{I_scalar_ren3}
may be 
substituted by 
the combination 
\be
Z_3 =\, \Ls(k) -  \alpha_{s}\!\! \int_\ell  \ell^2 \Delta^2(\ell) \,{K}(k,\ell) \,\Ls(\ell) \,. \label{Z3}
\ee
This procedure is typical  
when dealing with multiplicative renormalizability at the level of the SDEs, see, \eg \cite{Bjorken:1965zz}, 
and is intimately connected with the 
so-called ``skeleton expansion'' of the SDE kernel. 

The second observation is novel, 
and completely specific to the 
particular form of the BSE in \1eq{BSEhom_euc}. 
In particular, when the 
substitution of \1eq{Z3} 
is implemented,
the BSE of \1eq{BSEhom_euc} is triggered inside 
\1eq{IEu}, 
leading to a crucial  
cancellation, and, finally, to 
\1eq{Iomega}.

In order to convey 
the underlying idea with a simple  
example, consider two functions $g(x)$ and $f(x)$, 
satisfying the system of integral equations  
\be
g(x) = z+ \int \!\! dy \, \tilde g (y) K(x,y) \,, \qquad\qquad
f(x) = b^{-1}\!\int \!\! dy \,\tilde f(y) K(x,y) \,,
\label{gf}
\ee
where $\tilde g (y) := g(y) \, u(y)$ and 
$\tilde f (y) : = f(y) \,u(y)$, with $u(y)$ a 
well-behaved function. The parameters  
$z$ and $b$ are real numbers, with  
$b\neq 0,1$, and the limits of integration are 
arbitrary. Finally, 
the {\it common} kernel $K(x,y)$
satisfies 
the symmetry relation $K(x,y)=K(y,x)$.

Let us further assume that the  
value of a constant $\beta$ is given by the integral 
\be
\beta= z\!\int \!\! dx \, \tilde f(x)\,.
\label{beta}
\ee
The dependence of $\beta$ on the parameter 
$z$ may be eliminated in favor of 
the function $g(x)$ by appealing to 
the system of \1eq{gf}. In particular, one 
employs the following sequence of steps
\bea
\beta &=& \int \!\! dx \, z \, \tilde f(x) 
\nonumber\\
&=& \int \!\! dx \left[g(x) -\int \!\! dy \, \tilde g(y) 
K(x,y) \right] \tilde f(x)
\nonumber\\
&=& \int \!\! dx \, g(x) \tilde f(x) - 
\int \!\! dx \! \!\int \!\! dy \,\tilde g(y) K(x,y) \tilde f(x) 
\nonumber\\
&=&  
\int \!\! dx \, \tilde g(x)  \Bigg[f(x) - 
\underbrace{\int \!\! dy \,\tilde f(y)  K(x,y)}_{b f(x)} \Bigg] \,,
\label{steps}
\eea
where in the second line we used the 
first relation in \1eq{gf}, while in the last line  the 
relabelling $x \leftrightarrow y$ 
of the integration variables 
was carried out, the symmetry 
of the kernel $K(x,y)$ was exploited,
and the second relation in 
\1eq{gf} was invoked. 
Thus, one obtains 
\be
\beta=  
(1- b)\int \!\! dx \tilde f(x) g(x) \,,
\label{betaren}
\ee
where we used that 
$\tilde g(x) f(x)  = \tilde f(x) g(x)$.
Evidently, the dependence 
of $\beta$ on the  
parameter $z$ 
has been exchanged 
for the dependence 
on the function 
$g(x)$, which was absent from the 
original integral 
in \1eq{beta}. 
In Appendix 
\ref{sec:toy}, we present a concrete example, where the 
equivalence between 
\2eqs{beta}{betaren} may be worked out exactly.

\begin{figure}[!ht]
\centering
\includegraphics[width=0.7\linewidth]{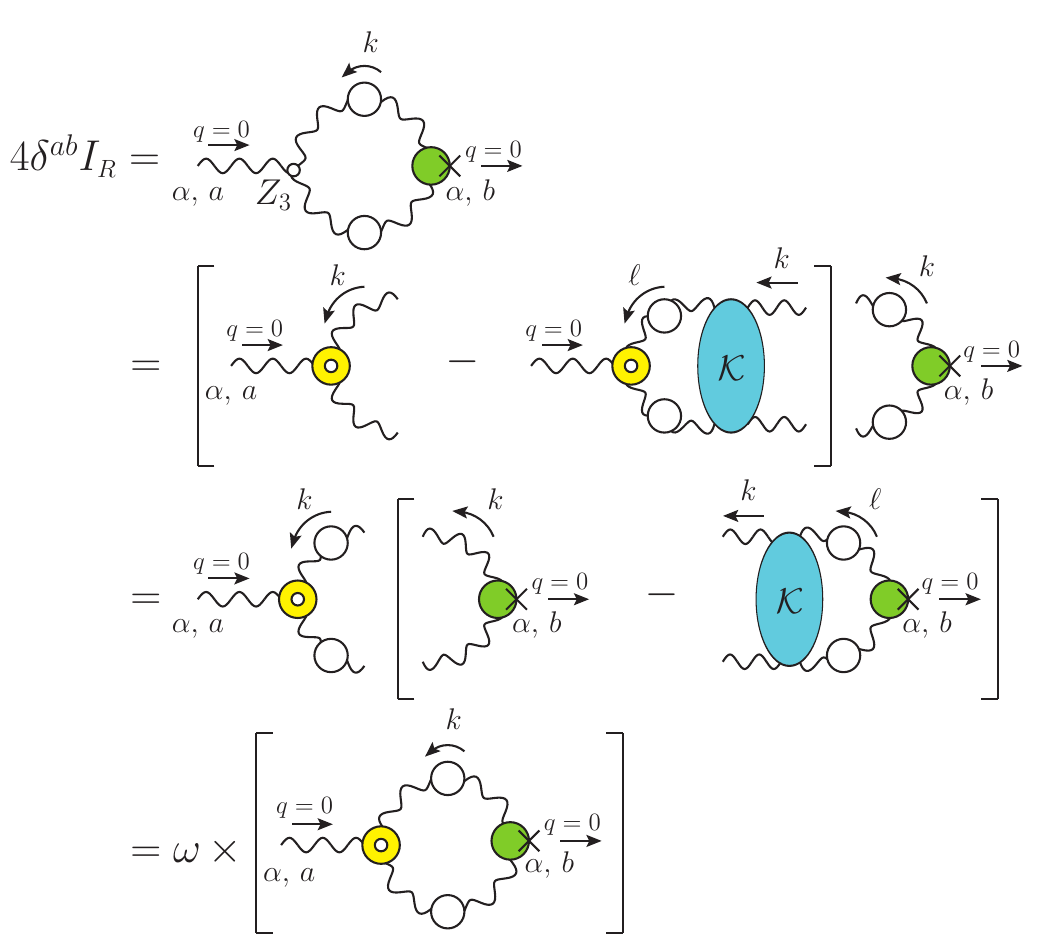}
\caption{ Diagrammatic illustration of the renormalization of $I$. Note that the diagrams containing the kernel ${\cal K}$ undergo a relabeling of integration momenta, $k\leftrightarrow \ell$, from the second to the third line.}
\label{fig:mass_trick}
\end{figure}

At this point it is straightforward to repeat the
construction leading to  
\1eq{betaren} for the system of integral equations 
given by \2eqs{BSEhom_euc}{BSE_Lren_euc}, in conjunction with \1eq{IEu}.  
To that end, all we need is to establish the correspondence 
\begin{align}
\{g(x), \,f(x), \, u(x), \, K(x,y)\} \,&\leftrightarrow \, 
\{\Ls(r), \,\Bfat (r),\, k^2 \Delta^2(k), \,
\alpha_{s} K(r,k)\} \,, \,\, 
\nonumber\\
\{\beta,\, z, \,b, \,dx,\, dy \} 
\,&\leftrightarrow \,
\{2I/3 C_{\rm A}, \,Z_3, \,t, \,d^4 k,  \,d^4 \ell\} \,. 
\end{align}
In particular, with the above identifications
and \1eq{thet}, we find that 
the analogue of \1eq{betaren} 
is precisely \1eq{Iomega}, which is 
the announced result. The diagrammatic representation of the 
steps described in \1eq{steps} 
is shown in \fig{fig:mass_trick}; 
note that, in doing so, we employ  
the representation of the BSE 
given in \fig{fig:B_BSE}, 
whose main building block 
was introduced in \fig{fig:B_derivative}.



The above construction reveals a 
striking fact: if the parameter 
$\omega$ is set to zero, the cancellation 
described in \1eq{steps} 
[\fig{fig:mass_trick}] 
is perfect; thus, even if a non-trivial
$\Bfat$ is obtained from \1eq{BSEhom_euc}, the 
renormalized transition amplitude $I$, and therefore, the gluon mass $m$, vanish. 
In that sense, the gluon mass 
emerges due to the   
mismatch between the kernels in the  equations 
for $\Bfat$ and  
$\Ls(r)$, [see \2eqs{BSEhom_euc}{BSE_Lren_euc}],
produced by the 
presence of a $t\neq 1$
in the former but not in the latter. In diagrammatic terms, 
the non-vanishing of the gluon mass
hinges on the 
difference between the 
${\cal T}$ appearing in 
\fig{fig:B_BSE} and the ${\cal K}$
in \fig{fig:L_BSE}, namely 
the component 
${\cal M}$ [see \fig{fig:Kern_pole}].
Even though, on intuitive grounds, the importance of this term seems evident, given that it carries precisely the information about the 
emergence 
of a massless scalar, the quantitative 
understanding attained through the 
above construction is most valuable.

Quite interestingly, 
the cancellations
described in \1eq{steps} are 
not coincidental, 
but 
are rather 
driven by an 
underlying 
mathematical principle. 
In particular,  
as we discuss in detail in the next section,
our SDE-BSE analysis may be interpreted as 
a rather subtle  
application of 
the so-called ``Fredholm 
alternatives theorem''.

\section{Fredholm alternatives theorem and its evasion}\label{sec:Fredholm}

Let us consider the inhomogeneous and homogeneous Fredholm equations of the second kind, given by~\cite{vladimirov1971equations,polyanin2008handbook}
\be 
f_1(x) = f_2(x) + \lambda \int_a^b K(x,y) f_1(y) dy \,, \label{fred_inhom}
\ee
and
\be 
f_3(x) = \lambda \int_a^b K(x,y) f_3(y) dy \,, \label{fred_hom}
\ee
respectively. Here, $f_1$ and $f_3$ are the unknown functions, while $f_2$ is a known inhomogeneous term. The kernel, $K(x,y)$, is assumed to be continuous and square-integrable in the interval $[a,b]$, in which case it is said to be a Hilbert-Schmidt integral operator.

The Fredholm alternatives theorem imposes restrictions on the existence of simultaneous solutions of \2eqs{fred_inhom}{fred_hom}. For the purposes of the present work, it suffices to consider the special case of the theorem when $K(x,y)$ is real and symmetric in $x\leftrightarrow y$~\footnote{For the generalization of the theorem for a complex and non-symmetric kernel see, \eg \cite{vladimirov1971equations,polyanin2008handbook}.}.  
Under these simplifications, the Fredholm alternatives theorem can be stated as follows~\cite{polyanin2008handbook}:

\begin{enumerate}
\item[({\itshape a})] If $\lambda$ is not an eigenvalue of $K(x,y)$, \ie if the homogeneous equation has only the trivial solution, $f_3(x) = 0$, then the inhomogeneous \1eq{fred_inhom} has a solution for \emph{any} nonzero $f_2(x)$.

\item[({\itshape b})] If $\lambda$ is an eigenvalue of $K(x,y)$, such that $f_3(x)$ is nonvanishing, then \1eq{fred_inhom} has solutions \emph{if and only if}
\be 
\int_a^b f_2(y) f_3(y) dy = 0 \,. \label{ortho}
\ee

\end{enumerate}

In order to connect the Fredholm alternative theorem to the system of equations satisfied by $\Ls(r)$ and $\Bfat(r)$, we need to perform certain transformations that will bring \2eqs{BSEhom_euc}{BSE_Lren_euc} to a form similar to \2eqs{fred_inhom}{fred_hom}.

We begin by rewriting \2eqs{BSEhom_euc}{BSE_Lren_euc} in spherical coordinates. To this end, we define the variables
\be
x := r^2 \,,  \qquad\qquad y := k^2 \,, \label{spherical}
\ee
and the angle $\theta$ between $r$ and $k$, \ie $r\cdot k = |r||k|c_\theta$, where we use the shorthand notation $c_\theta := \cos\theta$ and $s_\theta := \sin\theta$. Note that the kernel $K(r,k)$ is a function of $x$, $y$, and $\theta$, \ie $K(r,k)\equiv K(x,y,\theta)$.

Moreover, we introduce an ultraviolet momentum cutoff, $\Lambda$, to regularize potential divergences\footnote{For the sake of simplicity,
a hard cutoff has been employed; we have confirmed that exactly the same conclusions are reached when the calculation is carried out using 
dimensional regularization. }. In this case, the integral measure takes the form
\be 
\int_{k} := \frac{1}{(2\pi)^3}\int_0^{\Lambda^2} \!\! dy \, y\int_0^\pi \!\! d\theta \, s_\theta^2 \,. \label{K_ang}
\ee

Now, the only dependence of the integrands of \2eqs{BSEhom_euc}{BSE_Lren_euc} on the angle is through $K(x,y,\theta)$. Then, we can define an ``angle-integrated kernel'', $\Khat(x,y)$, by
\be 
\Khat(x,y) = \frac{1}{(2\pi)^3}\!\int_0^\pi \!\! d\theta \, s_\theta^2 \, K(x,y,\theta)\,. \label{K_ang_int}
\ee

With the above definitions, \2eqs{BSEhom_euc}{BSE_Lren_euc} are recast as
\begin{align} 
\Bfat(x) =&\, t^{-1} 
\alpha_s
\!\int_0^{\Lambda^2} \!\! dy \, {\cal Z}^2(y) \Khat(x,y) \Bfat(y) \,, \nonumber\\ 
\Ls(x) =&\, Z_3 + \alpha_{s}\! \int_0^{\Lambda^2} \!\! dy \, {\cal Z}^2(y) \Khat(x,y) \Ls(y) \,, \label{BSE_spherical}
\end{align}
where we introduced the gluon dressing function, ${\cal Z}(x) := x\Delta(x)$.

At this point, we note that since $K(r,k)$ is symmetric under the exchange of $r\leftrightarrow k$, then $\Khat(x,y) = \Khat(y,x)$. However, the complete kernel of \1eq{BSE_spherical}, namely ${\cal Z}^2(y)\Khat(x,y)$, is not symmetric under $x\leftrightarrow y$, due to the factor of ${\cal Z}^2(y)$. Nevertheless, it can be transformed into an equivalent system of equations with a symmetric kernel.

To this end, we multiply \1eq{BSE_spherical} by ${\cal Z}(x)$, and define
\be 
{\widetilde \Bfat}(x) := {\cal Z}(x)\Bfat(x) \,, \qquad {\widetilde L}_{sg}(x) := {\cal Z}(x)\Ls(x) \,, \qquad {\widetilde K}(x,y) := {\cal Z}(x){\cal Z}(y)\Khat(x,y) \,.
\ee
Then, \1eq{BSE_spherical} is equivalent to
\begin{align} 
{\widetilde \Bfat}(x) =&\, t^{-1} 
\alpha_s
\!\int_0^{\Lambda^2} \!\! dy \, {\widetilde K}(x,y) {\widetilde \Bfat}(y) \,, \nonumber\\ 
{\widetilde L}_{sg}(x) =&\, Z_3 {\cal Z}(x) + \alpha_{s}\! \int_0^{\Lambda^2} \!\! dy \, {\widetilde K}(x,y) {\widetilde L}_{sg}(y) \,, \label{BSE_sym}
\end{align}
whose kernel ${\widetilde K}(x,y)$ is symmetric. 

Finally, the \1eq{IEu} for $I$ can be recast as
\be 
I = \frac{3 C_{\rm A}Z_3}{32\pi^2}\!\int_0^{\Lambda^2}\!\! dy \, {\cal Z}(y) \, {\widetilde \Bfat}(y) \,. \label{I_spherical}
\ee

The implications of the Fredholm alternatives theorem for \1eq{BSE_sym} are now straightforward to establish. Specifically, suppose that $\omega = 0$, such that $t = 1$. Then, we have a direct correspondence between \1eq{BSE_sym} and the \2eqs{fred_inhom}{fred_hom} through the identification
\begin{align}
\{f_1(x), \,f_2(x), \, f_3(x), \, K(x,y) \} \,&\leftrightarrow \, 
\{ {\widetilde L}_{sg}(x), \, Z_3{\cal Z}(x),\, {\widetilde \Bfat}(x), \,
{\widetilde K}(x,y) \} \,, \nonumber\\
\{ \lambda, \, a, \, b \} \,&\leftrightarrow \,  \{ \alpha_s, \, 0, \, {\Lambda^2} \} \,.
\end{align}

Hence, for both $\Ls(x)$ and $\Bfat(x)$ to be nonzero, the Fredholm alternatives theorem implies that
\be 
Z_3 \int_0^{\Lambda^2} \!\! dy \, {\cal Z}(y) \, {\widetilde \Bfat}(y) = 0 \,.
\ee
Comparison to \1eq{I_spherical} then yields $I = 0$, and therefore,
due to \1eq{glmf}, 
$m^2 = 0$.

We conclude that the 
generation of a gluon mass through the SM 
hinges on $\omega\neq 0$,  $t \neq 1$, which leads to the evasion of 
Fredholm alternatives theorem, by relaxing the assumption of equal kernels 
in \1eq{BSE_sym}. 
Otherwise, if $\omega =0$, 
the theorem imposes a vanishing gluon mass, even in the presence 
of a 
nonvanishing $\Bfat$,
because, quite remarkably, the 
condition of \1eq{ortho}
is {\it precisely} the
equation for the transition amplitude $I$.

Let us finally discuss a subtlety related to the applicability 
of Fredholm's theorem in the present context. 
Specifically, the theorem applies to a system of \emph{linear} equations, whereas the kernel $K(r,k)$ depends on $\Ls(r)$ [see \fig{fig:Kern_diags} and \1eq{planar}]. In that sense, even for $\omega = 0$, the equation for $\Ls(r)$ is nonlinear. Nevertheless, our main conclusion, namely that for $\omega = 0$ the gluon mass vanishes, persists.

To see this, suppose there exists a solution, $L_{0}(r)$ and $\Bfat_{0}(r)$, to the full nonlinear system of equations. Then, let $K_0(r,k)$ be the value of $K(r,k)$ obtained by the substitution of $\Ls\to L_0$ in its 
expression\footnote{As can be seen from \fig{fig:Kern_diags}, the 
$\Ls$ enters in $K(r,k)$ in 
such a way that the symmetry 
under $r \longleftrightarrow k$
is preserved.}. 
Now, $L_{0}(r)$ and $\Bfat_{0}(r)$ must also be a solution of 
\begin{align}
\Ls(r) =&\, Z_3 + \alpha_{s}\! \int_{k}  k^2 \Delta^2(k^2) \,{K}_0(r,k) \,\Ls(k) \,, \nonumber\\
\Bfat(r) =&\, t^{-1}\alpha_{s}\! \int_{k}  k^2 \Delta^2(k^2) \,{K}_0(r,k) \,\Bfat(k) \,, \label{system_linear}
\end{align}
since setting $\lbrace \Ls,\, \Bfat\rbrace \to \lbrace L_0,\, \Bfat_0\rbrace$ in the above equation recovers the original, nonlinear, system. But the Fredholm alternatives theorem applies to \1eq{system_linear} with $\omega = 0$, in which case the arguments of Sec.~\ref{sec:Fredholm} lead to $m = 0$. Hence, the nonlinear nature of the equations cannot by itself evade the Fredholm alternatives theorem when $\omega = 0$ and the conclusion of our analysis is unaffected.

\section{Numerical analysis}\label{sec:res}

In this section we
explore the numerical 
implications of the main results 
presented so far, placing particular 
emphasis on the value of the gluon mass $m$, and the shape and size of the 
displacement function $\Cfat(r)$. 
In doing so, we will use two 
important results as 
benchmarks for these quantities.
In particular,  
we identify as the optimal value 
for the gluon mass 
the inverse of the saturation point 
of the lattice gluon propagator
at the origin. Evidently, this is a renormalization-point dependent mass scale, and does not admit a direct physical interpretation\footnote{An RGI gluon mass of about 450 MeV may be obtained following the procedure described in~\cite{Binosi:2016nme,Cui:2019dwv}.}; when the lattice curve 
has been renormalized such that 
$\Delta^{-1}(\mu^2) = \mu^2$ 
at \mbox{$\mu =4.3$~GeV}, one 
obtains 
the value  $m_{\srm{lat}} = 354$~MeV~\cite{Aguilar:2021okw}. 
As for the displacement function, 
we will compare our results 
with the curve shown in the left panel of 
\fig{fig:Cfat_vs_mass}, which was obtained 
following the 
method outlined in 
Sec.~\ref{subsec:disp}; for details, see~\cite{Aguilar:2022thg}.

The main steps of the numerical procedure followed may be 
summarized as follows: 

({\it i})
Since the four-gluon kernel 
$K(r,k)$ entering in BSE of \1eq{BSEhom_euc} 
is not known, we resort to 
models for it, using its  
one-gluon exchange approximation,
$\Ko(r,k)$, as our point of departure.
Specifically, at the beginning of the 
numerical procedure, 
an {\it Ansatz} for  
$K(r,k)$ is constructed, 
by varying  
the parameters
 of $\Ko(r,k)$.

({\it ii})
The $K(r,k)$ from the previous step  
is fed into the 
BSE of \1eq{BSEhom_euc}, which is 
solved as an eigenvalue problem,
yielding a solution for 
$\Bfat (r)$. 
The corresponding eigenvalue $\alpha_\star$
is used in \1eq{scale_setting} 
to determine the value of 
$\sigma$, thus fixing 
the scale of $\Bfat (r)$. 
In addition, the value of $\omega$ is obtained 
directly from \1eq{cond}. 

({\it iii})
The above solution for $\Bfat (r)$ is substituted 
into \1eq{Iomega} to yield the value of $I$. 
Then, using the two relations in \1eq{C_from_B_euc}, we 
get the value of the gluon mass and the form of the 
displacement function, and compare them with the 
benchmark results.

({\it vi})
The procedure is repeated, using a 
different model for the kernel $K(r,k)$.

\subsection{One-gluon exchange approximation of the BSE kernel}\label{subsec:trunc}

For practical calculations, the kernel ${\cal K}_{\rho\sigma\mu\nu}^{mnbc}(-k,k,r,-r)$ of the BSE must be truncated. To this end, we start with the skeleton expansion of ${\cal K}_{\rho\sigma\mu\nu}^{mnbc}(-k,k,r,-r)$, shown in \fig{fig:Kern_diags}, which we write as
\be 
{\cal K}_{\rho\sigma\mu\nu}^{mnbc}(-k,k,r,-r) = \sum_{i = 1}^\infty (b_i)_{\rho\sigma\mu\nu}^{mnbc}(-k,k,r,-r) \,. \label{Kcal_skeleton}
\ee

Substituting \1eq{Kcal_skeleton} into \1eq{K_def_euc}, we can compute the contribution of each diagram $(b_i)$ to the Euclidean $K(r,k)$ entering \1eq{BSEhom_euc}. As it turns out, the diagram $(b_2)$, which is simply the tree-level four-gluon vertex, vanishes when inserted in \1eq{K_def_euc}~\cite{Aguilar:2011xe,Aguilar:2017dco}, \ie
\be 
f^{abc}f^{amn} P^{\mu\nu}(r)P^{\rho\sigma}(k)\,
(b_2)^{mnbc}_{\rho\sigma\mu\nu}(-k,k,r,-r) = 0 \,.
\ee
Hence, $K(r,k)$ reads
\be 
K(r,k) = \Ko(r,k) + K_{\rm ho}(r,k) \,, \label{K_full}
\ee
where $\Ko(r,k)$ is the contribution of the one-gluon exchange diagram, $(b_1)$, and $K_{\rm ho}(r,k)$ represent the higher-order diagrams $(b_i)$ with $i\geq3$.

Next, we focus on the one-gluon exchange contribution, $\Ko(r,k)$. From diagram $(b_1)$ of \fig{fig:Kern_diags} and \1eq{K_def_euc}, one obtains in the Landau gauge
\be 
\Ko(r,k) = \frac{ 2 \pi C_{\rm A} }{3} \Delta(k-r) \left[ \gb_{\mu\rho\sigma}(r,-k,k-r) \gb^{\mu\rho\sigma}(-r,k,r - k) \right]_{\srm E}\,, \label{Ko_Gbar}
\ee
where 
\be 
\gb_{\mu\rho\sigma}(r,-k,k-r) := P^{\mu'}_\mu(r)P^{\rho'}_\rho(k)P^{\sigma'}_\sigma(k-r)\fatg_{\mu'\rho'\sigma'}(r,-k,k-r) \,, \label{gbar_def}
\ee
is the transversely projected three-gluon vertex~\cite{Eichmann:2014xya,Blum:2014gna,Mitter:2014wpa,Huber:2018ned,Huber:2020keu,Aguilar:2021lke,
Pinto-Gomez:2022brg,Aguilar:2023qqd,Ferreira:2023fva}.

In principle, the $\gb_{\mu\rho\sigma}$ appearing in \1eq{Ko_Gbar} consists of four independent tensor structures, each accompanied by a form factor that depends on three kinematic variables. However, as has been shown in numerous recent works
\cite{
Eichmann:2014xya,Blum:2014gna,Mitter:2014wpa,Huber:2018ned,Huber:2020keu,Aguilar:2021lke,Pawlowski:2022oyq,Pinto-Gomez:2022brg,Aguilar:2023qqd,Ferreira:2023fva}, the classical tensor structure of $\gb_{\mu\rho\sigma}$ is dominant, 
and the associated form factor $\Ls$
may be accurately described as a function 
of a single special kinematic variable, 
denoted by $s^2$, namely 
\be 
\gb^{\mu\rho\sigma}(r,-k,k-r) = \gb_{\!0}^{\mu\rho\sigma}(r,-k,k-r) \Ls(s) \,, \qquad s^2 := \frac{1}{2}[ r^2 + k^2 + ( k - r )^2 ] \,,\label{planar}
\ee
where $\gb_{\!0}^{\mu\rho\sigma}$ is the tree-level value of $\gb^{\mu\rho\sigma}$, obtained by substituting the $\fatg^{\mu'\rho'\sigma'}$ in \1eq{gbar_def} by the $\fatg_{\!0}^{\mu'\rho'\sigma'}$ of \1eq{bare3g}.

Then, using \1eq{planar}, employing spherical coordinates, and introducing the variables defined in \1eq{spherical}, one obtains 
\be 
\Ko(x,y,\theta) = \kin(x,y, \theta)\Ro(u,s) \,, \label{K_oge}
\ee
where $u^2 := ( k - r )^2 = x + y - 2 c_\theta\sqrt{xy}$, $s^2 = x + y - c_\theta\sqrt{xy}$,
\be 
\kin(x,y,\theta) := \frac{8\pi C_{\rm A}}{3u^2\sqrt{xy}}\left\lbrace c_\theta s_\theta^2\left[ (c_\theta^2 + 8 )xy - 6c_\theta\sqrt{xy}(x + y ) + 3(x^2 + y^2 ) \right] \right\rbrace \,,
\ee
and
\be 
\Ro(u,s) := {\Delta(u)\Ls^2(s)} \,.
\ee

\begin{figure}[!ht]
  \centering
\includegraphics[width=0.45
\textwidth]{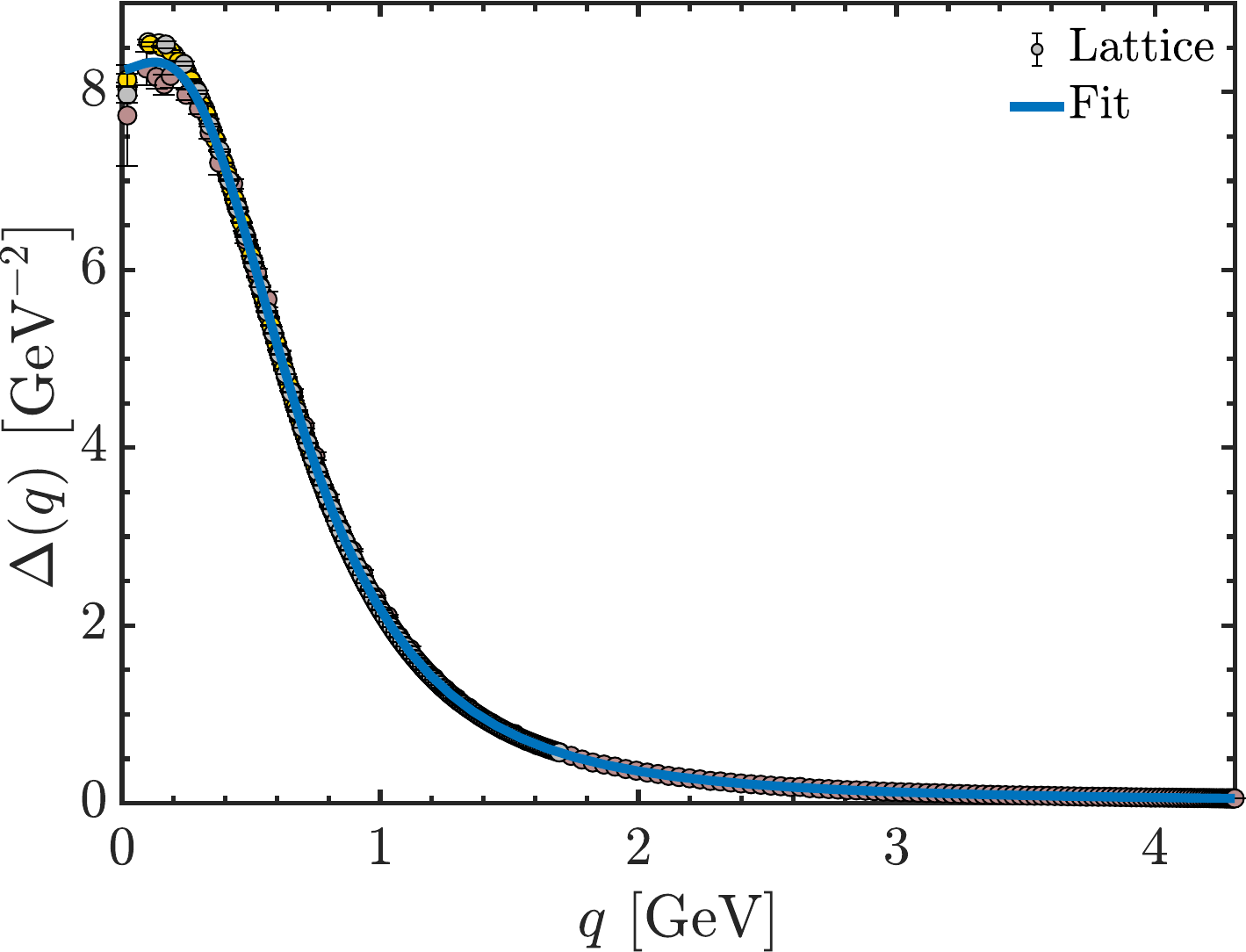}\hfil\includegraphics[width=0.45
\textwidth]{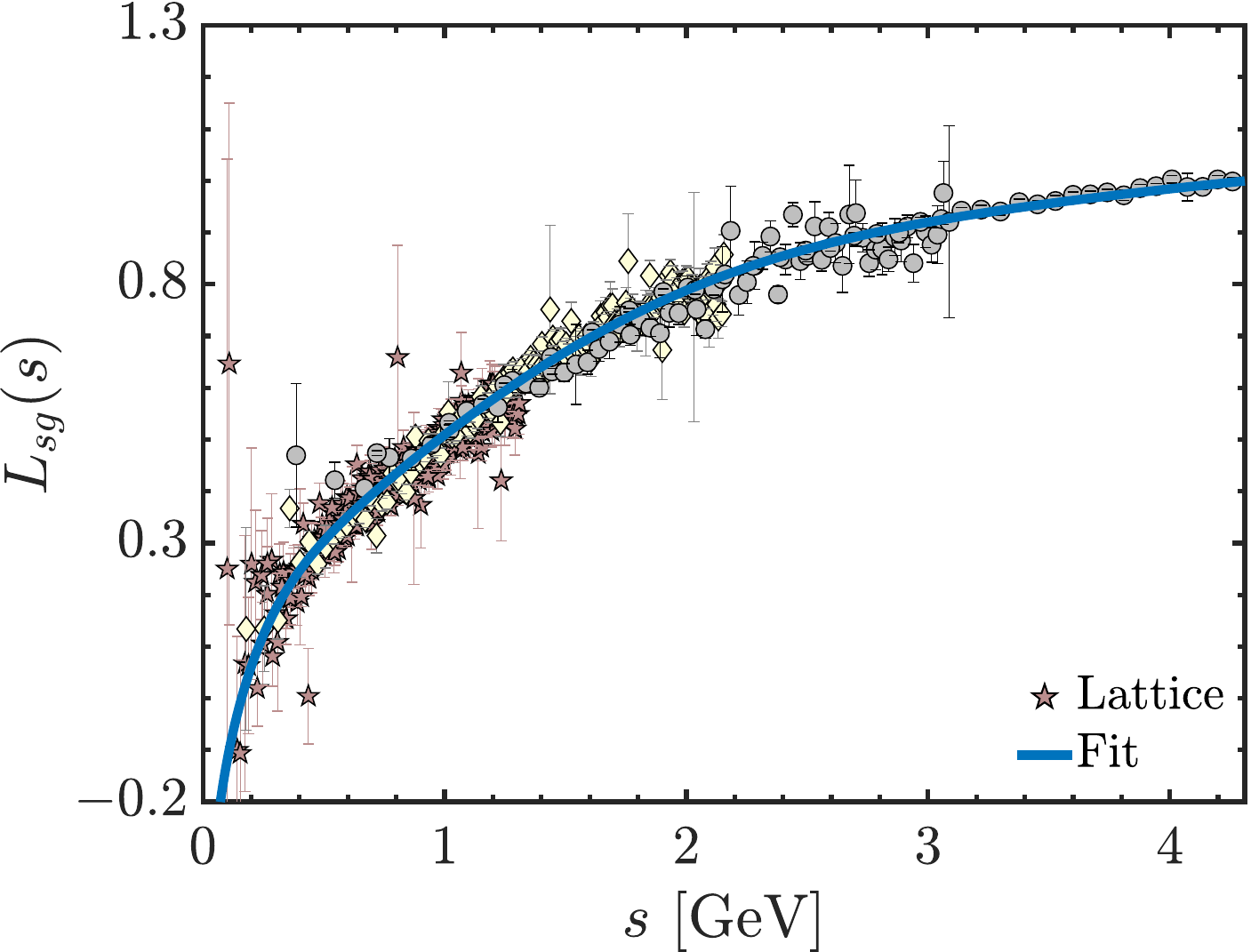}
\caption{Lattice data (points) for the gluon propagator from \cite{Aguilar:2021okw} (left) and the three-gluon vertex form factor, $\Ls(s)$, from \cite{Aguilar:2021lke} (right). The blue continuous curves represent the corresponding fits given by Eqs.~(C11) and (C12) of \cite{Aguilar:2021uwa}.}\label{fig:inputs}
\end{figure}

The ingredients entering in the $\Ko(x,y,\theta)$ of \1eq{K_oge} are all accurately known from lattice simulations. In particular, we use for $\Delta(u)$ and $\Ls(s)$ the fits 
to the lattice results of \cite{Aguilar:2021okw}, given 
by Eqs.~(C.11) and~(C.12) of \cite{Aguilar:2021uwa}, respectively;
the corresponding curves are shown 
in \fig{fig:inputs}.
Note that these ingredients are renormalized in the so-called ``asymmetric MOM scheme''~\cite{Athenodorou:2016oyh,Boucaud:2017obn,Aguilar:2021lke,Aguilar:2021okw,Aguilar:2023qqd}, defined by the renormalization condition
\be 
\Delta^{-1}(\mu) = \mu^2 \,, \qquad \Ls(\mu) = 1 \,,
\ee
with $\mu = 4.3$~GeV denoting the renormalization point. For this renormalization scheme and point, the corresponding strong charge takes the value $\alpha_s = 0.27$~\cite{Boucaud:2017obn}, which we adopt from now on.

\subsection{Asymptotic behavior of \texorpdfstring{$\Bfat(r)$}{B(r)} }\label{subsec:asympt}

For a given kernel, $K(r,k)$, the eigenvalue problem of \1eq{BSE_hom2} can be solved with standard procedures, such as \emph{Nystr\"om}'s method~\cite{Press:1992zz}. This determines the eigenvalue, $\alpha_\star$, and an eigenfunction, $\Bfat_0(r)$, arbitrarily normalized such that its global maximum is $1$.
In order to fix the physical scale of $\Bfat(r)$ and compute the gluon mass, we then need to compute $\omega$ and $I$ through \2eqs{omegaEu}{Iomega}. This task requires an analysis of the ultraviolet behavior of the integrals in the latter equations, which we describe in detail below. For simplicity, we will illustrate the procedure with the case of the one-gluon exchange kernel, $\Ko$, but it applies equally to the kernels constructed through \1eq{K_ansatz}.

Let us begin by writing \2eqs{omegaEu}{Iomega} in spherical coordinates,
\begin{align}
\omega =&\, \lambda  \int_0^{\infty} \! \!\! dy \, y^2 {\Delta}^2(y) \, \Bfat^2(y) \,, \label{omega_spherical} \\
I =&\, \lambda\omega \int_0^{\infty} \! \!\! dy \, y^2 {\Delta}^2(y) \, \Ls(y) \, \Bfat(y) \,, \label{I_spherical2}
\end{align}
where $\lambda := 3 C_{\rm A}/(32\pi^2)$.

In addition, let us recall that at large momenta $\Delta(y)$ and $\Ls(y)$ approach their one-loop resumed perturbative behaviors~\cite{Altarelli:1981ax,Roberts:1994dr,vonSmekal:1997ern,Fischer:2002eq,Pennington:2011xs,Huber:2018ned}
\be 
\Delta(y) \sim y^{-1} L_{\srm{UV}}^{-13/22}(y) \,, \qquad \Ls(y) \sim L_{\srm{UV}}^{17/44}(y) \,, \qquad L_{\srm{UV}}(y) := c\ln(y/\Lambda^2_{\srm{MOM}}) \,, \label{Delta_L_asympt}
\ee
where $c = 1/\ln(\mu^2/\Lambda^2_{\srm{MOM}})$, and $\Lambda_{\srm{MOM}}$ is the (quenched) QCD mass-scale in the MOM scheme~\cite{Celmaster:1979dm,Celmaster:1979km,Alles:1996ka,Boucaud:2008gn}; we employ the one-loop result $\Lambda^2_{\srm{MOM}} = \mu^2\exp[-12\pi/(11C_{\rm A}\alpha_s)]$, for which $\Lambda = 520$~MeV and $c = 0.236$. Note that the fits of Eqs.~(C.11) and~(C.12) of \cite{Aguilar:2021uwa} for $\Delta(r)$ and $\Ls(r)$ reproduce \1eq{Delta_L_asympt} by construction.

\begin{figure}[ht]
\includegraphics[width=0.45\textwidth]{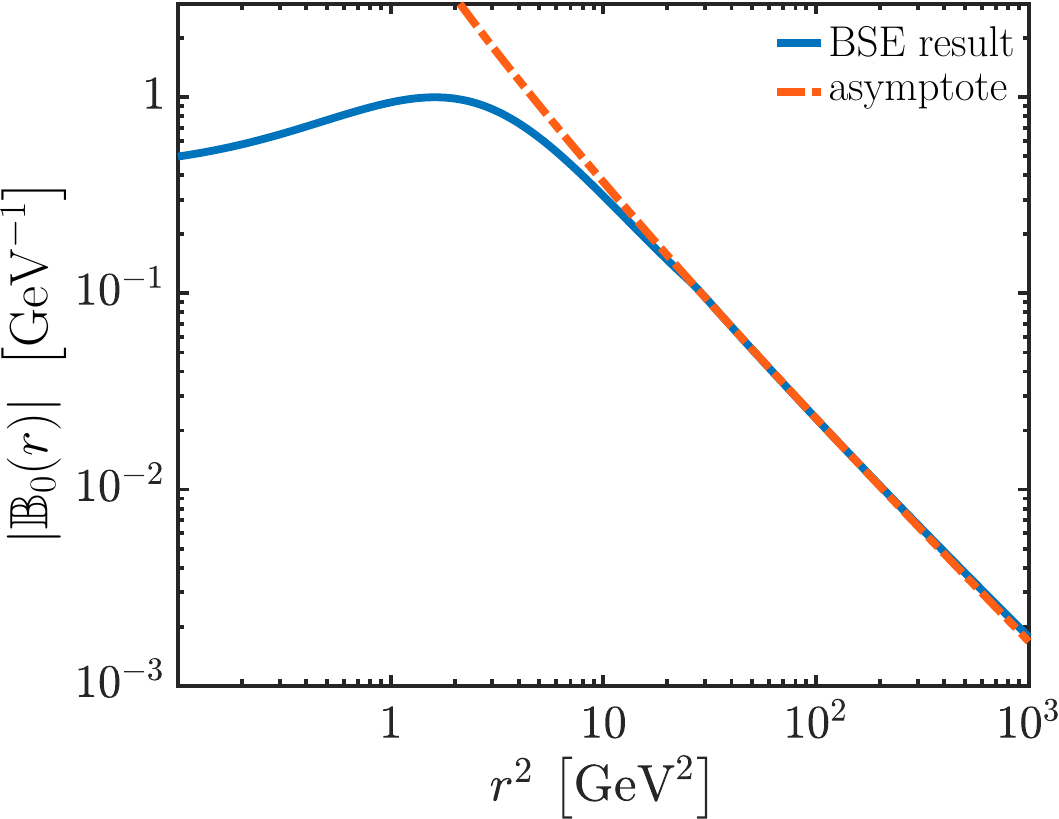}
\caption{Absolute value of the eigenfunction $\Bfat_0(r)$ obtained with the one-gluon exchange kernel, $\Ko(r,k)$, normalized such that it has $1$ as global maximum (blue continuous). The orange dot-dashed line corresponds to the asymptotic form of \1eq{B_asymp}, and accurately approximates $\Bfat_0(r)$ for $r^2 > 25$~GeV$^2$.}
\label{fig:Bfat_asympt}
\end{figure}

Now, the asymptotic behavior of $\Bfat_0(r)$ is not known analytically. Nevertheless, the numerical solutions for $\Bfat_0(r)$ are found to be accurately approximated by the asymptotic form
\be 
\lim_{r \to \infty}\Bfat_0(r) = \frac{a}{r^2 L_{\srm{UV}}^{\kappa}(r^2)} \,, \label{B_asymp}
\ee
for $ r> 5$~GeV, where the constants $a$ and $\kappa$ can be determined by fitting the $\Bfat_0(r)$ data in this range. In \fig{fig:Bfat_asympt} this is shown explicitly for the case when $K(r,k) = \Ko(r,k)$, with the blue continuous curve showing the full solution $\Bfat_0(r)$, whereas the orange dot-dashed represents \1eq{B_asymp} with $a = 3.17$ and $\kappa = 0.958$. The corresponding eigenvalue is $\alpha_\star = 0.685$.

Next, to get $\omega_0$, we set $\Bfat\to\Bfat_0$ into \1eq{omega_spherical}. With the aid of \2eqs{Delta_L_asympt}{B_asymp}, 
it is easy to establish that the 
integral converges rapidly, and its numerical 
evaluation is rather straightforward,
yielding 
$\omega_0 = 0.389$.

In contrast, the expression for $I$ given in \1eq{I_spherical2} is found to converge slowly, requiring 
a more 
careful treatment.

To analyze the ultraviolet convergence of \1eq{I_spherical2}, we first separate 
the integration interval
in two parts, by means of a large momentum scale, $M^2$, \ie
\be
I = \underbrace{ \lambda\omega\int_0^{M^2} \! \!\! dy \, y^2 {\Delta}^2(y) \, \Ls(y) \, \Bfat(y)}_{I_{\srm{IR}}} + \underbrace{ \lambda\omega \int_{M^2}^{\infty} \! \!\! dy \, y^2 {\Delta}^2(y) \, \Ls(y) \, \Bfat(y) }_{I_{\srm{UV}}} \,. \label{I_break}
\ee
For $M$ sufficiently large, all of the inputs in $I_{\srm{UV}}$ can be substituted by their asymptotic forms, namely \2eqs{Delta_L_asympt}{B_asymp}. The resulting integral can then be done analytically,
\begin{align} 
I_{\srm{UV}} \sim&\, \sigma a\lambda\omega \int_{M^2}^{\infty} \! \!\! dy \, \frac{1}{y L^{\kappa+35/44}_{\srm{UV}}(y)} = \frac{\sigma a \lambda\omega}{c^{\kappa + 35/44}(\kappa-9/44)\ln^{\kappa-9/44}(M^2/\Lambda^2_{\srm{MOM}})}  \,, \label{IUV}
\end{align}
provided that $\kappa > 9/44 = 0.204$, which we find to be satisfied by our numerical solutions for $\Bfat(r)$. 

However, the convergence of \1eq{IUV} is slow, and direct numerical integration of \1eq{I_spherical2} becomes unreliable. Instead, in what follows we use \1eq{I_break} with $M = 5$~GeV, and compute $I$ in two steps: first, $I_{\srm{IR}}$ is calculated by direct numerical integration of the interpolated data for $\Bfat(r)$; then, $I_{\srm{UV}}$ is approximated by \1eq{IUV}, and evaluated with the previously determined values of  $a$, $\kappa$ and $\sigma$.
Finally, this procedure yields $I = \pm0.689$~GeV.

Then, using \1eq{C_from_B_euc}, we obtain \mbox{$m = 1.27$~GeV}, and the resulting displacement function $\Cfat(r)$ is shown as a yellow dashed curve in \fig{fig:Cfat_vs_mass}. Evidently, both findings are 
considerably larger than the results obtained 
from the lattice analysis 
of~\cite{Aguilar:2021okw,Aguilar:2022thg} for $\{m_{\srm{lat}},\Cfat_{\srm{lat}}(r)\}$. 
This, in turn, motivates 
the study of the potential impact 
that the form of the kernel may have on these quantities. 

\subsection{Varying the  BSE kernel}\label{subsec:eff_K}

In this subsection, we 
modify the form of the BSE kernel 
for the purpose 
of reaching a better agreement with the lattice results. 

It is clear that, 
in order to determine the 
effect of the higher-order diagrams 
comprising 
$K_{\rm ho}(r,k)$,  one has to carry out 
extensive calculations. Instead, in 
the present analysis we simply treat the full $K(r,k)$ as a deformation of 
the one-gluon exchange $\Ko(r,k)$, 
obtained by varying the function $\Delta(u)$ entering in \1eq{K_oge}. 
We emphasize that this is the only place 
where this function is varied; 
everywhere else, $\Delta$ 
retains its 
standard form.

Specifically, 
in the kernel of the 
BSE for $\Bfat(r)$,  
see \1eq{K_ang}, 
we set 
\be 
K(x,y,\theta) \to K_{\rm eff}(x,y,\theta) = \kin(x,y, \theta){\cal R}_{\rm eff}(u,s) \,, \label{K_ansatz}
\ee
where the function ${\cal R}_{\rm eff}(u,s)$ is given by
\be 
{\cal R}_{\rm eff}(u,s) = \Deltaeff(u) \Ls(s)  \,, \label{R_eff}
\ee
with 
$\Deltaeff(u)$ parametrized 
as 
\be 
\Deltaeff(u) = \Delta(u)\times\left[  1 + \frac{c_0 u^2}{1 + c_1 u^2 + c_2 u^4 } \right] \,. \label{Delta_eff_fit}
\ee
Note that 
$\Ko$ is recovered
by setting $c_0 = 0$ in \1eq{Delta_eff_fit}.  
Moreover, at large momenta, $\Deltaeff(u)$
reduces to $\Delta(u)$, 
such that the $K_{\rm eff}$ reduce to $\Ko$.

We next vary the 
$c_i$ in   
\1eq{Delta_eff_fit}
within certain intervals, and 
consider the resulting 
behaviour of $\{m,\,\Cfat\}$. 
Upon systematic inspection of the parameter space, 
we conclude that the best values for  
$c_1$ and $c_2$ are $c_1 = 0.00667$~GeV$^{-2}$ and $c_2 = 0.0486$~GeV$^{-4}$; we therefore keep $c_1$ and $c_2$
fixed at these values, and only vary $c_0$.

\begin{figure}[!ht]
  \centering
  \includegraphics[width=0.45
\textwidth]{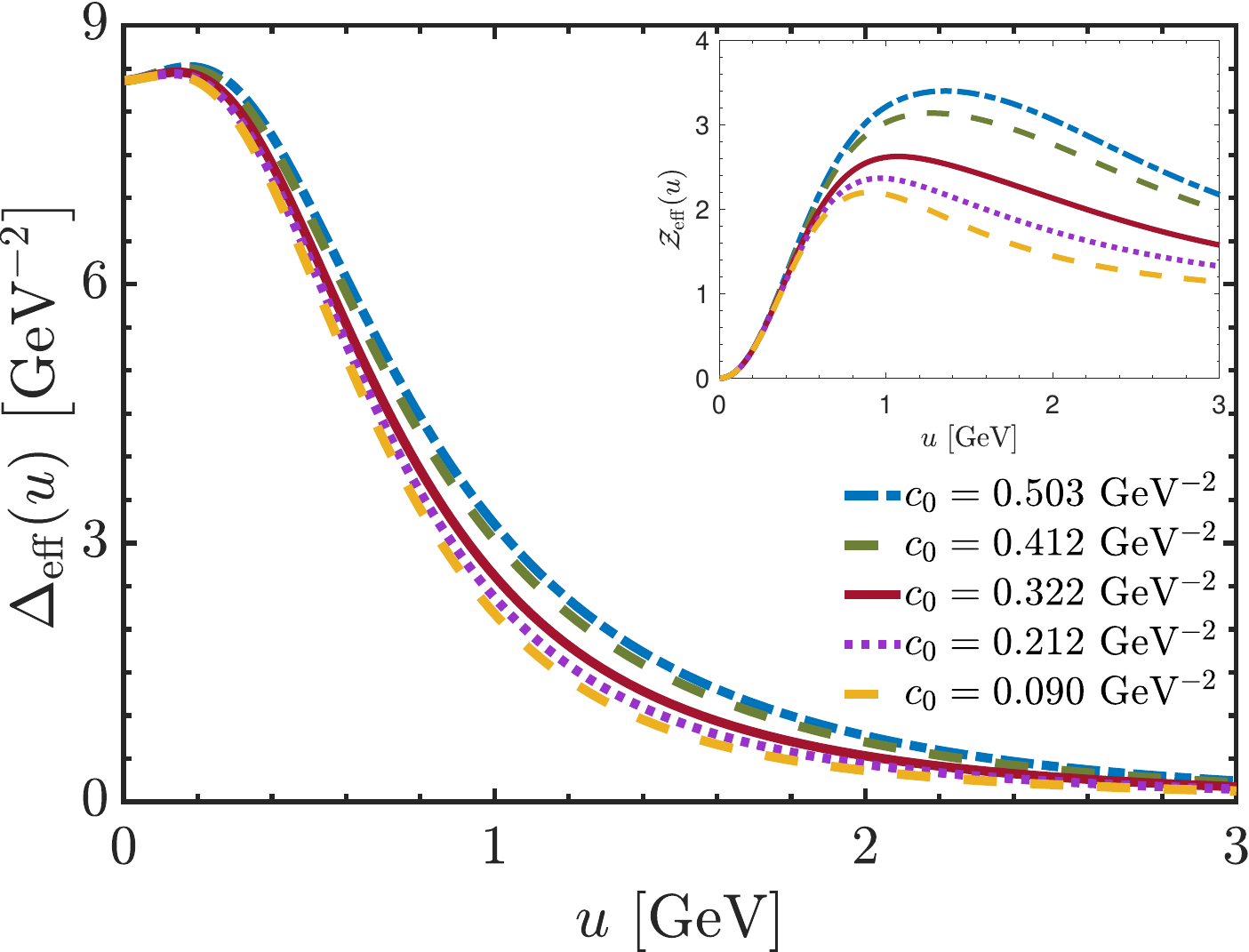}\hfil\includegraphics[width=0.45
\textwidth]{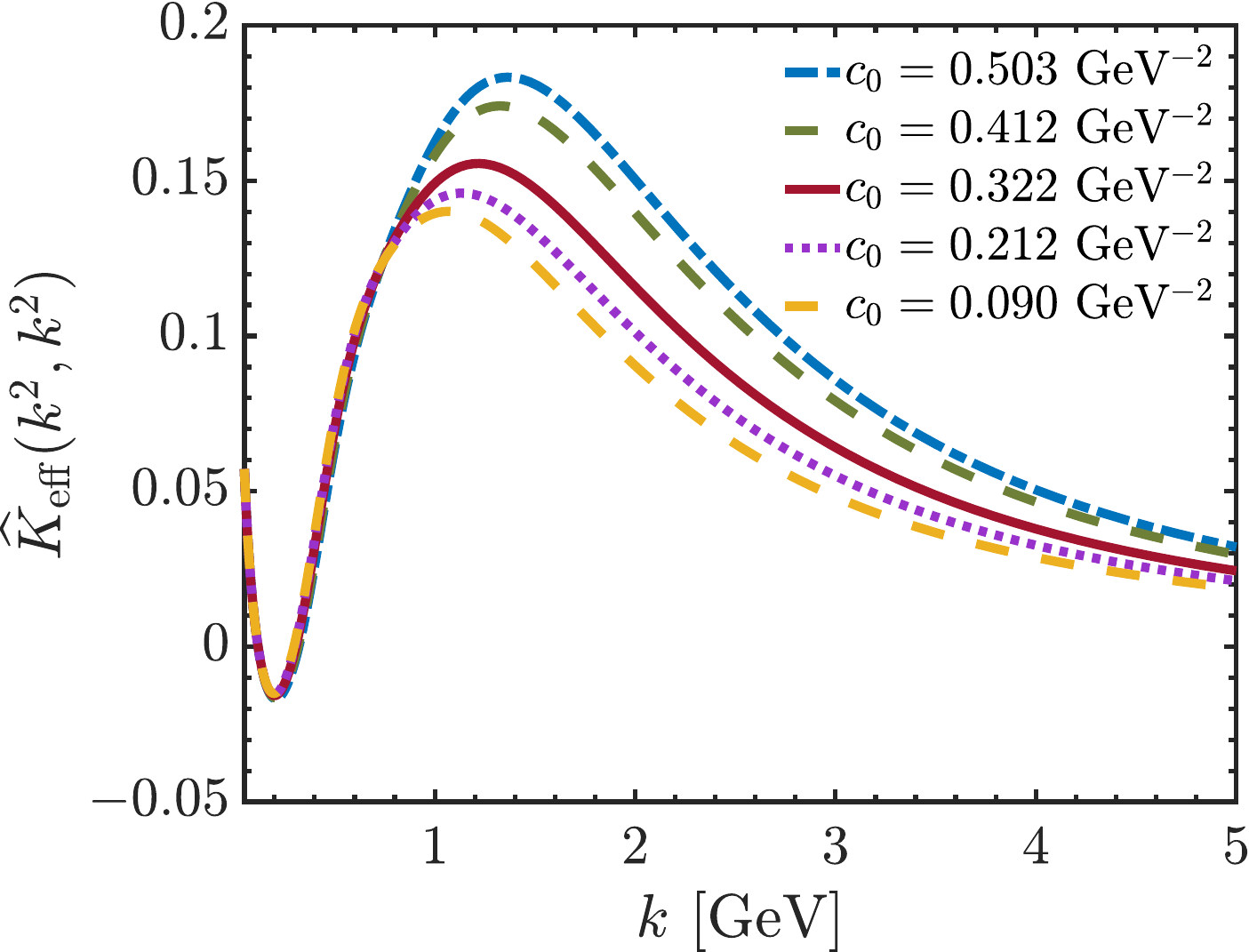}\\
\includegraphics[width=0.45
\textwidth]{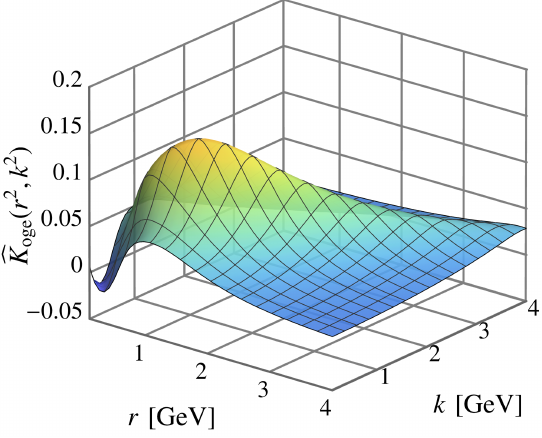}\hfil  \includegraphics[width=0.45
\textwidth]{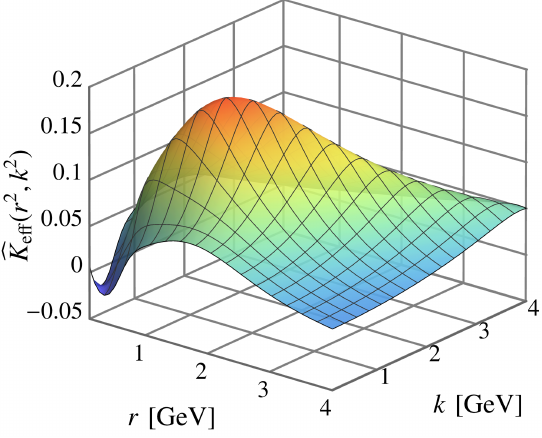}
\caption{ Top left: A family of variations of $\Deltaeff(u)$, obtained by varying the parameter $c_0$ in \1eq{Delta_eff_fit}. The inset shows the corresponding dressing functions, ${\cal Z}_{\rm{eff}}(q) := q^2\Deltaeff(q)$. Top right: Diagonal $\Khat_{\rm{eff}}(k^2,k^2)$ of the angular-integrated kernel $\Khat_{\rm{eff}}(r^2,k^2)$, defined by \1eq{K_ang_int}, corresponding to the family of $\Deltaeff(u)$ on the left panel. Bottom: General kinematics $\Khat(r^2,k^2)$ for $K\to\Ko$ (left), $K\to K_{\rm{eff}}$ (right). Note that 
the maximum of $K_{\rm{eff}}$ is enhanced by about $30\%$ with respect to that of $\Khat_{\rm{oge}}$.}
 \label{fig:kern_vs_mass}
\end{figure}

On the top left panel 
of \fig{fig:kern_vs_mass}
we show the family of  $\Deltaeff(r)$ obtained by varying $c_0$ within the interval $[0,0.503]$~GeV$^{-2}$,
while 
on the top right panel
the diagonals ($r^2 = k^2$) of the resulting $\Khat_{\rm{eff}}(r^2,k^2)$ are displayed. As we see,  enhancing $\Deltaeff(u)$ leads to an enhancement of $\Khat_{\rm{eff}}(r^2,k^2)$.
In the bottom 
panel one may see  
how the entire shape 
of the angular-integrated kernel gets modified 
when $c_0$ varies from 
$c_0 =0$, corresponding to 
the $\Khat_{\rm{oge}}(r^2,k^2)$,
to the case $c_0 = 0.503$~GeV$^{-2}$.
Note, in particular, that 
the maximum of $\Khat_{\rm{eff}}(r^2,k^2)$
is enhanced by about $30\%$ with respect to that of $\Khat_{\rm{oge}}$. 

\begin{figure}[!ht]
  \centering
  \includegraphics[width=0.45
\textwidth]{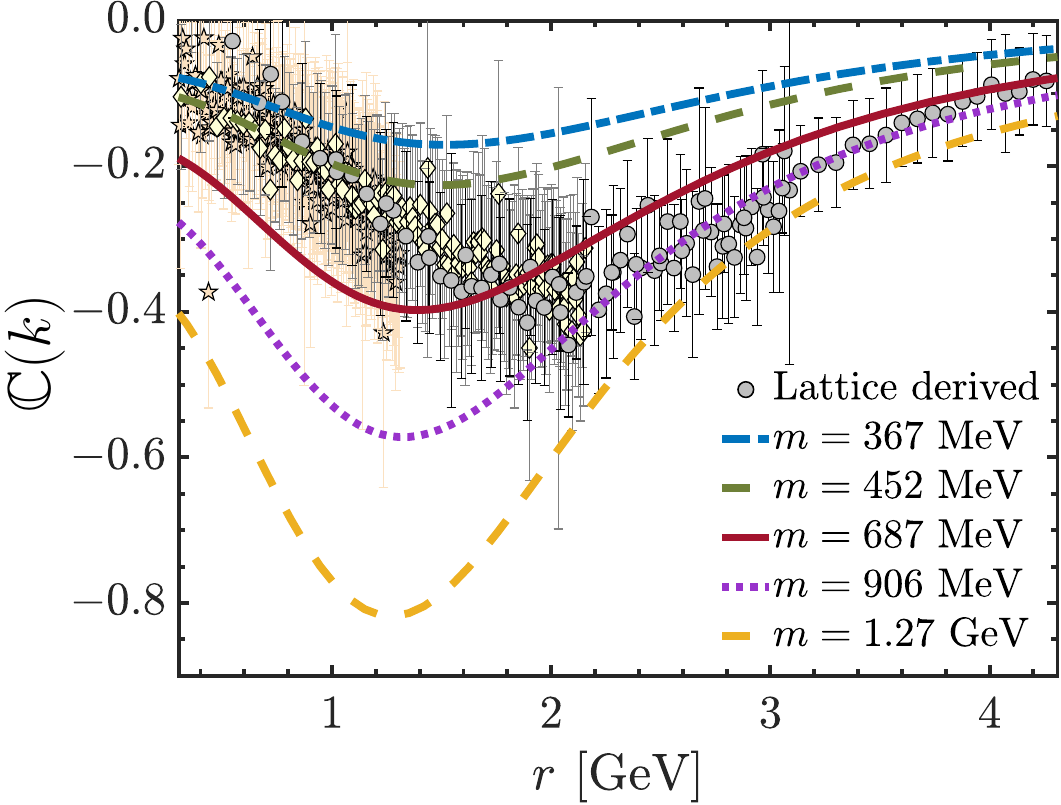}\hfil\includegraphics[width=0.45
\textwidth]{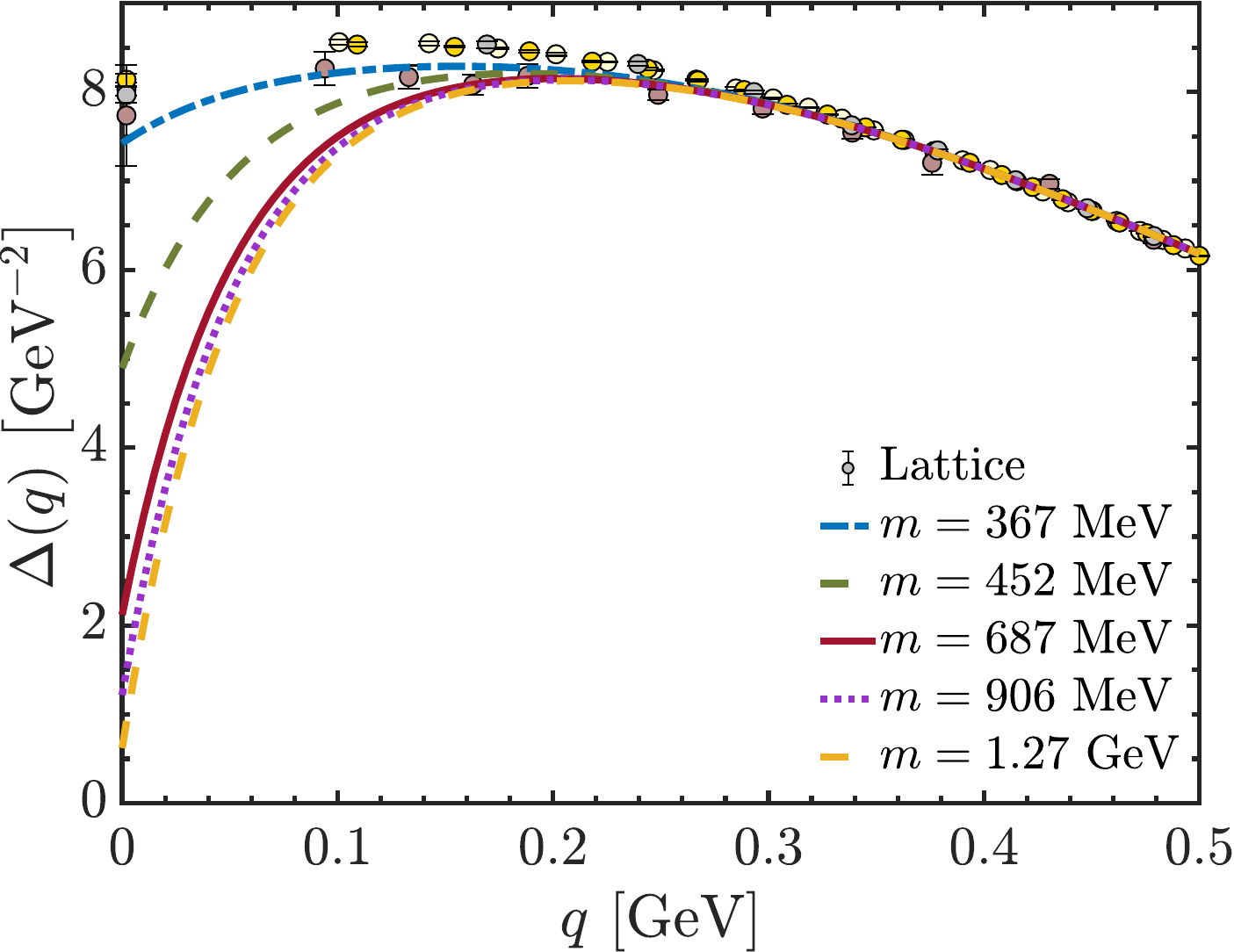}
\caption{ Left: Displacement amplitude, $\Cfat(r)$, for different values of the parameter $c_0$ (see legends of \fig{fig:kern_vs_mass}) of \1eq{Delta_eff_fit}. The lattice-derived results of \cite{Aguilar:2022thg} are shown as points for comparison. Note that the red dashed line lies entirely within the errors of $\Cfat_{\srm{lat}}(r)$. Right: Gluon propagator, $\Delta(q)$, for different masses, assuming the functional form of \1eq{Delta_family}, compared to the lattice data of \cite{Aguilar:2021okw}.  }\label{fig:Cfat_vs_mass}
\end{figure}

Next, we compute the values of $m$ and $\Cfat(r)$ corresponding to the family of $\Deltaeff(u)$ of \fig{fig:kern_vs_mass}. The resulting displacement amplitudes are shown on the left panel of \fig{fig:Cfat_vs_mass}, with the corresponding values of $m$ given in the legend. 

Note that, for $c_0 = 0.212$~GeV$^{-2}$ (red continuous curve on the panel of \fig{fig:Cfat_vs_mass}), the resulting $\Cfat(r)$ is statistically consistent with $\Cfat_{\srm{lat}}(r)$; however, for this value of $c_0$, the mass turns out to be $m = 687$~MeV, nearly twice the value of $m_{\srm{lat}}$. Variation of the remaining parameters of \1eq{Delta_eff_fit} show the same pattern: improving the agreement between $\Cfat(r)$ and $\Cfat_{\srm{lat}}(r)$ causes $m$ to exceed $m_{\srm{lat}}$.  

Lastly, we illustrate the effect of the masses obtained in the \fig{fig:Cfat_vs_mass} on the gluon propagator. Evidently, determining $\Delta(q)$ self-consistently requires solving the coupled system of equations comprised by the gluon propagator and three-gluon vertex SDEs, which is beyond the scope of the present study. Instead, we assume that $\Delta(q)$ is modified only near the origin by the value of $m$, recovering the lattice results for large $q$. A functional form that implements this assumption is
\be 
\Delta(q) = \Delta_{\srm{fit}}(q) - [\Delta_{\srm{fit}}(0) - 1/m^2 ]\exp(- q/\nu ) \,, \label{Delta_family}
\ee
where $\nu = 0.05$~GeV, and $\Delta_{\srm{fit}}(q)$ is the fit to the lattice data of \cite{Aguilar:2021okw} given by Eq.~(C.11) of \cite{Aguilar:2021uwa}. The resulting curves are shown on the right panel of \fig{fig:Cfat_vs_mass}.

\section{Discussion and conclusions}\label{sec:conc}

In the present work we have 
addressed important open   
issues related to
the implementation of the 
SM in a 
non-Abelian context.
In particular, the question of 
how the scale of the gluon mass may be
fixed has been solved, at least in principle, by introducing a 
nonlinear component into 
the BSE that controls the 
bound state amplitude, $\Bfat(r)$,
of the massless pole formation.
In addition, we have put forth 
a novel approach for dealing 
with the problem of 
the multiplicative renormalizability 
of the gluon mass equation, thus 
reaching a manifestly finite answer for 
the gluon mass. The success of this renormalization procedure hinges on 
the implementation of a 
key cancellation, which proceeds through the judicious combination of the 
SDE for the three-gluon vertex 
and the BSE for $\Bfat(r)$. 
It turns out that the mathematical origin of  
this cancellation may be traced back to  
Fredholm's alternatives theorem, operating 
at the level of the SDE-BSE system that we consider. The nonlinear character of the 
BSE is crucial for evading this theorem, 
which would 
otherwise 
enforce the strict vanishing of the gluon mass. 

For the purpose of simplifying this rather  technical subject, 
in our presentation we have considered only the minimum 
of diagrams required for 
exposing the crucial cancellations [see Sec.~\ref{sec:trick}]
and their 
connection to Fredholm's theorem. 
It is, however, of the utmost importance to 
emphasize 
that, as we have explicitly confirmed, the omitted contributions undergo themselves 
completely analogous cancellations, 
which proceed in exactly the same way, 
and for precisely the same reason. 
Actually, the construction related to the renormalization 
may be repeated, involving the ghost-gluon and four-gluon vertices, their SDE-BSE systems, and the corresponding 
vertex renormalization constants associated with them. 
In fact, 
the only contributions that survive, 
and may induce modifications to  
the results presented here originate 
from the non-linear components  
of the additional BSEs that control the formation of 
poles in the ghost-gluon and four-gluon vertices.
The detailed elaboration of these 
issues will be presented in a  
forthcoming communication. 

The numerical treatment presented in 
Sec.~\ref{sec:res} focused on two principal   
quantities,  namely the value of the gluon mass,
$m$, and the shape and size of the displacement 
function, $\Cfat(r)$. 
The procedure followed consists in producing a sequence of 
models for 
$K(r,k)$, in an attempt to optimize the resulting set $\{m,\,\Cfat(r)\}$ 
with respect to the lattice-based benchmarks for these quantities. 
The analysis reveals 
a preference towards gluon masses 
that are larger than what the saturation point 
of the lattice gluon propagator would indicate. 
In particular, the 
red continuous curve in \fig{fig:Cfat_vs_mass},   
which shows the 
best coincidence 
with the $\Cfat(r)$ of~\cite{Aguilar:2022thg}, 
is obtained for $m=687$ MeV, 
which is to be contrasted to the $m_{\srm{lat}} = 354$~MeV~\cite{Aguilar:2021okw}.
As we discuss below,  
there are 
at least three reasons that could account for   
this relative tension between 
$m$ and $\Cfat(r)$. 

First, for our numerical analysis we have essentially 
resorted to models for the kernel, using the one-gluon exchange 
version as our reference. Evidently, 
the next important task in this context 
would be to actually compute the modifications  
induced to $K(r,k)$ by the inclusions of the one-loop dressed 
diagrams ($b_3$)-($b_6$) appearing in the skeleton expansion of \fig{fig:Kern_diags}. 
The result of such a computation 
is likely to modify the shape of 
$\Bfat(r)$, in ways not captured 
by the modeling described by \2eqs{K_ansatz}{Delta_eff_fit}, thus improving the matching between  
$m$ and $\Cfat(r)$. Calculations in this direction 
are already underway; we hope to report on the 
results of this study in the near future.

Second, the indirect determination of $\Cfat(r)$
presented in~\cite{Aguilar:2022thg}
involves a particular partial derivative 
 of the ghost-gluon kernel, $H_{\nu\mu}(p,q,r)$, 
appearing in the STI of \1eq{STI}. This derivative 
has never been simulated on the lattice; 
in the analysis of~\cite{Aguilar:2022thg} it 
was estimated with the aid of the ``one-loop dressed'' SDE 
that controls the evolution of  
$H_{\nu\mu}(p,q,r)$. 
Preliminary analysis indicates that a relatively small change in the functional form of 
this derivative would suffice for bringing $m$ and 
$\Cfat(r)$ to a closer agreement. It is 
therefore 
important to examine the modifications  
induced to this quantity when further contributions to the 
SDE are included; in particular, 
such a study requires the 
inclusion into the aforementioned SDE 
of the ghost-ghost-gluon-gluon vertex, first 
explored in~\cite{Huber:2017txg}.

Third, as mentioned above, the inclusion of 
the remaining fundamental vertices of the theory, 
namely the ghost-gluon and four-gluon vertices, 
is likely to modify the expression for $I$ given in
\1eq{Iomega}.
In particular, the non-linearities in the 
BSEs that control the formation of the 
composite scalar through the fusion of a 
ghost-antighost pair or three gluons 
will provide additional finite contributions to the $I$,
thus affecting the value of the gluon mass. 
The inclusion of the ghost-gluon vertex and the associated 
pole dynamics is well within our grasp, and 
various of its main aspects have already been addressed 
in the literature~\cite{Aguilar:2016vin,Aguilar:2017dco,Aguilar:2021uwa,Eichmann:2021zuv}. On the other hand, the 
pole content and structure of the four-gluon vertex are
largely unexplored, and quantitative information is 
rather difficult to obtain~\cite{Binosi:2014kka,Cyrol:2014kca,Pawlowski:2022zhh,Huber:2018ned,Colaco:2024gmt,Aguilar:2024fen}; nonetheless, the 
understanding acquired from the present analysis 
is expected to facilitate future attempts in this direction.

\section{Acknowledgments}
\label{sec:acknowledgments}

The work of M.~N.~F. is supported by the National Natural Science Foundation of China (grant no. 
12135007). The research of 
J.~P. is supported by the Spanish MICINN grant PID2020-113334GB-I00,
the Generalitat Valenciana grant CIPROM/2022/66, and in part by the EMMI visiting grant of 
the ExtreMe Matter Institute EMMI
at the GSI Helmholtzzentrum f\"ur Schwerionenforschung GmbH, Darmstadt, 
Germany. J.~P. thanks J.~M.~Pawlowski 
for his kind hospitality at the 
Institute for Theoretical Physics, University of 
Heidelberg, where part of this work was completed,
and for numerous stimulating discussions.

\appendix

\section{A toy model}\label{sec:toy}

In this Appendix we present a concrete 
example of the mathematical construction 
described in Sec.\ref{sec:trick}, leading from 
\1eq{gf} to \1eq{betaren}.

In order to make the analogy with the equations for $\Ls(r)$, $\Bfat(r)$, and $I$ as realistic as possible, 
we use an integration interval from $0$ to $\Lambda^2$, with $\Lambda$ an ultraviolet cutoff, which will eventually be taken to infinity.

For simplicity, we consider the case of a separable kernel, $K(x,y) = A(x)B(y)$, for which \1eq{gf} can be solved formally. Then, the symmetry of the kernel under the exchange of $x$ and $y$ implies $B(y) = A(y)$. Moreover, we set $u(x) = 1$. In this case, \1eq{gf} simplifies to
\be
g(x) = z + A(x)\int_{0}^{\Lambda^2} \!\!\!dy \, g (y) A(y) \,, \qquad 
f(x) = b^{-1}A(x)\!\int_{0}^{\Lambda^2} \!\!\! dy \, f(y) A(y) \,.
\label{gf_cutoff}
\ee

To fix the ideas, let us first ignore convergence issues of the integrals in \1eq{gf_cutoff} as $\Lambda\to \infty$. Then, the formal solution of \1eq{gf_cutoff} for $g(x)$ and $f(x)$ is read off immediately, 
\be
g(x) = z + c A(x) \,, \qquad\qquad f(x) = d A(x) \,, 
\label{gf_sol}
\ee
where $c$ and $d$ are so-far unknown constants, defined as
\be 
c = \int_{0}^{\Lambda^2} \!\!\!dy \, g (y) A(y) \,, \qquad\qquad d = b^{-1}\!\int_{0}^{\Lambda^2} \!\!\! dy \, f(y) A(y) \,. \label{cd_const}
\ee

Next, the constant $c$ can be determined by substituting \1eq{gf_sol} into the first of \1eq{cd_const} to eliminate $g(y)$. This yields a linear equation for $c$, whose solution reads
\be 
c = z\left[ 1 - \int_{0}^{\Lambda^2} \!\!\!dw \, A^2(w) \right]^{-1} \int_{0}^{\Lambda^2} \!\!\!dy \, A(y) \,.\label{c_sol}
\ee

On the other hand, substituting \1eq{gf_sol} into the second of \1eq{cd_const} leads to
\be
d = d b^{-1}\!\int_{0}^{\Lambda^2} \!\!\! dy \, A^2(y) \,, \label{b_eq}
\ee
which leaves $d$ undetermined. That is to be expected, since the equation for $f(x)$ is homogeneous; its solutions can only be determined up to a multiplicative constant. Instead, \1eq{b_eq} determines $b$, which plays the role of eigenvalue. Specifically,
\be
b = \int_{0}^{\Lambda^2} \!\!\! dy \, A^2(y) \,. \label{b_sol}
\ee

Now we consider the two forms of $\beta$, given by \2eqs{beta}{betaren}. Using \1eq{gf_sol}, the \2eqs{beta}{betaren} reduce to
\be
\beta= z \, d \! \int_{0}^{\Lambda^2} \!\! dx \, A(x)\,, \qquad\qquad \beta = d\,(1 - b)\!\int_{0}^{\Lambda^2} \!\! dx \, A(x)[ z + c A(x)] \,.
\label{beta_sep}
\ee
Then, using \2eqs{c_sol}{b_sol} for $c$ and $b$, one finds by simple algebra that the two expressions yield the same result.

To make this example more concrete, let $A(x)=e^{-x}$. With this kernel, all of the above integrals converge in the limit $\Lambda\to \infty$, which we can take directly. Then, we can set $z =1$, such that \1eq{gf} reduces to
\be
g(x) = 1 + \int_{0}^{\infty} \!\!\!dy \, g (y) e^ {-x-y} \,, \qquad\qquad 
f(x) = b^{-1}\!\int_{0}^{\infty} \!\!\! dy \, f(y) e^ {-x-y} \,.
\label{gf_conc}
\ee
The solution is found from Eqs.~\eqref{gf_sol},~\eqref{c_sol}, and~\eqref{b_sol}, which yield $b = 1/2$, $c = 2$, and
\be
g(x) = 1 + 2 e^{-x} \,, \qquad\qquad f(x) = d e^{-x} \,,
\label{gf_sol_conc}
\ee
with $d$ an undetermined constant. These results are readily verified to satisfy \1eq{gf_conc}. 
Finally, both forms of \1eq{beta_sep} yield $\beta=d$.

Now, the actual equations for $\Ls(r)$ and $\Bfat(r)$ contain ultraviolet divergences that are cured by renormalization. To emulate this situation within our toy model, let us choose 
\be 
A(x) = \frac{a}{x + \sat^2} \,, \label{A_power}
\ee
with $a$ and $h$ constant. With this kernel, no infrared divergences arise, but the treatment of the $\Lambda\to \infty$ limit requires additional care.

The homogeneous equation for $f(x)$ does not present a problem; its solution is still given by \1eq{gf_sol_conc}, with the ultraviolet finite eigenvalue
\be
b = \int_{0}^{\Lambda^2} \!\!\! dy \, A^2(y) =  \frac{a^2\Lambda^2}{\sat^2(\Lambda^2 + \sat^2)}\,, \qquad\qquad \lim_{\Lambda\to \infty} b = \frac{a^2}{\sat^2} \,.\label{b_sol_ren}
\ee

However, if $z$ were 
cutoff-independent, say $z = 1$, the inhomogeneous equation for $g(x)$ in \1eq{gf_cutoff} would lead to a divergent result. Specifically, from \1eq{c_sol},
\be 
c = \frac{a \sat^2 (\Lambda^2 + \sat^2 ) }{ ( \sat^2 - a^2 )\Lambda^2 + \sat^4} \ln\left( \frac{\Lambda^2 + \sat^2}{\sat^2} \right) \,,
\ee
such that
\be 
\lim_{\Lambda \to \infty} c = \frac{a\sat^2}{\sat^2 - a^2}\ln\left( \frac{\Lambda^2}{\sat^2} \right) \,,
\ee
\ie $c$, and therefore $g(x)$, diverge logarithmically, in close analogy to the unrenormalized $\Ls(x)$. 

Let us then choose $z$ in \1eq{gf_cutoff} to be 
cutoff-dependent, in such a way that it cancels the divergence of the integral in the equation for $g(x)$. To this end, we impose a ``renormalization condition'', $g(\mu^2)=1$, at some renormalization point, $\mu^2$. This condition determines $z$ as
\be 
z = 1 - A(\mu^2)\int_{0}^{\Lambda^2} \!\!\!dy \, g (y) A(y) \,. \label{z}
\ee
Then, the first of \1eq{gf} becomes
\be 
g(x) = 1 + \left[ A(x) - A(\mu^2)\right]\int_0^{\Lambda^2}\!\!\!dy \, g (y) A(y) \,,
\ee
while the equation for $f(x)$ is unchanged.

The solution of the system is reached through the same steps used to obtain \2eqs{gf_sol}{cd_const}. Specifically, we find
\be
g(x) = 1 + c_{\srm R} [A(x) - A(\mu^2)] \,, \qquad\qquad f(x) = d_{\srm R} A(x) \,, 
\label{gf_sol_ren}
\ee
with
\begin{align} 
c_{\srm R} =&\, \left\lbrace 1 - \int_{0}^{\Lambda^2} \!\!\!dw \,A(w)\left[ A(w) - A(\mu^2)\right] \right\rbrace^{-1} \int_{0}^{\Lambda^2} \!\!\!dy \,A(y) \nonumber\\
=&\, \frac{a \sat^2 \left(\Lambda^2+\sat^2\right)
   \left(\mu^2+\sat^2\right) \ln \left(1 + \Lambda^2/\sat^2\right)}{\left(\mu^2+\sat^2\right) \left[ \Lambda^2( \sat^2 - a^2 ) + \sat^4\right] + a^2 \sat^2
   \left(\Lambda^2+\sat^2\right) \ln \left(1 + \Lambda^2/\sat^2\right)} \,, \label{c_sol_ren}
\end{align}
$d_{\srm R}$ arbitrary, and the eigenvalue $b$ given by \1eq{b_sol_ren}.

Now, the constant $c_{\srm R}$, and hence $g(x)$, is finite. In fact, for $\Lambda\to \infty$, $c_{\srm R}$ takes the simple form
\be 
\lim_{\Lambda\to \infty}c_{\srm R} = \frac{\mu^2 + \sat^2}{a} \,.
\ee

Next, we consider the two expressions for $\beta$ in \1eq{beta_sep}. To calculate the first form, we first need to know the explicit value of $z$; this is achieved by substituting the solution given in \2eqs{gf_sol_ren}{c_sol_ren} for $g(x)$ into \1eq{z}, which furnishes
\be 
z = \frac{\left(\mu^2+\sat^2\right) \left[ \Lambda^2( \sat^2 - a^2 ) + \sat^4\right]}{\left(\mu^2+\sat^2\right) \left[ \Lambda^2( \sat^2 - a^2 ) + \sat^4\right] + a^2 \sat^2 \left(\Lambda ^2+\sat^2\right) \ln \left(1 + \Lambda^2/\sat^2\right)} \,.
\ee
Then, with a little algebra, both expressions in \1eq{beta_sep} can be shown to yield the same result, namely
\be 
\beta = \frac{a d \left(\mu^2+\sat^2\right) \left[ \Lambda^2( \sat^2 - a^2 ) + \sat^4\right] \ln \left(1 + \Lambda^2/\sat^2\right)}{\left(\mu^2+\sat^2\right) \left[ \Lambda^2( \sat^2 - a^2 ) + \sat^4\right] + a^2 \sat^2 \left(\Lambda^2+\sat^2\right) \ln \left(1 + \Lambda^2/\sat^2\right)} \,,
\ee
which for $\Lambda\to \infty$ has the finite value 
\be 
\lim_{\Lambda\to\infty}\beta = \frac{d(\mu^2 + \sat^2)(\sat^2 - a^2)}{a\sat^2} \,.
\ee

We conclude this exercise by considering the scale-fixing of the homogeneous equation; this can be achieved by imposing an additional relation between $b$ and $f$. 

To retain a close analogy with the situation encountered in the equations for $\Bfat(r)$, we recall that, in our toy model, $b$ plays the role of the $t$ appearing in \1eq{BSEhomren}. Then, by analogy to \1eq{thet}, we impose that $b$ must satisfy
\be 
b = 1 - \int_{0}^{\Lambda^2} \!\!\!dy \, f^2(y) \,, \label{scale_toy}
\ee
with the integral on the r.h.s. emulating the $\omega$ of \1eq{thet}. Note that since \1eq{scale_toy} is quadratic in $f(y)$, \2eqs{A_power}{gf_sol_ren} yield an ultraviolet finite $b$, namely
\be 
b = 1 - \frac{d_{\srm R}^2 a^2\Lambda^2}{\sat^2(\Lambda^2 + \sat^2)} \,, \qquad\qquad \lim_{\Lambda\to\infty} b = 1 - \frac{d_{\srm R}^2a^2}{\sat^2} \,. \label{b_sol_scale}
\ee

Now, \2eqs{b_sol_ren}{b_sol_scale} must yield the same value of $b$. Hence, equating them fixes the scale of $d_{\rm R}$ to
\be 
d_{\srm R} = \pm\sqrt{ \frac{\sat^2(\Lambda^2 + \sat^2)}{a^2\Lambda^2} - 1} \,, \qquad\qquad \lim_{\Lambda\to\infty}d_{\srm R} = \pm\sqrt{\frac{\sat^2}{a^2} - 1} \,.
\ee
Since \1eq{scale_toy} is quadratic in $f(y)$, the sign of  $d_{\srm R}$ is left undetermined, similarly to the sign of $\sigma$ in \1eq{scale_setting}.

%


\begin{thebibliography}{146}%
\makeatletter
\providecommand \@ifxundefined [1]{%
 \@ifx{#1\undefined}
}%
\providecommand \@ifnum [1]{%
 \ifnum #1\expandafter \@firstoftwo
 \else \expandafter \@secondoftwo
 \fi
}%
\providecommand \@ifx [1]{%
 \ifx #1\expandafter \@firstoftwo
 \else \expandafter \@secondoftwo
 \fi
}%
\providecommand \natexlab [1]{#1}%
\providecommand \enquote  [1]{``#1''}%
\providecommand \bibnamefont  [1]{#1}%
\providecommand \bibfnamefont [1]{#1}%
\providecommand \citenamefont [1]{#1}%
\providecommand \href@noop [0]{\@secondoftwo}%
\providecommand \href [0]{\begingroup \@sanitize@url \@href}%
\providecommand \@href[1]{\@@startlink{#1}\@@href}%
\providecommand \@@href[1]{\endgroup#1\@@endlink}%
\providecommand \@sanitize@url [0]{\catcode `\\12\catcode `\$12\catcode
  `\&12\catcode `\#12\catcode `\^12\catcode `\_12\catcode `\%12\relax}%
\providecommand \@@startlink[1]{}%
\providecommand \@@endlink[0]{}%
\providecommand \url  [0]{\begingroup\@sanitize@url \@url }%
\providecommand \@url [1]{\endgroup\@href {#1}{\urlprefix }}%
\providecommand \urlprefix  [0]{URL }%
\providecommand \Eprint [0]{\href }%
\providecommand \doibase [0]{https://doi.org/}%
\providecommand \selectlanguage [0]{\@gobble}%
\providecommand \bibinfo  [0]{\@secondoftwo}%
\providecommand \bibfield  [0]{\@secondoftwo}%
\providecommand \translation [1]{[#1]}%
\providecommand \BibitemOpen [0]{}%
\providecommand \bibitemStop [0]{}%
\providecommand \bibitemNoStop [0]{.\EOS\space}%
\providecommand \EOS [0]{\spacefactor3000\relax}%
\providecommand \BibitemShut  [1]{\csname bibitem#1\endcsname}%
\let\auto@bib@innerbib\@empty
\bibitem [{\citenamefont {Yang}\ and\ \citenamefont
  {Mills}(1954)}]{Yang:1954ek}%
  \BibitemOpen
  \bibfield  {author} {\bibinfo {author} {\bibfnamefont {C.-N.}\ \bibnamefont
  {Yang}}\ and\ \bibinfo {author} {\bibfnamefont {R.~L.}\ \bibnamefont
  {Mills}},\ }\href {https://doi.org/10.1103/PhysRev.96.191} {\bibfield
  {journal} {\bibinfo  {journal} {Phys. Rev.}\ }\textbf {\bibinfo {volume}
  {96}},\ \bibinfo {pages} {191} (\bibinfo {year} {1954})}\BibitemShut
  {NoStop}%
\bibitem [{\citenamefont {Gross}\ and\ \citenamefont
  {Wilczek}(1973)}]{Gross:1973id}%
  \BibitemOpen
  \bibfield  {author} {\bibinfo {author} {\bibfnamefont {D.~J.}\ \bibnamefont
  {Gross}}\ and\ \bibinfo {author} {\bibfnamefont {F.}~\bibnamefont
  {Wilczek}},\ }\href {https://doi.org/10.1103/PhysRevLett.30.1343} {\bibfield
  {journal} {\bibinfo  {journal} {Phys. Rev. Lett.}\ }\textbf {\bibinfo
  {volume} {30}},\ \bibinfo {pages} {1343} (\bibinfo {year}
  {1973})}\BibitemShut {NoStop}%
\bibitem [{\citenamefont {Politzer}(1973)}]{Politzer:1973fx}%
  \BibitemOpen
  \bibfield  {author} {\bibinfo {author} {\bibfnamefont {H.~D.}\ \bibnamefont
  {Politzer}},\ }\href {https://doi.org/10.1103/PhysRevLett.30.1346} {\bibfield
   {journal} {\bibinfo  {journal} {Phys. Rev. Lett.}\ }\textbf {\bibinfo
  {volume} {30}},\ \bibinfo {pages} {1346} (\bibinfo {year}
  {1973})}\BibitemShut {NoStop}%
\bibitem [{\citenamefont {Marciano}\ and\ \citenamefont
  {Pagels}(1978)}]{Marciano:1977su}%
  \BibitemOpen
  \bibfield  {author} {\bibinfo {author} {\bibfnamefont {W.~J.}\ \bibnamefont
  {Marciano}}\ and\ \bibinfo {author} {\bibfnamefont {H.}~\bibnamefont
  {Pagels}},\ }\href {https://doi.org/10.1016/0370-1573(78)90208-9} {\bibfield
  {journal} {\bibinfo  {journal} {Phys. Rept.}\ }\textbf {\bibinfo {volume}
  {36}},\ \bibinfo {pages} {137} (\bibinfo {year} {1978})}\BibitemShut
  {NoStop}%
\bibitem [{\citenamefont {Faddeev}\ and\ \citenamefont
  {Popov}(1967)}]{Faddeev:1967fc}%
  \BibitemOpen
  \bibfield  {author} {\bibinfo {author} {\bibfnamefont {L.~D.}\ \bibnamefont
  {Faddeev}}\ and\ \bibinfo {author} {\bibfnamefont {V.~N.}\ \bibnamefont
  {Popov}},\ }\href {https://doi.org/10.1016/0370-2693(67)90067-6} {\bibfield
  {journal} {\bibinfo  {journal} {Phys. Lett. B}\ }\textbf {\bibinfo {volume}
  {25}},\ \bibinfo {pages} {29} (\bibinfo {year} {1967})}\BibitemShut {NoStop}%
\bibitem [{\citenamefont {Becchi}\ \emph {et~al.}(1975)\citenamefont {Becchi},
  \citenamefont {Rouet},\ and\ \citenamefont {Stora}}]{Becchi:1974md}%
  \BibitemOpen
  \bibfield  {author} {\bibinfo {author} {\bibfnamefont {C.}~\bibnamefont
  {Becchi}}, \bibinfo {author} {\bibfnamefont {A.}~\bibnamefont {Rouet}},\ and\
  \bibinfo {author} {\bibfnamefont {R.}~\bibnamefont {Stora}},\ }\href
  {https://doi.org/10.1007/BF01614158} {\bibfield  {journal} {\bibinfo
  {journal} {Commun. Math. Phys.}\ }\textbf {\bibinfo {volume} {42}},\ \bibinfo
  {pages} {127} (\bibinfo {year} {1975})}\BibitemShut {NoStop}%
\bibitem [{\citenamefont {Tyutin}(1975)}]{Tyutin:1975qk}%
  \BibitemOpen
  \bibfield  {author} {\bibinfo {author} {\bibfnamefont {I.~V.}\ \bibnamefont
  {Tyutin}},\ }\href@noop {} {\bibfield  {journal} {\bibinfo  {journal}
  {LEBEDEV-75-39}\ } (\bibinfo {year} {1975})}\BibitemShut {NoStop}%
\bibitem [{\citenamefont {Becchi}\ \emph {et~al.}(1976)\citenamefont {Becchi},
  \citenamefont {Rouet},\ and\ \citenamefont {Stora}}]{Becchi:1975nq}%
  \BibitemOpen
  \bibfield  {author} {\bibinfo {author} {\bibfnamefont {C.}~\bibnamefont
  {Becchi}}, \bibinfo {author} {\bibfnamefont {A.}~\bibnamefont {Rouet}},\ and\
  \bibinfo {author} {\bibfnamefont {R.}~\bibnamefont {Stora}},\ }\href
  {https://doi.org/10.1016/0003-4916(76)90156-1} {\bibfield  {journal}
  {\bibinfo  {journal} {Annals Phys.}\ }\textbf {\bibinfo {volume} {98}},\
  \bibinfo {pages} {287} (\bibinfo {year} {1976})}\BibitemShut {NoStop}%
\bibitem [{\citenamefont {'t~Hooft}\ and\ \citenamefont
  {Veltman}(1972)}]{tHooft:1972tcz}%
  \BibitemOpen
  \bibfield  {author} {\bibinfo {author} {\bibfnamefont {G.}~\bibnamefont
  {'t~Hooft}}\ and\ \bibinfo {author} {\bibfnamefont {M.~J.~G.}\ \bibnamefont
  {Veltman}},\ }\href {https://doi.org/10.1016/0550-3213(72)90279-9} {\bibfield
   {journal} {\bibinfo  {journal} {Nucl. Phys. B}\ }\textbf {\bibinfo {volume}
  {44}},\ \bibinfo {pages} {189} (\bibinfo {year} {1972})}\BibitemShut
  {NoStop}%
\bibitem [{\citenamefont {Bollini}\ and\ \citenamefont
  {Giambiagi}(1972)}]{Bollini:1972ui}%
  \BibitemOpen
  \bibfield  {author} {\bibinfo {author} {\bibfnamefont {C.~G.}\ \bibnamefont
  {Bollini}}\ and\ \bibinfo {author} {\bibfnamefont {J.~J.}\ \bibnamefont
  {Giambiagi}},\ }\href {https://doi.org/10.1007/BF02895558} {\bibfield
  {journal} {\bibinfo  {journal} {Nuovo Cim. B}\ }\textbf {\bibinfo {volume}
  {12}},\ \bibinfo {pages} {20} (\bibinfo {year} {1972})}\BibitemShut {NoStop}%
\bibitem [{\citenamefont {Cornwall}(1982)}]{Cornwall:1981zr}%
  \BibitemOpen
  \bibfield  {author} {\bibinfo {author} {\bibfnamefont {J.~M.}\ \bibnamefont
  {Cornwall}},\ }\href {https://doi.org/10.1103/PhysRevD.26.1453} {\bibfield
  {journal} {\bibinfo  {journal} {Phys. Rev. D}\ }\textbf {\bibinfo {volume}
  {26}},\ \bibinfo {pages} {1453} (\bibinfo {year} {1982})}\BibitemShut
  {NoStop}%
\bibitem [{\citenamefont {Schwinger}(1962{\natexlab{a}})}]{Schwinger:1962tn}%
  \BibitemOpen
  \bibfield  {author} {\bibinfo {author} {\bibfnamefont {J.~S.}\ \bibnamefont
  {Schwinger}},\ }\href {https://doi.org/10.1103/PhysRev.125.397} {\bibfield
  {journal} {\bibinfo  {journal} {Phys. Rev.}\ }\textbf {\bibinfo {volume}
  {125}},\ \bibinfo {pages} {397} (\bibinfo {year}
  {1962}{\natexlab{a}})}\BibitemShut {NoStop}%
\bibitem [{\citenamefont {Schwinger}(1962{\natexlab{b}})}]{Schwinger:1962tp}%
  \BibitemOpen
  \bibfield  {author} {\bibinfo {author} {\bibfnamefont {J.~S.}\ \bibnamefont
  {Schwinger}},\ }\href {https://doi.org/10.1103/PhysRev.128.2425} {\bibfield
  {journal} {\bibinfo  {journal} {Phys. Rev.}\ }\textbf {\bibinfo {volume}
  {128}},\ \bibinfo {pages} {2425} (\bibinfo {year}
  {1962}{\natexlab{b}})}\BibitemShut {NoStop}%
\bibitem [{\citenamefont {Eichten}\ and\ \citenamefont
  {Feinberg}(1974)}]{Eichten:1974et}%
  \BibitemOpen
  \bibfield  {author} {\bibinfo {author} {\bibfnamefont {E.}~\bibnamefont
  {Eichten}}\ and\ \bibinfo {author} {\bibfnamefont {F.}~\bibnamefont
  {Feinberg}},\ }\href {https://doi.org/10.1103/PhysRevD.10.3254} {\bibfield
  {journal} {\bibinfo  {journal} {Phys. Rev. D}\ }\textbf {\bibinfo {volume}
  {10}},\ \bibinfo {pages} {3254} (\bibinfo {year} {1974})}\BibitemShut
  {NoStop}%
\bibitem [{\citenamefont {Smit}(1974)}]{Smit:1974je}%
  \BibitemOpen
  \bibfield  {author} {\bibinfo {author} {\bibfnamefont {J.}~\bibnamefont
  {Smit}},\ }\href {https://doi.org/10.1103/PhysRevD.10.2473} {\bibfield
  {journal} {\bibinfo  {journal} {Phys. Rev. D}\ }\textbf {\bibinfo {volume}
  {10}},\ \bibinfo {pages} {2473} (\bibinfo {year} {1974})}\BibitemShut
  {NoStop}%
\bibitem [{\citenamefont {Parisi}\ and\ \citenamefont
  {Petronzio}(1980)}]{Parisi:1980jy}%
  \BibitemOpen
  \bibfield  {author} {\bibinfo {author} {\bibfnamefont {G.}~\bibnamefont
  {Parisi}}\ and\ \bibinfo {author} {\bibfnamefont {R.}~\bibnamefont
  {Petronzio}},\ }\href {https://doi.org/10.1016/0370-2693(80)90822-9}
  {\bibfield  {journal} {\bibinfo  {journal} {Phys. Lett. B}\ }\textbf
  {\bibinfo {volume} {94}},\ \bibinfo {pages} {51} (\bibinfo {year}
  {1980})}\BibitemShut {NoStop}%
\bibitem [{\citenamefont {Bernard}(1982)}]{Bernard:1981pg}%
  \BibitemOpen
  \bibfield  {author} {\bibinfo {author} {\bibfnamefont {C.~W.}\ \bibnamefont
  {Bernard}},\ }\href {https://doi.org/10.1016/0370-2693(82)91228-X} {\bibfield
   {journal} {\bibinfo  {journal} {Phys. Lett. B}\ }\textbf {\bibinfo {volume}
  {108}},\ \bibinfo {pages} {431} (\bibinfo {year} {1982})}\BibitemShut
  {NoStop}%
\bibitem [{\citenamefont {Bernard}(1983)}]{Bernard:1982my}%
  \BibitemOpen
  \bibfield  {author} {\bibinfo {author} {\bibfnamefont {C.~W.}\ \bibnamefont
  {Bernard}},\ }\href {https://doi.org/10.1016/0550-3213(83)90645-4} {\bibfield
   {journal} {\bibinfo  {journal} {Nucl. Phys. B}\ }\textbf {\bibinfo {volume}
  {219}},\ \bibinfo {pages} {341} (\bibinfo {year} {1983})}\BibitemShut
  {NoStop}%
\bibitem [{\citenamefont {Donoghue}(1984)}]{Donoghue:1983fy}%
  \BibitemOpen
  \bibfield  {author} {\bibinfo {author} {\bibfnamefont {J.~F.}\ \bibnamefont
  {Donoghue}},\ }\href {https://doi.org/10.1103/PhysRevD.29.2559} {\bibfield
  {journal} {\bibinfo  {journal} {Phys. Rev. D}\ }\textbf {\bibinfo {volume}
  {29}},\ \bibinfo {pages} {2559} (\bibinfo {year} {1984})}\BibitemShut
  {NoStop}%
\bibitem [{\citenamefont {Mandula}\ and\ \citenamefont
  {Ogilvie}(1987)}]{Mandula:1987rh}%
  \BibitemOpen
  \bibfield  {author} {\bibinfo {author} {\bibfnamefont {J.}~\bibnamefont
  {Mandula}}\ and\ \bibinfo {author} {\bibfnamefont {M.}~\bibnamefont
  {Ogilvie}},\ }\href {https://doi.org/10.1016/0370-2693(87)91541-3} {\bibfield
   {journal} {\bibinfo  {journal} {Phys. Lett. B}\ }\textbf {\bibinfo {volume}
  {185}},\ \bibinfo {pages} {127} (\bibinfo {year} {1987})}\BibitemShut
  {NoStop}%
\bibitem [{\citenamefont {Cornwall}\ and\ \citenamefont
  {Papavassiliou}(1989)}]{Cornwall:1989gv}%
  \BibitemOpen
  \bibfield  {author} {\bibinfo {author} {\bibfnamefont {J.~M.}\ \bibnamefont
  {Cornwall}}\ and\ \bibinfo {author} {\bibfnamefont {J.}~\bibnamefont
  {Papavassiliou}},\ }\href {https://doi.org/10.1103/PhysRevD.40.3474}
  {\bibfield  {journal} {\bibinfo  {journal} {Phys. Rev. D}\ }\textbf {\bibinfo
  {volume} {40}},\ \bibinfo {pages} {3474} (\bibinfo {year}
  {1989})}\BibitemShut {NoStop}%
\bibitem [{\citenamefont {Lavelle}(1991)}]{Lavelle:1991ve}%
  \BibitemOpen
  \bibfield  {author} {\bibinfo {author} {\bibfnamefont {M.}~\bibnamefont
  {Lavelle}},\ }\href {https://doi.org/10.1103/PhysRevD.44.R26} {\bibfield
  {journal} {\bibinfo  {journal} {Phys. Rev. D}\ }\textbf {\bibinfo {volume}
  {44}},\ \bibinfo {pages} {26} (\bibinfo {year} {1991})}\BibitemShut {NoStop}%
\bibitem [{\citenamefont {Halzen}\ \emph {et~al.}(1993)\citenamefont {Halzen},
  \citenamefont {Krein},\ and\ \citenamefont {Natale}}]{Halzen:1992vd}%
  \BibitemOpen
  \bibfield  {author} {\bibinfo {author} {\bibfnamefont {F.}~\bibnamefont
  {Halzen}}, \bibinfo {author} {\bibfnamefont {G.~I.}\ \bibnamefont {Krein}},\
  and\ \bibinfo {author} {\bibfnamefont {A.~A.}\ \bibnamefont {Natale}},\
  }\href {https://doi.org/10.1103/PhysRevD.47.295} {\bibfield  {journal}
  {\bibinfo  {journal} {Phys. Rev. D}\ }\textbf {\bibinfo {volume} {47}},\
  \bibinfo {pages} {295} (\bibinfo {year} {1993})}\BibitemShut {NoStop}%
\bibitem [{\citenamefont {Wilson}\ \emph {et~al.}(1994)\citenamefont {Wilson},
  \citenamefont {Walhout}, \citenamefont {Harindranath}, \citenamefont {Zhang},
  \citenamefont {Perry},\ and\ \citenamefont {Glazek}}]{Wilson:1994fk}%
  \BibitemOpen
  \bibfield  {author} {\bibinfo {author} {\bibfnamefont {K.~G.}\ \bibnamefont
  {Wilson}}, \bibinfo {author} {\bibfnamefont {T.~S.}\ \bibnamefont {Walhout}},
  \bibinfo {author} {\bibfnamefont {A.}~\bibnamefont {Harindranath}}, \bibinfo
  {author} {\bibfnamefont {W.-M.}\ \bibnamefont {Zhang}}, \bibinfo {author}
  {\bibfnamefont {R.~J.}\ \bibnamefont {Perry}},\ and\ \bibinfo {author}
  {\bibfnamefont {S.~D.}\ \bibnamefont {Glazek}},\ }\href
  {https://doi.org/10.1103/PhysRevD.49.6720} {\bibfield  {journal} {\bibinfo
  {journal} {Phys. Rev.}\ }\textbf {\bibinfo {volume} {D49}},\ \bibinfo {pages}
  {6720} (\bibinfo {year} {1994})}\BibitemShut {NoStop}%
\bibitem [{\citenamefont {Mihara}\ and\ \citenamefont
  {Natale}(2000)}]{Mihara:2000wf}%
  \BibitemOpen
  \bibfield  {author} {\bibinfo {author} {\bibfnamefont {A.}~\bibnamefont
  {Mihara}}\ and\ \bibinfo {author} {\bibfnamefont {A.~A.}\ \bibnamefont
  {Natale}},\ }\href {https://doi.org/10.1016/S0370-2693(00)00546-3} {\bibfield
   {journal} {\bibinfo  {journal} {Phys. Lett. B}\ }\textbf {\bibinfo {volume}
  {482}},\ \bibinfo {pages} {378} (\bibinfo {year} {2000})}\BibitemShut
  {NoStop}%
\bibitem [{\citenamefont {Kondo}(2001)}]{Kondo:2001nq}%
  \BibitemOpen
  \bibfield  {author} {\bibinfo {author} {\bibfnamefont {K.-I.}\ \bibnamefont
  {Kondo}},\ }\href {https://doi.org/10.1016/S0370-2693(01)00817-6} {\bibfield
  {journal} {\bibinfo  {journal} {Phys. Lett. B}\ }\textbf {\bibinfo {volume}
  {514}},\ \bibinfo {pages} {335} (\bibinfo {year} {2001})}\BibitemShut
  {NoStop}%
\bibitem [{\citenamefont {Philipsen}(2002)}]{Philipsen:2001ip}%
  \BibitemOpen
  \bibfield  {author} {\bibinfo {author} {\bibfnamefont {O.}~\bibnamefont
  {Philipsen}},\ }\href {https://doi.org/10.1016/S0550-3213(02)00089-5}
  {\bibfield  {journal} {\bibinfo  {journal} {Nucl. Phys.}\ }\textbf {\bibinfo
  {volume} {B628}},\ \bibinfo {pages} {167} (\bibinfo {year}
  {2002})}\BibitemShut {NoStop}%
\bibitem [{\citenamefont {Aguilar}\ \emph {et~al.}(2003)\citenamefont
  {Aguilar}, \citenamefont {Natale},\ and\ \citenamefont {Rodrigues~da
  Silva}}]{Aguilar:2002tc}%
  \BibitemOpen
  \bibfield  {author} {\bibinfo {author} {\bibfnamefont {A.~C.}\ \bibnamefont
  {Aguilar}}, \bibinfo {author} {\bibfnamefont {A.~A.}\ \bibnamefont
  {Natale}},\ and\ \bibinfo {author} {\bibfnamefont {P.~S.}\ \bibnamefont
  {Rodrigues~da Silva}},\ }\href
  {https://doi.org/10.1103/PhysRevLett.90.152001} {\bibfield  {journal}
  {\bibinfo  {journal} {Phys. Rev. Lett.}\ }\textbf {\bibinfo {volume} {90}},\
  \bibinfo {pages} {152001} (\bibinfo {year} {2003})}\BibitemShut {NoStop}%
\bibitem [{\citenamefont {Pawlowski}\ \emph {et~al.}(2004)\citenamefont
  {Pawlowski}, \citenamefont {Litim}, \citenamefont {Nedelko},\ and\
  \citenamefont {von Smekal}}]{Pawlowski:2003hq}%
  \BibitemOpen
  \bibfield  {author} {\bibinfo {author} {\bibfnamefont {J.~M.}\ \bibnamefont
  {Pawlowski}}, \bibinfo {author} {\bibfnamefont {D.~F.}\ \bibnamefont
  {Litim}}, \bibinfo {author} {\bibfnamefont {S.}~\bibnamefont {Nedelko}},\
  and\ \bibinfo {author} {\bibfnamefont {L.}~\bibnamefont {von Smekal}},\
  }\href {https://doi.org/10.1103/PhysRevLett.93.152002} {\bibfield  {journal}
  {\bibinfo  {journal} {Phys. Rev. Lett.}\ }\textbf {\bibinfo {volume} {93}},\
  \bibinfo {pages} {152002} (\bibinfo {year} {2004})}\BibitemShut {NoStop}%
\bibitem [{\citenamefont {Aguilar}\ and\ \citenamefont
  {Natale}(2004)}]{Aguilar:2004sw}%
  \BibitemOpen
  \bibfield  {author} {\bibinfo {author} {\bibfnamefont {A.~C.}\ \bibnamefont
  {Aguilar}}\ and\ \bibinfo {author} {\bibfnamefont {A.~A.}\ \bibnamefont
  {Natale}},\ }\href {https://doi.org/10.1088/1126-6708/2004/08/057} {\bibfield
   {journal} {\bibinfo  {journal} {{JHEP}}\ }\textbf {\bibinfo {volume} {08}},\
  \bibinfo {pages} {057} (\bibinfo {year} {2004})}\BibitemShut {NoStop}%
\bibitem [{\citenamefont {Aguilar}\ and\ \citenamefont
  {Papavassiliou}(2006)}]{Aguilar:2006gr}%
  \BibitemOpen
  \bibfield  {author} {\bibinfo {author} {\bibfnamefont {A.~C.}\ \bibnamefont
  {Aguilar}}\ and\ \bibinfo {author} {\bibfnamefont {J.}~\bibnamefont
  {Papavassiliou}},\ }\href {https://doi.org/10.1088/1126-6708/2006/12/012}
  {\bibfield  {journal} {\bibinfo  {journal} {{JHEP}}\ }\textbf {\bibinfo
  {volume} {12}},\ \bibinfo {pages} {012} (\bibinfo {year} {2006})}\BibitemShut
  {NoStop}%
\bibitem [{\citenamefont {Epple}\ \emph {et~al.}(2008)\citenamefont {Epple},
  \citenamefont {Reinhardt}, \citenamefont {Schleifenbaum},\ and\ \citenamefont
  {Szczepaniak}}]{Epple:2007ut}%
  \BibitemOpen
  \bibfield  {author} {\bibinfo {author} {\bibfnamefont {D.}~\bibnamefont
  {Epple}}, \bibinfo {author} {\bibfnamefont {H.}~\bibnamefont {Reinhardt}},
  \bibinfo {author} {\bibfnamefont {W.}~\bibnamefont {Schleifenbaum}},\ and\
  \bibinfo {author} {\bibfnamefont {A.~P.}\ \bibnamefont {Szczepaniak}},\
  }\href {https://doi.org/10.1103/PhysRevD.77.085007} {\bibfield  {journal}
  {\bibinfo  {journal} {Phys. Rev.}\ }\textbf {\bibinfo {volume} {D77}},\
  \bibinfo {pages} {085007} (\bibinfo {year} {2008})}\BibitemShut {NoStop}%
\bibitem [{\citenamefont {Aguilar}\ \emph {et~al.}(2008)\citenamefont
  {Aguilar}, \citenamefont {Binosi},\ and\ \citenamefont
  {Papavassiliou}}]{Aguilar:2008xm}%
  \BibitemOpen
  \bibfield  {author} {\bibinfo {author} {\bibfnamefont {A.~C.}\ \bibnamefont
  {Aguilar}}, \bibinfo {author} {\bibfnamefont {D.}~\bibnamefont {Binosi}},\
  and\ \bibinfo {author} {\bibfnamefont {J.}~\bibnamefont {Papavassiliou}},\
  }\href {https://doi.org/10.1103/PhysRevD.78.025010} {\bibfield  {journal}
  {\bibinfo  {journal} {Phys. Rev.}\ }\textbf {\bibinfo {volume} {D78}},\
  \bibinfo {pages} {025010} (\bibinfo {year} {2008})}\BibitemShut {NoStop}%
\bibitem [{\citenamefont {Braun}\ \emph {et~al.}(2010)\citenamefont {Braun},
  \citenamefont {Gies},\ and\ \citenamefont {Pawlowski}}]{Braun:2007bx}%
  \BibitemOpen
  \bibfield  {author} {\bibinfo {author} {\bibfnamefont {J.}~\bibnamefont
  {Braun}}, \bibinfo {author} {\bibfnamefont {H.}~\bibnamefont {Gies}},\ and\
  \bibinfo {author} {\bibfnamefont {J.~M.}\ \bibnamefont {Pawlowski}},\ }\href
  {https://doi.org/10.1016/j.physletb.2010.01.009} {\bibfield  {journal}
  {\bibinfo  {journal} {Phys. Lett.}\ }\textbf {\bibinfo {volume} {B684}},\
  \bibinfo {pages} {262} (\bibinfo {year} {2010})}\BibitemShut {NoStop}%
\bibitem [{\citenamefont {Tissier}\ and\ \citenamefont
  {Wschebor}(2010)}]{Tissier:2010ts}%
  \BibitemOpen
  \bibfield  {author} {\bibinfo {author} {\bibfnamefont {M.}~\bibnamefont
  {Tissier}}\ and\ \bibinfo {author} {\bibfnamefont {N.}~\bibnamefont
  {Wschebor}},\ }\href {https://doi.org/10.1103/PhysRevD.82.101701} {\bibfield
  {journal} {\bibinfo  {journal} {Phys. Rev. D}\ }\textbf {\bibinfo {volume}
  {82}},\ \bibinfo {pages} {101701} (\bibinfo {year} {2010})}\BibitemShut
  {NoStop}%
\bibitem [{\citenamefont {Campagnari}\ and\ \citenamefont
  {Reinhardt}(2010)}]{Campagnari:2010wc}%
  \BibitemOpen
  \bibfield  {author} {\bibinfo {author} {\bibfnamefont {D.~R.}\ \bibnamefont
  {Campagnari}}\ and\ \bibinfo {author} {\bibfnamefont {H.}~\bibnamefont
  {Reinhardt}},\ }\href {https://doi.org/10.1103/PhysRevD.82.105021} {\bibfield
   {journal} {\bibinfo  {journal} {Phys. Rev.}\ }\textbf {\bibinfo {volume}
  {D82}},\ \bibinfo {pages} {105021} (\bibinfo {year} {2010})}\BibitemShut
  {NoStop}%
\bibitem [{\citenamefont {Tissier}\ and\ \citenamefont
  {Wschebor}(2011)}]{Tissier:2011ey}%
  \BibitemOpen
  \bibfield  {author} {\bibinfo {author} {\bibfnamefont {M.}~\bibnamefont
  {Tissier}}\ and\ \bibinfo {author} {\bibfnamefont {N.}~\bibnamefont
  {Wschebor}},\ }\href {https://doi.org/10.1103/PhysRevD.84.045018} {\bibfield
  {journal} {\bibinfo  {journal} {Phys. Rev.}\ }\textbf {\bibinfo {volume}
  {D84}},\ \bibinfo {pages} {045018} (\bibinfo {year} {2011})}\BibitemShut
  {NoStop}%
\bibitem [{\citenamefont {Serreau}\ and\ \citenamefont
  {Tissier}(2012)}]{Serreau:2012cg}%
  \BibitemOpen
  \bibfield  {author} {\bibinfo {author} {\bibfnamefont {J.}~\bibnamefont
  {Serreau}}\ and\ \bibinfo {author} {\bibfnamefont {M.}~\bibnamefont
  {Tissier}},\ }\href {https://doi.org/10.1016/j.physletb.2012.04.041}
  {\bibfield  {journal} {\bibinfo  {journal} {Phys. Lett.}\ }\textbf {\bibinfo
  {volume} {B712}},\ \bibinfo {pages} {97} (\bibinfo {year}
  {2012})}\BibitemShut {NoStop}%
\bibitem [{\citenamefont {Fagundes}\ \emph {et~al.}(2012)\citenamefont
  {Fagundes}, \citenamefont {Luna}, \citenamefont {Menon},\ and\ \citenamefont
  {Natale}}]{Fagundes:2011zx}%
  \BibitemOpen
  \bibfield  {author} {\bibinfo {author} {\bibfnamefont {D.~A.}\ \bibnamefont
  {Fagundes}}, \bibinfo {author} {\bibfnamefont {E.~G.~S.}\ \bibnamefont
  {Luna}}, \bibinfo {author} {\bibfnamefont {M.~J.}\ \bibnamefont {Menon}},\
  and\ \bibinfo {author} {\bibfnamefont {A.~A.}\ \bibnamefont {Natale}},\
  }\href {https://doi.org/10.1016/j.nuclphysa.2012.05.002} {\bibfield
  {journal} {\bibinfo  {journal} {Nucl. Phys. A}\ }\textbf {\bibinfo {volume}
  {886}},\ \bibinfo {pages} {48} (\bibinfo {year} {2012})}\BibitemShut
  {NoStop}%
\bibitem [{\citenamefont {Binosi}\ \emph {et~al.}(2012)\citenamefont {Binosi},
  \citenamefont {Iba\~nez},\ and\ \citenamefont
  {Papavassiliou}}]{Binosi:2012sj}%
  \BibitemOpen
  \bibfield  {author} {\bibinfo {author} {\bibfnamefont {D.}~\bibnamefont
  {Binosi}}, \bibinfo {author} {\bibfnamefont {D.}~\bibnamefont {Iba\~nez}},\
  and\ \bibinfo {author} {\bibfnamefont {J.}~\bibnamefont {Papavassiliou}},\
  }\href {https://doi.org/10.1103/PhysRevD.86.085033} {\bibfield  {journal}
  {\bibinfo  {journal} {Phys. Rev.}\ }\textbf {\bibinfo {volume} {D86}},\
  \bibinfo {pages} {085033} (\bibinfo {year} {2012})}\BibitemShut {NoStop}%
\bibitem [{\citenamefont {Maas}(2013)}]{Maas:2011se}%
  \BibitemOpen
  \bibfield  {author} {\bibinfo {author} {\bibfnamefont {A.}~\bibnamefont
  {Maas}},\ }\href {https://doi.org/10.1016/j.physrep.2012.11.002} {\bibfield
  {journal} {\bibinfo  {journal} {Phys. Rept.}\ }\textbf {\bibinfo {volume}
  {524}},\ \bibinfo {pages} {203} (\bibinfo {year} {2013})}\BibitemShut
  {NoStop}%
\bibitem [{\citenamefont {Siringo}(2016)}]{Siringo:2015wtx}%
  \BibitemOpen
  \bibfield  {author} {\bibinfo {author} {\bibfnamefont {F.}~\bibnamefont
  {Siringo}},\ }\href {https://doi.org/10.1016/j.nuclphysb.2016.04.028}
  {\bibfield  {journal} {\bibinfo  {journal} {Nucl. Phys.}\ }\textbf {\bibinfo
  {volume} {B907}},\ \bibinfo {pages} {572} (\bibinfo {year}
  {2016})}\BibitemShut {NoStop}%
\bibitem [{\citenamefont {Cyrol}\ \emph {et~al.}(2016)\citenamefont {Cyrol},
  \citenamefont {Fister}, \citenamefont {Mitter}, \citenamefont {Pawlowski},\
  and\ \citenamefont {Strodthoff}}]{Cyrol:2016tym}%
  \BibitemOpen
  \bibfield  {author} {\bibinfo {author} {\bibfnamefont {A.~K.}\ \bibnamefont
  {Cyrol}}, \bibinfo {author} {\bibfnamefont {L.}~\bibnamefont {Fister}},
  \bibinfo {author} {\bibfnamefont {M.}~\bibnamefont {Mitter}}, \bibinfo
  {author} {\bibfnamefont {J.~M.}\ \bibnamefont {Pawlowski}},\ and\ \bibinfo
  {author} {\bibfnamefont {N.}~\bibnamefont {Strodthoff}},\ }\href
  {https://doi.org/10.1103/PhysRevD.94.054005} {\bibfield  {journal} {\bibinfo
  {journal} {Phys. Rev.}\ }\textbf {\bibinfo {volume} {D94}},\ \bibinfo {pages}
  {054005} (\bibinfo {year} {2016})}\BibitemShut {NoStop}%
\bibitem [{\citenamefont {Glazek}\ \emph {et~al.}(2017)\citenamefont {Glazek},
  \citenamefont {G\'omez-Rocha}, \citenamefont {More},\ and\ \citenamefont
  {Serafin}}]{Glazek:2017rwe}%
  \BibitemOpen
  \bibfield  {author} {\bibinfo {author} {\bibfnamefont {S.~D.}\ \bibnamefont
  {Glazek}}, \bibinfo {author} {\bibfnamefont {M.}~\bibnamefont
  {G\'omez-Rocha}}, \bibinfo {author} {\bibfnamefont {J.}~\bibnamefont
  {More}},\ and\ \bibinfo {author} {\bibfnamefont {K.}~\bibnamefont
  {Serafin}},\ }\href {https://doi.org/10.1016/j.physletb.2017.08.018}
  {\bibfield  {journal} {\bibinfo  {journal} {Phys. Lett.}\ }\textbf {\bibinfo
  {volume} {B773}},\ \bibinfo {pages} {172} (\bibinfo {year}
  {2017})}\BibitemShut {NoStop}%
\bibitem [{\citenamefont {Cyrol}\ \emph {et~al.}(2018)\citenamefont {Cyrol},
  \citenamefont {Pawlowski}, \citenamefont {Rothkopf},\ and\ \citenamefont
  {Wink}}]{Cyrol:2018xeq}%
  \BibitemOpen
  \bibfield  {author} {\bibinfo {author} {\bibfnamefont {A.~K.}\ \bibnamefont
  {Cyrol}}, \bibinfo {author} {\bibfnamefont {J.~M.}\ \bibnamefont
  {Pawlowski}}, \bibinfo {author} {\bibfnamefont {A.}~\bibnamefont
  {Rothkopf}},\ and\ \bibinfo {author} {\bibfnamefont {N.}~\bibnamefont
  {Wink}},\ }\href {https://doi.org/10.21468/SciPostPhys.5.6.065} {\bibfield
  {journal} {\bibinfo  {journal} {SciPost Phys.}\ }\textbf {\bibinfo {volume}
  {5}},\ \bibinfo {pages} {065} (\bibinfo {year} {2018})}\BibitemShut {NoStop}%
\bibitem [{\citenamefont {Gracey}\ \emph
  {et~al.}(2019{\natexlab{a}})\citenamefont {Gracey}, \citenamefont {Pel\'aez},
  \citenamefont {Reinosa},\ and\ \citenamefont {Tissier}}]{Gracey:2019xom}%
  \BibitemOpen
  \bibfield  {author} {\bibinfo {author} {\bibfnamefont {J.~A.}\ \bibnamefont
  {Gracey}}, \bibinfo {author} {\bibfnamefont {M.}~\bibnamefont {Pel\'aez}},
  \bibinfo {author} {\bibfnamefont {U.}~\bibnamefont {Reinosa}},\ and\ \bibinfo
  {author} {\bibfnamefont {M.}~\bibnamefont {Tissier}},\ }\href
  {https://doi.org/10.1103/PhysRevD.100.034023} {\bibfield  {journal} {\bibinfo
   {journal} {Phys. Rev. D}\ }\textbf {\bibinfo {volume} {100}},\ \bibinfo
  {pages} {034023} (\bibinfo {year} {2019}{\natexlab{a}})}\BibitemShut
  {NoStop}%
\bibitem [{\citenamefont {Roberts}(2020)}]{Roberts:2020hiw}%
  \BibitemOpen
  \bibfield  {author} {\bibinfo {author} {\bibfnamefont {C.~D.}\ \bibnamefont
  {Roberts}},\ }\href {https://doi.org/10.3390/sym12091468} {\bibfield
  {journal} {\bibinfo  {journal} {Symmetry}\ }\textbf {\bibinfo {volume}
  {12}},\ \bibinfo {pages} {1468} (\bibinfo {year} {2020})}\BibitemShut
  {NoStop}%
\bibitem [{\citenamefont {Pel\'aez}\ \emph {et~al.}(2021)\citenamefont
  {Pel\'aez}, \citenamefont {Reinosa}, \citenamefont {Serreau}, \citenamefont
  {Tissier},\ and\ \citenamefont {Wschebor}}]{Pelaez:2021tpq}%
  \BibitemOpen
  \bibfield  {author} {\bibinfo {author} {\bibfnamefont {M.}~\bibnamefont
  {Pel\'aez}}, \bibinfo {author} {\bibfnamefont {U.}~\bibnamefont {Reinosa}},
  \bibinfo {author} {\bibfnamefont {J.}~\bibnamefont {Serreau}}, \bibinfo
  {author} {\bibfnamefont {M.}~\bibnamefont {Tissier}},\ and\ \bibinfo {author}
  {\bibfnamefont {N.}~\bibnamefont {Wschebor}},\ }\href
  {https://doi.org/10.1088/1361-6633/ac36b8} {\bibfield  {journal} {\bibinfo
  {journal} {Rept. Prog. Phys.}\ }\textbf {\bibinfo {volume} {84}},\ \bibinfo
  {pages} {124202} (\bibinfo {year} {2021})}\BibitemShut {NoStop}%
\bibitem [{\citenamefont {Eichmann}\ \emph {et~al.}(2021)\citenamefont
  {Eichmann}, \citenamefont {Pawlowski},\ and\ \citenamefont
  {Silva}}]{Eichmann:2021zuv}%
  \BibitemOpen
  \bibfield  {author} {\bibinfo {author} {\bibfnamefont {G.}~\bibnamefont
  {Eichmann}}, \bibinfo {author} {\bibfnamefont {J.~M.}\ \bibnamefont
  {Pawlowski}},\ and\ \bibinfo {author} {\bibfnamefont {J.~a.~M.}\ \bibnamefont
  {Silva}},\ }\href {https://doi.org/10.1103/PhysRevD.104.114016} {\bibfield
  {journal} {\bibinfo  {journal} {Phys. Rev. D}\ }\textbf {\bibinfo {volume}
  {104}},\ \bibinfo {pages} {114016} (\bibinfo {year} {2021})}\BibitemShut
  {NoStop}%
\bibitem [{\citenamefont {Horak}\ \emph {et~al.}(2022)\citenamefont {Horak},
  \citenamefont {Ihssen}, \citenamefont {Papavassiliou}, \citenamefont
  {Pawlowski}, \citenamefont {Weber},\ and\ \citenamefont
  {Wetterich}}]{Horak:2022aqx}%
  \BibitemOpen
  \bibfield  {author} {\bibinfo {author} {\bibfnamefont {J.}~\bibnamefont
  {Horak}}, \bibinfo {author} {\bibfnamefont {F.}~\bibnamefont {Ihssen}},
  \bibinfo {author} {\bibfnamefont {J.}~\bibnamefont {Papavassiliou}}, \bibinfo
  {author} {\bibfnamefont {J.~M.}\ \bibnamefont {Pawlowski}}, \bibinfo {author}
  {\bibfnamefont {A.}~\bibnamefont {Weber}},\ and\ \bibinfo {author}
  {\bibfnamefont {C.}~\bibnamefont {Wetterich}},\ }\href
  {https://doi.org/10.21468/SciPostPhys.13.2.042} {\bibfield  {journal}
  {\bibinfo  {journal} {SciPost Phys.}\ }\textbf {\bibinfo {volume} {13}},\
  \bibinfo {pages} {042} (\bibinfo {year} {2022})}\BibitemShut {NoStop}%
\bibitem [{\citenamefont {Papavassiliou}(2022)}]{Papavassiliou:2022wrb}%
  \BibitemOpen
  \bibfield  {author} {\bibinfo {author} {\bibfnamefont {J.}~\bibnamefont
  {Papavassiliou}},\ }\href {https://doi.org/10.1088/1674-1137/ac84ca}
  {\bibfield  {journal} {\bibinfo  {journal} {Chin. Phys. C}\ }\textbf
  {\bibinfo {volume} {46}},\ \bibinfo {pages} {112001} (\bibinfo {year}
  {2022})}\BibitemShut {NoStop}%
\bibitem [{\citenamefont {Ding}\ \emph {et~al.}(2023)\citenamefont {Ding},
  \citenamefont {Roberts},\ and\ \citenamefont {Schmidt}}]{Ding:2022ows}%
  \BibitemOpen
  \bibfield  {author} {\bibinfo {author} {\bibfnamefont {M.}~\bibnamefont
  {Ding}}, \bibinfo {author} {\bibfnamefont {C.~D.}\ \bibnamefont {Roberts}},\
  and\ \bibinfo {author} {\bibfnamefont {S.~M.}\ \bibnamefont {Schmidt}},\
  }\href {https://doi.org/10.3390/particles6010004} {\bibfield  {journal}
  {\bibinfo  {journal} {Particles}\ }\textbf {\bibinfo {volume} {6}},\ \bibinfo
  {pages} {57} (\bibinfo {year} {2023})}\BibitemShut {NoStop}%
\bibitem [{\citenamefont {Ferreira}\ and\ \citenamefont
  {Papavassiliou}(2023)}]{Ferreira:2023fva}%
  \BibitemOpen
  \bibfield  {author} {\bibinfo {author} {\bibfnamefont {M.~N.}\ \bibnamefont
  {Ferreira}}\ and\ \bibinfo {author} {\bibfnamefont {J.}~\bibnamefont
  {Papavassiliou}},\ }\href {https://doi.org/10.3390/particles6010017}
  {\bibfield  {journal} {\bibinfo  {journal} {Particles}\ }\textbf {\bibinfo
  {volume} {6}},\ \bibinfo {pages} {312} (\bibinfo {year} {2023})}\BibitemShut
  {NoStop}%
\bibitem [{\citenamefont {Poggio}\ \emph {et~al.}(1975)\citenamefont {Poggio},
  \citenamefont {Tomboulis},\ and\ \citenamefont {Tye}}]{Poggio:1974qs}%
  \BibitemOpen
  \bibfield  {author} {\bibinfo {author} {\bibfnamefont {E.~C.}\ \bibnamefont
  {Poggio}}, \bibinfo {author} {\bibfnamefont {E.}~\bibnamefont {Tomboulis}},\
  and\ \bibinfo {author} {\bibfnamefont {S.~H.~H.}\ \bibnamefont {Tye}},\
  }\href {https://doi.org/10.1103/PhysRevD.11.2839} {\bibfield  {journal}
  {\bibinfo  {journal} {Phys. Rev.}\ }\textbf {\bibinfo {volume} {D11}},\
  \bibinfo {pages} {2839} (\bibinfo {year} {1975})}\BibitemShut {NoStop}%
\bibitem [{\citenamefont {Aguilar}\ \emph {et~al.}(2012)\citenamefont
  {Aguilar}, \citenamefont {Ibanez}, \citenamefont {Mathieu},\ and\
  \citenamefont {Papavassiliou}}]{Aguilar:2011xe}%
  \BibitemOpen
  \bibfield  {author} {\bibinfo {author} {\bibfnamefont {A.~C.}\ \bibnamefont
  {Aguilar}}, \bibinfo {author} {\bibfnamefont {D.}~\bibnamefont {Ibanez}},
  \bibinfo {author} {\bibfnamefont {V.}~\bibnamefont {Mathieu}},\ and\ \bibinfo
  {author} {\bibfnamefont {J.}~\bibnamefont {Papavassiliou}},\ }\href
  {https://doi.org/10.1103/PhysRevD.85.014018} {\bibfield  {journal} {\bibinfo
  {journal} {Phys. Rev.}\ }\textbf {\bibinfo {volume} {D85}},\ \bibinfo {pages}
  {014018} (\bibinfo {year} {2012})}\BibitemShut {NoStop}%
\bibitem [{\citenamefont {Iba{\~n}ez}\ and\ \citenamefont
  {Papavassiliou}(2013)}]{Ibanez:2012zk}%
  \BibitemOpen
  \bibfield  {author} {\bibinfo {author} {\bibfnamefont {D.}~\bibnamefont
  {Iba{\~n}ez}}\ and\ \bibinfo {author} {\bibfnamefont {J.}~\bibnamefont
  {Papavassiliou}},\ }\href {https://doi.org/10.1103/PhysRevD.87.034008}
  {\bibfield  {journal} {\bibinfo  {journal} {Phys. Rev.}\ }\textbf {\bibinfo
  {volume} {D87}},\ \bibinfo {pages} {034008} (\bibinfo {year}
  {2013})}\BibitemShut {NoStop}%
\bibitem [{\citenamefont {Aguilar}\ \emph
  {et~al.}(2016{\natexlab{a}})\citenamefont {Aguilar}, \citenamefont {Binosi},\
  and\ \citenamefont {Papavassiliou}}]{Aguilar:2015bud}%
  \BibitemOpen
  \bibfield  {author} {\bibinfo {author} {\bibfnamefont {A.~C.}\ \bibnamefont
  {Aguilar}}, \bibinfo {author} {\bibfnamefont {D.}~\bibnamefont {Binosi}},\
  and\ \bibinfo {author} {\bibfnamefont {J.}~\bibnamefont {Papavassiliou}},\
  }\href {https://doi.org/10.1007/s11467-015-0517-6} {\bibfield  {journal}
  {\bibinfo  {journal} {Front. Phys.(Beijing)}\ }\textbf {\bibinfo {volume}
  {11}},\ \bibinfo {pages} {111203} (\bibinfo {year}
  {2016}{\natexlab{a}})}\BibitemShut {NoStop}%
\bibitem [{\citenamefont {Aguilar}\ \emph {et~al.}(2018)\citenamefont
  {Aguilar}, \citenamefont {Binosi}, \citenamefont {Figueiredo},\ and\
  \citenamefont {Papavassiliou}}]{Aguilar:2017dco}%
  \BibitemOpen
  \bibfield  {author} {\bibinfo {author} {\bibfnamefont {A.~C.}\ \bibnamefont
  {Aguilar}}, \bibinfo {author} {\bibfnamefont {D.}~\bibnamefont {Binosi}},
  \bibinfo {author} {\bibfnamefont {C.~T.}\ \bibnamefont {Figueiredo}},\ and\
  \bibinfo {author} {\bibfnamefont {J.}~\bibnamefont {Papavassiliou}},\ }\href
  {https://doi.org/10.1140/epjc/s10052-018-5679-2} {\bibfield  {journal}
  {\bibinfo  {journal} {Eur. Phys. J.}\ }\textbf {\bibinfo {volume} {C78}},\
  \bibinfo {pages} {181} (\bibinfo {year} {2018})}\BibitemShut {NoStop}%
\bibitem [{\citenamefont {Aguilar}\ \emph {et~al.}(2022)\citenamefont
  {Aguilar}, \citenamefont {Ferreira},\ and\ \citenamefont
  {Papavassiliou}}]{Aguilar:2021uwa}%
  \BibitemOpen
  \bibfield  {author} {\bibinfo {author} {\bibfnamefont {A.~C.}\ \bibnamefont
  {Aguilar}}, \bibinfo {author} {\bibfnamefont {M.~N.}\ \bibnamefont
  {Ferreira}},\ and\ \bibinfo {author} {\bibfnamefont {J.}~\bibnamefont
  {Papavassiliou}},\ }\href {https://doi.org/10.1103/PhysRevD.105.014030}
  {\bibfield  {journal} {\bibinfo  {journal} {Phys. Rev. D}\ }\textbf {\bibinfo
  {volume} {105}},\ \bibinfo {pages} {014030} (\bibinfo {year}
  {2022})}\BibitemShut {NoStop}%
\bibitem [{\citenamefont {Roberts}\ and\ \citenamefont
  {Williams}(1994)}]{Roberts:1994dr}%
  \BibitemOpen
  \bibfield  {author} {\bibinfo {author} {\bibfnamefont {C.~D.}\ \bibnamefont
  {Roberts}}\ and\ \bibinfo {author} {\bibfnamefont {A.~G.}\ \bibnamefont
  {Williams}},\ }\href {https://doi.org/10.1016/0146-6410(94)90049-3}
  {\bibfield  {journal} {\bibinfo  {journal} {Prog. Part. Nucl. Phys.}\
  }\textbf {\bibinfo {volume} {33}},\ \bibinfo {pages} {477} (\bibinfo {year}
  {1994})}\BibitemShut {NoStop}%
\bibitem [{\citenamefont {Alkofer}\ and\ \citenamefont {von
  Smekal}(2001)}]{Alkofer:2000wg}%
  \BibitemOpen
  \bibfield  {author} {\bibinfo {author} {\bibfnamefont {R.}~\bibnamefont
  {Alkofer}}\ and\ \bibinfo {author} {\bibfnamefont {L.}~\bibnamefont {von
  Smekal}},\ }\href {https://doi.org/10.1016/S0370-1573(01)00010-2} {\bibfield
  {journal} {\bibinfo  {journal} {Phys. Rept.}\ }\textbf {\bibinfo {volume}
  {353}},\ \bibinfo {pages} {281} (\bibinfo {year} {2001})}\BibitemShut
  {NoStop}%
\bibitem [{\citenamefont {Maris}\ and\ \citenamefont
  {Roberts}(2003)}]{Maris:2003vk}%
  \BibitemOpen
  \bibfield  {author} {\bibinfo {author} {\bibfnamefont {P.}~\bibnamefont
  {Maris}}\ and\ \bibinfo {author} {\bibfnamefont {C.~D.}\ \bibnamefont
  {Roberts}},\ }\href {https://doi.org/10.1142/S0218301303001326} {\bibfield
  {journal} {\bibinfo  {journal} {Int. J. Mod. Phys.}\ }\textbf {\bibinfo
  {volume} {E12}},\ \bibinfo {pages} {297} (\bibinfo {year}
  {2003})}\BibitemShut {NoStop}%
\bibitem [{\citenamefont {Fischer}(2006)}]{Fischer:2006ub}%
  \BibitemOpen
  \bibfield  {author} {\bibinfo {author} {\bibfnamefont {C.~S.}\ \bibnamefont
  {Fischer}},\ }\href {https://doi.org/10.1088/0954-3899/32/8/R02} {\bibfield
  {journal} {\bibinfo  {journal} {J. Phys. G}\ }\textbf {\bibinfo {volume}
  {32}},\ \bibinfo {pages} {R253} (\bibinfo {year} {2006})}\BibitemShut
  {NoStop}%
\bibitem [{\citenamefont {Roberts}(2008)}]{Roberts:2007ji}%
  \BibitemOpen
  \bibfield  {author} {\bibinfo {author} {\bibfnamefont {C.~D.}\ \bibnamefont
  {Roberts}},\ }\href {https://doi.org/10.1016/j.ppnp.2007.12.034} {\bibfield
  {journal} {\bibinfo  {journal} {Prog. Part. Nucl. Phys.}\ }\textbf {\bibinfo
  {volume} {61}},\ \bibinfo {pages} {50} (\bibinfo {year} {2008})}\BibitemShut
  {NoStop}%
\bibitem [{\citenamefont {Fischer}\ \emph {et~al.}(2009)\citenamefont
  {Fischer}, \citenamefont {Maas},\ and\ \citenamefont
  {Pawlowski}}]{Fischer:2008uz}%
  \BibitemOpen
  \bibfield  {author} {\bibinfo {author} {\bibfnamefont {C.~S.}\ \bibnamefont
  {Fischer}}, \bibinfo {author} {\bibfnamefont {A.}~\bibnamefont {Maas}},\ and\
  \bibinfo {author} {\bibfnamefont {J.~M.}\ \bibnamefont {Pawlowski}},\ }\href
  {https://doi.org/10.1016/j.aop.2009.07.009} {\bibfield  {journal} {\bibinfo
  {journal} {Annals Phys.}\ }\textbf {\bibinfo {volume} {324}},\ \bibinfo
  {pages} {2408} (\bibinfo {year} {2009})}\BibitemShut {NoStop}%
\bibitem [{\citenamefont {Binosi}\ and\ \citenamefont
  {Papavassiliou}(2009)}]{Binosi:2009qm}%
  \BibitemOpen
  \bibfield  {author} {\bibinfo {author} {\bibfnamefont {D.}~\bibnamefont
  {Binosi}}\ and\ \bibinfo {author} {\bibfnamefont {J.}~\bibnamefont
  {Papavassiliou}},\ }\href {https://doi.org/10.1016/j.physrep.2009.05.001}
  {\bibfield  {journal} {\bibinfo  {journal} {Phys. Rept.}\ }\textbf {\bibinfo
  {volume} {479}},\ \bibinfo {pages} {1} (\bibinfo {year} {2009})}\BibitemShut
  {NoStop}%
\bibitem [{\citenamefont {Bashir}\ \emph {et~al.}(2012)\citenamefont {Bashir},
  \citenamefont {Chang}, \citenamefont {Cloet}, \citenamefont {El-Bennich},
  \citenamefont {Liu} \emph {et~al.}}]{Bashir:2012fs}%
  \BibitemOpen
  \bibfield  {author} {\bibinfo {author} {\bibfnamefont {A.}~\bibnamefont
  {Bashir}}, \bibinfo {author} {\bibfnamefont {L.}~\bibnamefont {Chang}},
  \bibinfo {author} {\bibfnamefont {I.~C.}\ \bibnamefont {Cloet}}, \bibinfo
  {author} {\bibfnamefont {B.}~\bibnamefont {El-Bennich}}, \bibinfo {author}
  {\bibfnamefont {Y.-X.}\ \bibnamefont {Liu}}, \emph {et~al.},\ }\href
  {https://doi.org/10.1088/0253-6102/58/1/16} {\bibfield  {journal} {\bibinfo
  {journal} {Commun. Theor. Phys.}\ }\textbf {\bibinfo {volume} {58}},\
  \bibinfo {pages} {79} (\bibinfo {year} {2012})}\BibitemShut {NoStop}%
\bibitem [{\citenamefont {Fister}\ and\ \citenamefont
  {Pawlowski}(2013)}]{Fister:2013bh}%
  \BibitemOpen
  \bibfield  {author} {\bibinfo {author} {\bibfnamefont {L.}~\bibnamefont
  {Fister}}\ and\ \bibinfo {author} {\bibfnamefont {J.~M.}\ \bibnamefont
  {Pawlowski}},\ }\href {https://doi.org/10.1103/PhysRevD.88.045010} {\bibfield
   {journal} {\bibinfo  {journal} {Phys. Rev.}\ }\textbf {\bibinfo {volume}
  {D88}},\ \bibinfo {pages} {045010} (\bibinfo {year} {2013})}\BibitemShut
  {NoStop}%
\bibitem [{\citenamefont {Cloet}\ and\ \citenamefont
  {Roberts}(2014)}]{Cloet:2013jya}%
  \BibitemOpen
  \bibfield  {author} {\bibinfo {author} {\bibfnamefont {I.~C.}\ \bibnamefont
  {Cloet}}\ and\ \bibinfo {author} {\bibfnamefont {C.~D.}\ \bibnamefont
  {Roberts}},\ }\href {https://doi.org/10.1016/j.ppnp.2014.02.001} {\bibfield
  {journal} {\bibinfo  {journal} {Prog. Part. Nucl. Phys.}\ }\textbf {\bibinfo
  {volume} {77}},\ \bibinfo {pages} {1} (\bibinfo {year} {2014})}\BibitemShut
  {NoStop}%
\bibitem [{\citenamefont {Binosi}\ \emph {et~al.}(2015)\citenamefont {Binosi},
  \citenamefont {Chang}, \citenamefont {Papavassiliou},\ and\ \citenamefont
  {Roberts}}]{Binosi:2014aea}%
  \BibitemOpen
  \bibfield  {author} {\bibinfo {author} {\bibfnamefont {D.}~\bibnamefont
  {Binosi}}, \bibinfo {author} {\bibfnamefont {L.}~\bibnamefont {Chang}},
  \bibinfo {author} {\bibfnamefont {J.}~\bibnamefont {Papavassiliou}},\ and\
  \bibinfo {author} {\bibfnamefont {C.~D.}\ \bibnamefont {Roberts}},\ }\href
  {https://doi.org/10.1016/j.physletb.2015.01.031} {\bibfield  {journal}
  {\bibinfo  {journal} {Phys. Lett.}\ }\textbf {\bibinfo {volume} {B742}},\
  \bibinfo {pages} {183} (\bibinfo {year} {2015})}\BibitemShut {NoStop}%
\bibitem [{\citenamefont {Binosi}\ \emph {et~al.}(2016)\citenamefont {Binosi},
  \citenamefont {Chang}, \citenamefont {Papavassiliou}, \citenamefont {Qin},\
  and\ \citenamefont {Roberts}}]{Binosi:2016rxz}%
  \BibitemOpen
  \bibfield  {author} {\bibinfo {author} {\bibfnamefont {D.}~\bibnamefont
  {Binosi}}, \bibinfo {author} {\bibfnamefont {L.}~\bibnamefont {Chang}},
  \bibinfo {author} {\bibfnamefont {J.}~\bibnamefont {Papavassiliou}}, \bibinfo
  {author} {\bibfnamefont {S.-X.}\ \bibnamefont {Qin}},\ and\ \bibinfo {author}
  {\bibfnamefont {C.~D.}\ \bibnamefont {Roberts}},\ }\href
  {https://doi.org/10.1103/PhysRevD.93.096010} {\bibfield  {journal} {\bibinfo
  {journal} {Phys. Rev.}\ }\textbf {\bibinfo {volume} {D93}},\ \bibinfo {pages}
  {096010} (\bibinfo {year} {2016})}\BibitemShut {NoStop}%
\bibitem [{\citenamefont {Binosi}\ \emph {et~al.}(2017)\citenamefont {Binosi},
  \citenamefont {Mezrag}, \citenamefont {Papavassiliou}, \citenamefont
  {Roberts},\ and\ \citenamefont {Rodriguez-Quintero}}]{Binosi:2016nme}%
  \BibitemOpen
  \bibfield  {author} {\bibinfo {author} {\bibfnamefont {D.}~\bibnamefont
  {Binosi}}, \bibinfo {author} {\bibfnamefont {C.}~\bibnamefont {Mezrag}},
  \bibinfo {author} {\bibfnamefont {J.}~\bibnamefont {Papavassiliou}}, \bibinfo
  {author} {\bibfnamefont {C.~D.}\ \bibnamefont {Roberts}},\ and\ \bibinfo
  {author} {\bibfnamefont {J.}~\bibnamefont {Rodriguez-Quintero}},\ }\href
  {https://doi.org/10.1103/PhysRevD.96.054026} {\bibfield  {journal} {\bibinfo
  {journal} {Phys. Rev.}\ }\textbf {\bibinfo {volume} {D96}},\ \bibinfo {pages}
  {054026} (\bibinfo {year} {2017})}\BibitemShut {NoStop}%
\bibitem [{\citenamefont {Huber}(2020{\natexlab{a}})}]{Huber:2018ned}%
  \BibitemOpen
  \bibfield  {author} {\bibinfo {author} {\bibfnamefont {M.~Q.}\ \bibnamefont
  {Huber}},\ }\href {https://doi.org/10.1016/j.physrep.2020.04.004} {\bibfield
  {journal} {\bibinfo  {journal} {Phys. Rept.}\ }\textbf {\bibinfo {volume}
  {879}},\ \bibinfo {pages} {1} (\bibinfo {year}
  {2020}{\natexlab{a}})}\BibitemShut {NoStop}%
\bibitem [{\citenamefont {Huber}(2020{\natexlab{b}})}]{Huber:2020keu}%
  \BibitemOpen
  \bibfield  {author} {\bibinfo {author} {\bibfnamefont {M.~Q.}\ \bibnamefont
  {Huber}},\ }\href {https://doi.org/10.1103/PhysRevD.101.114009} {\bibfield
  {journal} {\bibinfo  {journal} {Phys. Rev. D}\ }\textbf {\bibinfo {volume}
  {101}},\ \bibinfo {pages} {114009} (\bibinfo {year}
  {2020}{\natexlab{b}})}\BibitemShut {NoStop}%
\bibitem [{\citenamefont {Aguilar}\ \emph
  {et~al.}(2016{\natexlab{b}})\citenamefont {Aguilar}, \citenamefont {Binosi},
  \citenamefont {Figueiredo},\ and\ \citenamefont
  {Papavassiliou}}]{Aguilar:2016vin}%
  \BibitemOpen
  \bibfield  {author} {\bibinfo {author} {\bibfnamefont {A.~C.}\ \bibnamefont
  {Aguilar}}, \bibinfo {author} {\bibfnamefont {D.}~\bibnamefont {Binosi}},
  \bibinfo {author} {\bibfnamefont {C.~T.}\ \bibnamefont {Figueiredo}},\ and\
  \bibinfo {author} {\bibfnamefont {J.}~\bibnamefont {Papavassiliou}},\ }\href
  {https://doi.org/10.1103/PhysRevD.94.045002} {\bibfield  {journal} {\bibinfo
  {journal} {Phys. Rev.}\ }\textbf {\bibinfo {volume} {D94}},\ \bibinfo {pages}
  {045002} (\bibinfo {year} {2016}{\natexlab{b}})}\BibitemShut {NoStop}%
\bibitem [{\citenamefont {Aguilar}\ \emph
  {et~al.}(2023{\natexlab{a}})\citenamefont {Aguilar}, \citenamefont {De~Soto},
  \citenamefont {Ferreira}, \citenamefont {Papavassiliou}, \citenamefont
  {Pinto-G\'omez}, \citenamefont {Roberts},\ and\ \citenamefont
  {Rodr\'\i{}guez-Quintero}}]{Aguilar:2022thg}%
  \BibitemOpen
  \bibfield  {author} {\bibinfo {author} {\bibfnamefont {A.~C.}\ \bibnamefont
  {Aguilar}}, \bibinfo {author} {\bibfnamefont {F.}~\bibnamefont {De~Soto}},
  \bibinfo {author} {\bibfnamefont {M.~N.}\ \bibnamefont {Ferreira}}, \bibinfo
  {author} {\bibfnamefont {J.}~\bibnamefont {Papavassiliou}}, \bibinfo {author}
  {\bibfnamefont {F.}~\bibnamefont {Pinto-G\'omez}}, \bibinfo {author}
  {\bibfnamefont {C.~D.}\ \bibnamefont {Roberts}},\ and\ \bibinfo {author}
  {\bibfnamefont {J.}~\bibnamefont {Rodr\'\i{}guez-Quintero}},\ }\href
  {https://doi.org/10.1016/j.physletb.2023.137906} {\bibfield  {journal}
  {\bibinfo  {journal} {Phys. Lett. B}\ }\textbf {\bibinfo {volume} {841}},\
  \bibinfo {pages} {137906} (\bibinfo {year} {2023}{\natexlab{a}})}\BibitemShut
  {NoStop}%
\bibitem [{\citenamefont {Alexandrou}\ \emph {et~al.}(2002)\citenamefont
  {Alexandrou}, \citenamefont {de~Forcrand},\ and\ \citenamefont
  {Follana}}]{Alexandrou:2001fh}%
  \BibitemOpen
  \bibfield  {author} {\bibinfo {author} {\bibfnamefont {C.}~\bibnamefont
  {Alexandrou}}, \bibinfo {author} {\bibfnamefont {P.}~\bibnamefont
  {de~Forcrand}},\ and\ \bibinfo {author} {\bibfnamefont {E.}~\bibnamefont
  {Follana}},\ }\href {https://doi.org/10.1103/PhysRevD.65.114508} {\bibfield
  {journal} {\bibinfo  {journal} {Phys. Rev. D}\ }\textbf {\bibinfo {volume}
  {65}},\ \bibinfo {pages} {114508} (\bibinfo {year} {2002})}\BibitemShut
  {NoStop}%
\bibitem [{\citenamefont {Cucchieri}\ and\ \citenamefont
  {Mendes}(2007)}]{Cucchieri:2007md}%
  \BibitemOpen
  \bibfield  {author} {\bibinfo {author} {\bibfnamefont {A.}~\bibnamefont
  {Cucchieri}}\ and\ \bibinfo {author} {\bibfnamefont {T.}~\bibnamefont
  {Mendes}},\ }\href {https://doi.org/10.22323/1.042.0297} {\bibfield
  {journal} {\bibinfo  {journal} {PoS}\ }\textbf {\bibinfo {volume}
  {LATTICE2007}},\ \bibinfo {pages} {297} (\bibinfo {year} {2007})}\BibitemShut
  {NoStop}%
\bibitem [{\citenamefont {Bogolubsky}\ \emph {et~al.}(2007)\citenamefont
  {Bogolubsky}, \citenamefont {Ilgenfritz}, \citenamefont {Muller-Preussker},\
  and\ \citenamefont {Sternbeck}}]{Bogolubsky:2007ud}%
  \BibitemOpen
  \bibfield  {author} {\bibinfo {author} {\bibfnamefont {I.}~\bibnamefont
  {Bogolubsky}}, \bibinfo {author} {\bibfnamefont {E.}~\bibnamefont
  {Ilgenfritz}}, \bibinfo {author} {\bibfnamefont {M.}~\bibnamefont
  {Muller-Preussker}},\ and\ \bibinfo {author} {\bibfnamefont {A.}~\bibnamefont
  {Sternbeck}},\ }\href {https://doi.org/10.22323/1.042.0290} {\bibfield
  {journal} {\bibinfo  {journal} {PoS}\ }\textbf {\bibinfo {volume}
  {LATTICE2007}},\ \bibinfo {pages} {290} (\bibinfo {year} {2007})}\BibitemShut
  {NoStop}%
\bibitem [{\citenamefont {Bowman}\ \emph {et~al.}(2007)\citenamefont {Bowman},
  \citenamefont {Heller}, \citenamefont {Leinweber}, \citenamefont
  {Parappilly}, \citenamefont {Sternbeck}, \citenamefont {von Smekal},
  \citenamefont {Williams},\ and\ \citenamefont {Zhang}}]{Bowman:2007du}%
  \BibitemOpen
  \bibfield  {author} {\bibinfo {author} {\bibfnamefont {P.~O.}\ \bibnamefont
  {Bowman}}, \bibinfo {author} {\bibfnamefont {U.~M.}\ \bibnamefont {Heller}},
  \bibinfo {author} {\bibfnamefont {D.~B.}\ \bibnamefont {Leinweber}}, \bibinfo
  {author} {\bibfnamefont {M.~B.}\ \bibnamefont {Parappilly}}, \bibinfo
  {author} {\bibfnamefont {A.}~\bibnamefont {Sternbeck}}, \bibinfo {author}
  {\bibfnamefont {L.}~\bibnamefont {von Smekal}}, \bibinfo {author}
  {\bibfnamefont {A.~G.}\ \bibnamefont {Williams}},\ and\ \bibinfo {author}
  {\bibfnamefont {J.-b.}\ \bibnamefont {Zhang}},\ }\href
  {https://doi.org/10.1103/PhysRevD.76.094505} {\bibfield  {journal} {\bibinfo
  {journal} {Phys. Rev. D}\ }\textbf {\bibinfo {volume} {76}},\ \bibinfo
  {pages} {094505} (\bibinfo {year} {2007})}\BibitemShut {NoStop}%
\bibitem [{\citenamefont {Kamleh}\ \emph {et~al.}(2007)\citenamefont {Kamleh},
  \citenamefont {Bowman}, \citenamefont {Leinweber}, \citenamefont {Williams},\
  and\ \citenamefont {Zhang}}]{Kamleh:2007ud}%
  \BibitemOpen
  \bibfield  {author} {\bibinfo {author} {\bibfnamefont {W.}~\bibnamefont
  {Kamleh}}, \bibinfo {author} {\bibfnamefont {P.~O.}\ \bibnamefont {Bowman}},
  \bibinfo {author} {\bibfnamefont {D.~B.}\ \bibnamefont {Leinweber}}, \bibinfo
  {author} {\bibfnamefont {A.~G.}\ \bibnamefont {Williams}},\ and\ \bibinfo
  {author} {\bibfnamefont {J.}~\bibnamefont {Zhang}},\ }\href
  {https://doi.org/10.1103/PhysRevD.76.094501} {\bibfield  {journal} {\bibinfo
  {journal} {Phys. Rev.}\ }\textbf {\bibinfo {volume} {D76}},\ \bibinfo {pages}
  {094501} (\bibinfo {year} {2007})}\BibitemShut {NoStop}%
\bibitem [{\citenamefont {Cucchieri}\ and\ \citenamefont
  {Mendes}(2008)}]{Cucchieri:2007rg}%
  \BibitemOpen
  \bibfield  {author} {\bibinfo {author} {\bibfnamefont {A.}~\bibnamefont
  {Cucchieri}}\ and\ \bibinfo {author} {\bibfnamefont {T.}~\bibnamefont
  {Mendes}},\ }\href {https://doi.org/10.1103/PhysRevLett.100.241601}
  {\bibfield  {journal} {\bibinfo  {journal} {Phys. Rev. Lett.}\ }\textbf
  {\bibinfo {volume} {100}},\ \bibinfo {pages} {241601} (\bibinfo {year}
  {2008})}\BibitemShut {NoStop}%
\bibitem [{\citenamefont {Bogolubsky}\ \emph {et~al.}(2009)\citenamefont
  {Bogolubsky}, \citenamefont {Ilgenfritz}, \citenamefont {Muller-Preussker},\
  and\ \citenamefont {Sternbeck}}]{Bogolubsky:2009dc}%
  \BibitemOpen
  \bibfield  {author} {\bibinfo {author} {\bibfnamefont {I.}~\bibnamefont
  {Bogolubsky}}, \bibinfo {author} {\bibfnamefont {E.}~\bibnamefont
  {Ilgenfritz}}, \bibinfo {author} {\bibfnamefont {M.}~\bibnamefont
  {Muller-Preussker}},\ and\ \bibinfo {author} {\bibfnamefont {A.}~\bibnamefont
  {Sternbeck}},\ }\href {https://doi.org/10.1016/j.physletb.2009.04.076}
  {\bibfield  {journal} {\bibinfo  {journal} {Phys. Lett.}\ }\textbf {\bibinfo
  {volume} {B676}},\ \bibinfo {pages} {69} (\bibinfo {year}
  {2009})}\BibitemShut {NoStop}%
\bibitem [{\citenamefont {Oliveira}\ and\ \citenamefont
  {Silva}(2009)}]{Oliveira:2009eh}%
  \BibitemOpen
  \bibfield  {author} {\bibinfo {author} {\bibfnamefont {O.}~\bibnamefont
  {Oliveira}}\ and\ \bibinfo {author} {\bibfnamefont {P.}~\bibnamefont
  {Silva}},\ }\href {https://doi.org/10.22323/1.091.0226} {\bibfield  {journal}
  {\bibinfo  {journal} {PoS}\ }\textbf {\bibinfo {volume} {LAT2009}},\ \bibinfo
  {pages} {226} (\bibinfo {year} {2009})}\BibitemShut {NoStop}%
\bibitem [{\citenamefont {Cucchieri}\ \emph {et~al.}(2009)\citenamefont
  {Cucchieri}, \citenamefont {Mendes},\ and\ \citenamefont
  {Santos}}]{Cucchieri:2009kk}%
  \BibitemOpen
  \bibfield  {author} {\bibinfo {author} {\bibfnamefont {A.}~\bibnamefont
  {Cucchieri}}, \bibinfo {author} {\bibfnamefont {T.}~\bibnamefont {Mendes}},\
  and\ \bibinfo {author} {\bibfnamefont {E.~M.~S.}\ \bibnamefont {Santos}},\
  }\href {https://doi.org/10.1103/PhysRevLett.103.141602} {\bibfield  {journal}
  {\bibinfo  {journal} {Phys. Rev. Lett.}\ }\textbf {\bibinfo {volume} {103}},\
  \bibinfo {pages} {141602} (\bibinfo {year} {2009})}\BibitemShut {NoStop}%
\bibitem [{\citenamefont {Cucchieri}\ and\ \citenamefont
  {Mendes}(2010)}]{Cucchieri:2009zt}%
  \BibitemOpen
  \bibfield  {author} {\bibinfo {author} {\bibfnamefont {A.}~\bibnamefont
  {Cucchieri}}\ and\ \bibinfo {author} {\bibfnamefont {T.}~\bibnamefont
  {Mendes}},\ }\href {https://doi.org/10.1103/PhysRevD.81.016005} {\bibfield
  {journal} {\bibinfo  {journal} {Phys. Rev.}\ }\textbf {\bibinfo {volume}
  {D81}},\ \bibinfo {pages} {016005} (\bibinfo {year} {2010})}\BibitemShut
  {NoStop}%
\bibitem [{\citenamefont {Cucchieri}\ \emph {et~al.}(2010)\citenamefont
  {Cucchieri}, \citenamefont {Mendes}, \citenamefont {Nakamura},\ and\
  \citenamefont {Santos}}]{Cucchieri:2011pp}%
  \BibitemOpen
  \bibfield  {author} {\bibinfo {author} {\bibfnamefont {A.}~\bibnamefont
  {Cucchieri}}, \bibinfo {author} {\bibfnamefont {T.}~\bibnamefont {Mendes}},
  \bibinfo {author} {\bibfnamefont {G.~M.}\ \bibnamefont {Nakamura}},\ and\
  \bibinfo {author} {\bibfnamefont {E.~M.~S.}\ \bibnamefont {Santos}},\ }\href
  {https://doi.org/10.22323/1.117.0026} {\bibfield  {journal} {\bibinfo
  {journal} {PoS}\ }\textbf {\bibinfo {volume} {FACESQCD}},\ \bibinfo {pages}
  {026} (\bibinfo {year} {2010})}\BibitemShut {NoStop}%
\bibitem [{\citenamefont {Oliveira}\ and\ \citenamefont
  {Bicudo}(2011)}]{Oliveira:2010xc}%
  \BibitemOpen
  \bibfield  {author} {\bibinfo {author} {\bibfnamefont {O.}~\bibnamefont
  {Oliveira}}\ and\ \bibinfo {author} {\bibfnamefont {P.}~\bibnamefont
  {Bicudo}},\ }\href {https://doi.org/10.1088/0954-3899/38/4/045003} {\bibfield
   {journal} {\bibinfo  {journal} {J. Phys. G}\ }\textbf {\bibinfo {volume}
  {G38}},\ \bibinfo {pages} {045003} (\bibinfo {year} {2011})}\BibitemShut
  {NoStop}%
\bibitem [{\citenamefont {Ayala}\ \emph {et~al.}(2012)\citenamefont {Ayala},
  \citenamefont {Bashir}, \citenamefont {Binosi}, \citenamefont
  {Cristoforetti},\ and\ \citenamefont {Rodriguez-Quintero}}]{Ayala:2012pb}%
  \BibitemOpen
  \bibfield  {author} {\bibinfo {author} {\bibfnamefont {A.}~\bibnamefont
  {Ayala}}, \bibinfo {author} {\bibfnamefont {A.}~\bibnamefont {Bashir}},
  \bibinfo {author} {\bibfnamefont {D.}~\bibnamefont {Binosi}}, \bibinfo
  {author} {\bibfnamefont {M.}~\bibnamefont {Cristoforetti}},\ and\ \bibinfo
  {author} {\bibfnamefont {J.}~\bibnamefont {Rodriguez-Quintero}},\ }\href
  {https://doi.org/10.1103/PhysRevD.86.074512} {\bibfield  {journal} {\bibinfo
  {journal} {Phys. Rev.}\ }\textbf {\bibinfo {volume} {D86}},\ \bibinfo {pages}
  {074512} (\bibinfo {year} {2012})}\BibitemShut {NoStop}%
\bibitem [{\citenamefont {Sternbeck}\ and\ \citenamefont
  {M\"uller-Preussker}(2013)}]{Sternbeck:2012mf}%
  \BibitemOpen
  \bibfield  {author} {\bibinfo {author} {\bibfnamefont {A.}~\bibnamefont
  {Sternbeck}}\ and\ \bibinfo {author} {\bibfnamefont {M.}~\bibnamefont
  {M\"uller-Preussker}},\ }\href
  {https://doi.org/10.1016/j.physletb.2013.08.017} {\bibfield  {journal}
  {\bibinfo  {journal} {Phys. Lett. B}\ }\textbf {\bibinfo {volume} {726}},\
  \bibinfo {pages} {396} (\bibinfo {year} {2013})}\BibitemShut {NoStop}%
\bibitem [{\citenamefont {Bicudo}\ \emph {et~al.}(2015)\citenamefont {Bicudo},
  \citenamefont {Binosi}, \citenamefont {Cardoso}, \citenamefont {Oliveira},\
  and\ \citenamefont {Silva}}]{Bicudo:2015rma}%
  \BibitemOpen
  \bibfield  {author} {\bibinfo {author} {\bibfnamefont {P.}~\bibnamefont
  {Bicudo}}, \bibinfo {author} {\bibfnamefont {D.}~\bibnamefont {Binosi}},
  \bibinfo {author} {\bibfnamefont {N.}~\bibnamefont {Cardoso}}, \bibinfo
  {author} {\bibfnamefont {O.}~\bibnamefont {Oliveira}},\ and\ \bibinfo
  {author} {\bibfnamefont {P.~J.}\ \bibnamefont {Silva}},\ }\href
  {https://doi.org/10.1103/PhysRevD.92.114514} {\bibfield  {journal} {\bibinfo
  {journal} {Phys. Rev.}\ }\textbf {\bibinfo {volume} {D92}},\ \bibinfo {pages}
  {114514} (\bibinfo {year} {2015})}\BibitemShut {NoStop}%
\bibitem [{\citenamefont {Duarte}\ \emph {et~al.}(2016)\citenamefont {Duarte},
  \citenamefont {Oliveira},\ and\ \citenamefont {Silva}}]{Duarte:2016iko}%
  \BibitemOpen
  \bibfield  {author} {\bibinfo {author} {\bibfnamefont {A.~G.}\ \bibnamefont
  {Duarte}}, \bibinfo {author} {\bibfnamefont {O.}~\bibnamefont {Oliveira}},\
  and\ \bibinfo {author} {\bibfnamefont {P.~J.}\ \bibnamefont {Silva}},\ }\href
  {https://doi.org/10.1103/PhysRevD.94.014502} {\bibfield  {journal} {\bibinfo
  {journal} {Phys. Rev. D}\ }\textbf {\bibinfo {volume} {94}},\ \bibinfo
  {pages} {014502} (\bibinfo {year} {2016})}\BibitemShut {NoStop}%
\bibitem [{\citenamefont {Dudal}\ \emph {et~al.}(2018)\citenamefont {Dudal},
  \citenamefont {Oliveira},\ and\ \citenamefont {Silva}}]{Dudal:2018cli}%
  \BibitemOpen
  \bibfield  {author} {\bibinfo {author} {\bibfnamefont {D.}~\bibnamefont
  {Dudal}}, \bibinfo {author} {\bibfnamefont {O.}~\bibnamefont {Oliveira}},\
  and\ \bibinfo {author} {\bibfnamefont {P.~J.}\ \bibnamefont {Silva}},\ }\href
  {https://doi.org/10.1016/j.aop.2018.08.019} {\bibfield  {journal} {\bibinfo
  {journal} {Annals Phys.}\ }\textbf {\bibinfo {volume} {397}},\ \bibinfo
  {pages} {351} (\bibinfo {year} {2018})}\BibitemShut {NoStop}%
\bibitem [{\citenamefont {Aguilar}\ \emph {et~al.}(2020)\citenamefont
  {Aguilar}, \citenamefont {De~Soto}, \citenamefont {Ferreira}, \citenamefont
  {Papavassiliou}, \citenamefont {Rodríguez-Quintero},\ and\ \citenamefont
  {Zafeiropoulos}}]{Aguilar:2019uob}%
  \BibitemOpen
  \bibfield  {author} {\bibinfo {author} {\bibfnamefont {A.~C.}\ \bibnamefont
  {Aguilar}}, \bibinfo {author} {\bibfnamefont {F.}~\bibnamefont {De~Soto}},
  \bibinfo {author} {\bibfnamefont {M.~N.}\ \bibnamefont {Ferreira}}, \bibinfo
  {author} {\bibfnamefont {J.}~\bibnamefont {Papavassiliou}}, \bibinfo {author}
  {\bibfnamefont {J.}~\bibnamefont {Rodríguez-Quintero}},\ and\ \bibinfo
  {author} {\bibfnamefont {S.}~\bibnamefont {Zafeiropoulos}},\ }\href
  {https://doi.org/10.1140/epjc/s10052-020-7741-0} {\bibfield  {journal}
  {\bibinfo  {journal} {Eur. Phys. J.}\ }\textbf {\bibinfo {volume} {C80}},\
  \bibinfo {pages} {154} (\bibinfo {year} {2020})}\BibitemShut {NoStop}%
\bibitem [{\citenamefont {Aguilar}\ \emph
  {et~al.}(2023{\natexlab{b}})\citenamefont {Aguilar}, \citenamefont
  {Ferreira}, \citenamefont {Iba\~nez},\ and\ \citenamefont
  {Papavassiliou}}]{Aguilar:2023mam}%
  \BibitemOpen
  \bibfield  {author} {\bibinfo {author} {\bibfnamefont {A.~C.}\ \bibnamefont
  {Aguilar}}, \bibinfo {author} {\bibfnamefont {M.~.~N.}\ \bibnamefont
  {Ferreira}}, \bibinfo {author} {\bibfnamefont {D.}~\bibnamefont {Iba\~nez}},\
  and\ \bibinfo {author} {\bibfnamefont {J.}~\bibnamefont {Papavassiliou}},\
  }\href {https://doi.org/10.1140/epjc/s10052-023-12103-8} {\bibfield
  {journal} {\bibinfo  {journal} {Eur. Phys. J. C}\ }\textbf {\bibinfo {volume}
  {83}},\ \bibinfo {pages} {967} (\bibinfo {year}
  {2023}{\natexlab{b}})}\BibitemShut {NoStop}%
\bibitem [{\citenamefont {Vladimirov}(1971)}]{vladimirov1971equations}%
  \BibitemOpen
  \bibfield  {author} {\bibinfo {author} {\bibfnamefont {V.}~\bibnamefont
  {Vladimirov}},\ }\href {https://books.google.com.hk/books?id=oAfvAAAAIAAJ}
  {\emph {\bibinfo {title} {Equations of Mathematical Physics}}},\ Monographs
  and textbooks in pure and applied mathematics\ (\bibinfo  {publisher} {M.
  Dekker},\ \bibinfo {year} {1971})\BibitemShut {NoStop}%
\bibitem [{\citenamefont {Polyanin}\ and\ \citenamefont
  {Manzhirov}(2008)}]{polyanin2008handbook}%
  \BibitemOpen
  \bibfield  {author} {\bibinfo {author} {\bibfnamefont {A.}~\bibnamefont
  {Polyanin}}\ and\ \bibinfo {author} {\bibfnamefont {A.}~\bibnamefont
  {Manzhirov}},\ }\href {https://books.google.com.hk/books?id=zQYamAEACAAJ}
  {\emph {\bibinfo {title} {Handbook of Integral Equations: Second Edition}}}\
  (\bibinfo  {publisher} {Taylor \& Francis},\ \bibinfo {year}
  {2008})\BibitemShut {NoStop}%
\bibitem [{\citenamefont {Zumino}(1965)}]{Zumino:1965rka}%
  \BibitemOpen
  \bibfield  {author} {\bibinfo {author} {\bibfnamefont {B.}~\bibnamefont
  {Zumino}},\ }\href {https://doi.org/10.1007/978-3-7091-7649-8_12} {\bibfield
  {journal} {\bibinfo  {journal} {Acta Phys. Austriaca Suppl.}\ }\textbf
  {\bibinfo {volume} {2}},\ \bibinfo {pages} {212} (\bibinfo {year}
  {1965})}\BibitemShut {NoStop}%
\bibitem [{\citenamefont {Jackiw}\ and\ \citenamefont
  {Johnson}(1973)}]{Jackiw:1973tr}%
  \BibitemOpen
  \bibfield  {author} {\bibinfo {author} {\bibfnamefont {R.}~\bibnamefont
  {Jackiw}}\ and\ \bibinfo {author} {\bibfnamefont {K.}~\bibnamefont
  {Johnson}},\ }\href {https://doi.org/10.1103/PhysRevD.8.2386} {\bibfield
  {journal} {\bibinfo  {journal} {Phys. Rev. D}\ }\textbf {\bibinfo {volume}
  {8}},\ \bibinfo {pages} {2386} (\bibinfo {year} {1973})}\BibitemShut
  {NoStop}%
\bibitem [{\citenamefont {Jackiw}(1973)}]{Jackiw:1973ha}%
  \BibitemOpen
  \bibfield  {author} {\bibinfo {author} {\bibfnamefont {R.}~\bibnamefont
  {Jackiw}},\ }\href@noop {} {\emph {\bibinfo {title} {{Proceedings, Laws Of
  Hadronic Matter, Erice,}}}}\ (\bibinfo  {publisher} {MIT, Cambridge, MA},\
  \bibinfo {year} {1973})\BibitemShut {NoStop}%
\bibitem [{\citenamefont {Cornwall}\ and\ \citenamefont
  {Norton}(1973)}]{Cornwall:1973ts}%
  \BibitemOpen
  \bibfield  {author} {\bibinfo {author} {\bibfnamefont {J.}~\bibnamefont
  {Cornwall}}\ and\ \bibinfo {author} {\bibfnamefont {R.}~\bibnamefont
  {Norton}},\ }\href {https://doi.org/10.1103/PhysRevD.8.3338} {\bibfield
  {journal} {\bibinfo  {journal} {Phys. Rev. D}\ }\textbf {\bibinfo {volume}
  {8}},\ \bibinfo {pages} {3338} (\bibinfo {year} {1973})}\BibitemShut
  {NoStop}%
\bibitem [{\citenamefont {Athenodorou}\ \emph {et~al.}(2016)\citenamefont
  {Athenodorou}, \citenamefont {Binosi}, \citenamefont {Boucaud}, \citenamefont
  {De~Soto}, \citenamefont {Papavassiliou}, \citenamefont
  {Rodriguez-Quintero},\ and\ \citenamefont
  {Zafeiropoulos}}]{Athenodorou:2016oyh}%
  \BibitemOpen
  \bibfield  {author} {\bibinfo {author} {\bibfnamefont {A.}~\bibnamefont
  {Athenodorou}}, \bibinfo {author} {\bibfnamefont {D.}~\bibnamefont {Binosi}},
  \bibinfo {author} {\bibfnamefont {P.}~\bibnamefont {Boucaud}}, \bibinfo
  {author} {\bibfnamefont {F.}~\bibnamefont {De~Soto}}, \bibinfo {author}
  {\bibfnamefont {J.}~\bibnamefont {Papavassiliou}}, \bibinfo {author}
  {\bibfnamefont {J.}~\bibnamefont {Rodriguez-Quintero}},\ and\ \bibinfo
  {author} {\bibfnamefont {S.}~\bibnamefont {Zafeiropoulos}},\ }\href
  {https://doi.org/10.1016/j.physletb.2016.08.065} {\bibfield  {journal}
  {\bibinfo  {journal} {Phys. Lett.}\ }\textbf {\bibinfo {volume} {B761}},\
  \bibinfo {pages} {444} (\bibinfo {year} {2016})}\BibitemShut {NoStop}%
\bibitem [{\citenamefont {Boucaud}\ \emph {et~al.}(2017)\citenamefont
  {Boucaud}, \citenamefont {De~Soto}, \citenamefont {Rodríguez-Quintero},\
  and\ \citenamefont {Zafeiropoulos}}]{Boucaud:2017obn}%
  \BibitemOpen
  \bibfield  {author} {\bibinfo {author} {\bibfnamefont {P.}~\bibnamefont
  {Boucaud}}, \bibinfo {author} {\bibfnamefont {F.}~\bibnamefont {De~Soto}},
  \bibinfo {author} {\bibfnamefont {J.}~\bibnamefont {Rodríguez-Quintero}},\
  and\ \bibinfo {author} {\bibfnamefont {S.}~\bibnamefont {Zafeiropoulos}},\
  }\href {https://doi.org/10.1103/PhysRevD.95.114503} {\bibfield  {journal}
  {\bibinfo  {journal} {Phys. Rev.}\ }\textbf {\bibinfo {volume} {D95}},\
  \bibinfo {pages} {114503} (\bibinfo {year} {2017})}\BibitemShut {NoStop}%
\bibitem [{\citenamefont {Sternbeck}\ \emph {et~al.}(2017)\citenamefont
  {Sternbeck}, \citenamefont {Balduf}, \citenamefont {Kizilersu}, \citenamefont
  {Oliveira}, \citenamefont {Silva}, \citenamefont {Skullerud},\ and\
  \citenamefont {Williams}}]{Sternbeck:2017ntv}%
  \BibitemOpen
  \bibfield  {author} {\bibinfo {author} {\bibfnamefont {A.}~\bibnamefont
  {Sternbeck}}, \bibinfo {author} {\bibfnamefont {P.-H.}\ \bibnamefont
  {Balduf}}, \bibinfo {author} {\bibfnamefont {A.}~\bibnamefont {Kizilersu}},
  \bibinfo {author} {\bibfnamefont {O.}~\bibnamefont {Oliveira}}, \bibinfo
  {author} {\bibfnamefont {P.~J.}\ \bibnamefont {Silva}}, \bibinfo {author}
  {\bibfnamefont {J.-I.}\ \bibnamefont {Skullerud}},\ and\ \bibinfo {author}
  {\bibfnamefont {A.~G.}\ \bibnamefont {Williams}},\ }\href
  {https://doi.org/10.22323/1.256.0349} {\bibfield  {journal} {\bibinfo
  {journal} {PoS}\ }\textbf {\bibinfo {volume} {LATTICE2016}},\ \bibinfo
  {pages} {349} (\bibinfo {year} {2017})}\BibitemShut {NoStop}%
\bibitem [{\citenamefont {Aguilar}\ \emph
  {et~al.}(2021{\natexlab{a}})\citenamefont {Aguilar}, \citenamefont {De~Soto},
  \citenamefont {Ferreira}, \citenamefont {Papavassiliou},\ and\ \citenamefont
  {Rodríguez-Quintero}}]{Aguilar:2021lke}%
  \BibitemOpen
  \bibfield  {author} {\bibinfo {author} {\bibfnamefont {A.~C.}\ \bibnamefont
  {Aguilar}}, \bibinfo {author} {\bibfnamefont {F.}~\bibnamefont {De~Soto}},
  \bibinfo {author} {\bibfnamefont {M.~N.}\ \bibnamefont {Ferreira}}, \bibinfo
  {author} {\bibfnamefont {J.}~\bibnamefont {Papavassiliou}},\ and\ \bibinfo
  {author} {\bibfnamefont {J.}~\bibnamefont {Rodríguez-Quintero}},\ }\href
  {https://doi.org/10.1016/j.physletb.2021.136352} {\bibfield  {journal}
  {\bibinfo  {journal} {Phys. Lett. B}\ }\textbf {\bibinfo {volume} {818}},\
  \bibinfo {pages} {136352} (\bibinfo {year} {2021}{\natexlab{a}})}\BibitemShut
  {NoStop}%
\bibitem [{\citenamefont {Aguilar}\ \emph
  {et~al.}(2021{\natexlab{b}})\citenamefont {Aguilar}, \citenamefont
  {Ambr\'osio}, \citenamefont {De~Soto}, \citenamefont {Ferreira},
  \citenamefont {Oliveira}, \citenamefont {Papavassiliou},\ and\ \citenamefont
  {Rodr\'\i{}guez-Quintero}}]{Aguilar:2021okw}%
  \BibitemOpen
  \bibfield  {author} {\bibinfo {author} {\bibfnamefont {A.~C.}\ \bibnamefont
  {Aguilar}}, \bibinfo {author} {\bibfnamefont {C.~O.}\ \bibnamefont
  {Ambr\'osio}}, \bibinfo {author} {\bibfnamefont {F.}~\bibnamefont {De~Soto}},
  \bibinfo {author} {\bibfnamefont {M.~N.}\ \bibnamefont {Ferreira}}, \bibinfo
  {author} {\bibfnamefont {B.~M.}\ \bibnamefont {Oliveira}}, \bibinfo {author}
  {\bibfnamefont {J.}~\bibnamefont {Papavassiliou}},\ and\ \bibinfo {author}
  {\bibfnamefont {J.}~\bibnamefont {Rodr\'\i{}guez-Quintero}},\ }\href
  {https://doi.org/10.1103/PhysRevD.104.054028} {\bibfield  {journal} {\bibinfo
   {journal} {Phys. Rev. D}\ }\textbf {\bibinfo {volume} {104}},\ \bibinfo
  {pages} {054028} (\bibinfo {year} {2021}{\natexlab{b}})}\BibitemShut
  {NoStop}%
\bibitem [{\citenamefont {Maas}\ and\ \citenamefont
  {Vujinovi\'c}(2022)}]{Maas:2020zjp}%
  \BibitemOpen
  \bibfield  {author} {\bibinfo {author} {\bibfnamefont {A.}~\bibnamefont
  {Maas}}\ and\ \bibinfo {author} {\bibfnamefont {M.}~\bibnamefont
  {Vujinovi\'c}},\ }\href {https://doi.org/10.21468/SciPostPhysCore.5.2.019}
  {\bibfield  {journal} {\bibinfo  {journal} {SciPost Phys. Core}\ }\textbf
  {\bibinfo {volume} {5}},\ \bibinfo {pages} {019} (\bibinfo {year}
  {2022})}\BibitemShut {NoStop}%
\bibitem [{\citenamefont {Catumba}\ \emph
  {et~al.}(2022{\natexlab{a}})\citenamefont {Catumba}, \citenamefont
  {Oliveira},\ and\ \citenamefont {Silva}}]{Catumba:2021hng}%
  \BibitemOpen
  \bibfield  {author} {\bibinfo {author} {\bibfnamefont {G.~T.~R.}\
  \bibnamefont {Catumba}}, \bibinfo {author} {\bibfnamefont {O.}~\bibnamefont
  {Oliveira}},\ and\ \bibinfo {author} {\bibfnamefont {P.~J.}\ \bibnamefont
  {Silva}},\ }\href {https://doi.org/10.22323/1.396.0467} {\bibfield  {journal}
  {\bibinfo  {journal} {PoS}\ }\textbf {\bibinfo {volume} {LATTICE2021}},\
  \bibinfo {pages} {467} (\bibinfo {year} {2022}{\natexlab{a}})}\BibitemShut
  {NoStop}%
\bibitem [{\citenamefont {Catumba}\ \emph
  {et~al.}(2022{\natexlab{b}})\citenamefont {Catumba}, \citenamefont
  {Oliveira},\ and\ \citenamefont {Silva}}]{Catumba:2021yly}%
  \BibitemOpen
  \bibfield  {author} {\bibinfo {author} {\bibfnamefont {G.~T.~R.}\
  \bibnamefont {Catumba}}, \bibinfo {author} {\bibfnamefont {O.}~\bibnamefont
  {Oliveira}},\ and\ \bibinfo {author} {\bibfnamefont {P.~J.}\ \bibnamefont
  {Silva}},\ }\href {https://doi.org/10.1051/epjconf/202225802008} {\bibfield
  {journal} {\bibinfo  {journal} {EPJ Web Conf.}\ }\textbf {\bibinfo {volume}
  {258}},\ \bibinfo {pages} {02008} (\bibinfo {year}
  {2022}{\natexlab{b}})}\BibitemShut {NoStop}%
\bibitem [{\citenamefont {Pinto-G\'omez}\ \emph {et~al.}(2023)\citenamefont
  {Pinto-G\'omez}, \citenamefont {De~Soto}, \citenamefont {Ferreira},
  \citenamefont {Papavassiliou},\ and\ \citenamefont
  {Rodr\'\i{}guez-Quintero}}]{Pinto-Gomez:2022brg}%
  \BibitemOpen
  \bibfield  {author} {\bibinfo {author} {\bibfnamefont {F.}~\bibnamefont
  {Pinto-G\'omez}}, \bibinfo {author} {\bibfnamefont {F.}~\bibnamefont
  {De~Soto}}, \bibinfo {author} {\bibfnamefont {M.~N.}\ \bibnamefont
  {Ferreira}}, \bibinfo {author} {\bibfnamefont {J.}~\bibnamefont
  {Papavassiliou}},\ and\ \bibinfo {author} {\bibfnamefont {J.}~\bibnamefont
  {Rodr\'\i{}guez-Quintero}},\ }\href
  {https://doi.org/10.1016/j.physletb.2023.137737} {\bibfield  {journal}
  {\bibinfo  {journal} {Phys. Lett. B}\ }\textbf {\bibinfo {volume} {838}},\
  \bibinfo {pages} {137737} (\bibinfo {year} {2023})}\BibitemShut {NoStop}%
\bibitem [{\citenamefont {Berges}(2004)}]{Berges:2004pu}%
  \BibitemOpen
  \bibfield  {author} {\bibinfo {author} {\bibfnamefont {J.}~\bibnamefont
  {Berges}},\ }\href {https://doi.org/10.1103/PhysRevD.70.105010} {\bibfield
  {journal} {\bibinfo  {journal} {Phys. Rev. D}\ }\textbf {\bibinfo {volume}
  {70}},\ \bibinfo {pages} {105010} (\bibinfo {year} {2004})}\BibitemShut
  {NoStop}%
\bibitem [{\citenamefont {Carrington}\ and\ \citenamefont
  {Guo}(2011)}]{Carrington:2010qq}%
  \BibitemOpen
  \bibfield  {author} {\bibinfo {author} {\bibfnamefont {M.~E.}\ \bibnamefont
  {Carrington}}\ and\ \bibinfo {author} {\bibfnamefont {Y.}~\bibnamefont
  {Guo}},\ }\href {https://doi.org/10.1103/PhysRevD.83.016006} {\bibfield
  {journal} {\bibinfo  {journal} {Phys. Rev. D}\ }\textbf {\bibinfo {volume}
  {83}},\ \bibinfo {pages} {016006} (\bibinfo {year} {2011})}\BibitemShut
  {NoStop}%
\bibitem [{\citenamefont {York}\ \emph {et~al.}(2012)\citenamefont {York},
  \citenamefont {Moore},\ and\ \citenamefont {Tassler}}]{York:2012ib}%
  \BibitemOpen
  \bibfield  {author} {\bibinfo {author} {\bibfnamefont {M.~C.~A.}\
  \bibnamefont {York}}, \bibinfo {author} {\bibfnamefont {G.~D.}\ \bibnamefont
  {Moore}},\ and\ \bibinfo {author} {\bibfnamefont {M.}~\bibnamefont
  {Tassler}},\ }\href {https://doi.org/10.1007/JHEP06(2012)077} {\bibfield
  {journal} {\bibinfo  {journal} {{JHEP}}\ }\textbf {\bibinfo {volume} {06}},\
  \bibinfo {pages} {077} (\bibinfo {year} {2012})}\BibitemShut {NoStop}%
\bibitem [{\citenamefont {Mueller}\ and\ \citenamefont
  {Pawlowski}(2015)}]{Mueller:2015fka}%
  \BibitemOpen
  \bibfield  {author} {\bibinfo {author} {\bibfnamefont {N.}~\bibnamefont
  {Mueller}}\ and\ \bibinfo {author} {\bibfnamefont {J.~M.}\ \bibnamefont
  {Pawlowski}},\ }\href {https://doi.org/10.1103/PhysRevD.91.116010} {\bibfield
   {journal} {\bibinfo  {journal} {Phys. Rev.}\ }\textbf {\bibinfo {volume}
  {D91}},\ \bibinfo {pages} {116010} (\bibinfo {year} {2015})}\BibitemShut
  {NoStop}%
\bibitem [{\citenamefont {Williams}\ \emph {et~al.}(2016)\citenamefont
  {Williams}, \citenamefont {Fischer},\ and\ \citenamefont
  {Heupel}}]{Williams:2015cvx}%
  \BibitemOpen
  \bibfield  {author} {\bibinfo {author} {\bibfnamefont {R.}~\bibnamefont
  {Williams}}, \bibinfo {author} {\bibfnamefont {C.~S.}\ \bibnamefont
  {Fischer}},\ and\ \bibinfo {author} {\bibfnamefont {W.}~\bibnamefont
  {Heupel}},\ }\href {https://doi.org/10.1103/PhysRevD.93.034026} {\bibfield
  {journal} {\bibinfo  {journal} {Phys. Rev.}\ }\textbf {\bibinfo {volume}
  {D93}},\ \bibinfo {pages} {034026} (\bibinfo {year} {2016})}\BibitemShut
  {NoStop}%
\bibitem [{\citenamefont {Aguilar}\ \emph
  {et~al.}(2023{\natexlab{c}})\citenamefont {Aguilar}, \citenamefont
  {Ferreira}, \citenamefont {Papavassiliou},\ and\ \citenamefont
  {Santos}}]{Aguilar:2023qqd}%
  \BibitemOpen
  \bibfield  {author} {\bibinfo {author} {\bibfnamefont {A.~C.}\ \bibnamefont
  {Aguilar}}, \bibinfo {author} {\bibfnamefont {M.~N.}\ \bibnamefont
  {Ferreira}}, \bibinfo {author} {\bibfnamefont {J.}~\bibnamefont
  {Papavassiliou}},\ and\ \bibinfo {author} {\bibfnamefont {L.~R.}\
  \bibnamefont {Santos}},\ }\href
  {https://doi.org/10.1140/epjc/s10052-023-11732-3} {\bibfield  {journal}
  {\bibinfo  {journal} {Eur. Phys. J. C}\ }\textbf {\bibinfo {volume} {83}},\
  \bibinfo {pages} {549} (\bibinfo {year} {2023}{\natexlab{c}})}\BibitemShut
  {NoStop}%
\bibitem [{\citenamefont {Huber}\ \emph {et~al.}(2020)\citenamefont {Huber},
  \citenamefont {Fischer},\ and\ \citenamefont
  {Sanchis-Alepuz}}]{Huber:2020ngt}%
  \BibitemOpen
  \bibfield  {author} {\bibinfo {author} {\bibfnamefont {M.~Q.}\ \bibnamefont
  {Huber}}, \bibinfo {author} {\bibfnamefont {C.~S.}\ \bibnamefont {Fischer}},\
  and\ \bibinfo {author} {\bibfnamefont {H.}~\bibnamefont {Sanchis-Alepuz}},\
  }\href {https://doi.org/10.1140/epjc/s10052-020-08649-6} {\bibfield
  {journal} {\bibinfo  {journal} {Eur. Phys. J. C}\ }\textbf {\bibinfo {volume}
  {80}},\ \bibinfo {pages} {1077} (\bibinfo {year} {2020})}\BibitemShut
  {NoStop}%
\bibitem [{\citenamefont {Huber}\ \emph {et~al.}(2021)\citenamefont {Huber},
  \citenamefont {Fischer},\ and\ \citenamefont
  {Sanchis-Alepuz}}]{Huber:2021yfy}%
  \BibitemOpen
  \bibfield  {author} {\bibinfo {author} {\bibfnamefont {M.~Q.}\ \bibnamefont
  {Huber}}, \bibinfo {author} {\bibfnamefont {C.~S.}\ \bibnamefont {Fischer}},\
  and\ \bibinfo {author} {\bibfnamefont {H.}~\bibnamefont {Sanchis-Alepuz}},\
  }\href {https://doi.org/10.1140/epjc/s10052-021-09864-5} {\bibfield
  {journal} {\bibinfo  {journal} {Eur. Phys. J. C}\ }\textbf {\bibinfo {volume}
  {81}},\ \bibinfo {pages} {1083} (\bibinfo {year} {2021})},\ \bibinfo {note}
  {[Erratum: Eur.Phys.J.C 82, 38 (2022)]}\BibitemShut {NoStop}%
\bibitem [{\citenamefont {Taylor}(1971)}]{Taylor:1971ff}%
  \BibitemOpen
  \bibfield  {author} {\bibinfo {author} {\bibfnamefont {J.}~\bibnamefont
  {Taylor}},\ }\href {https://doi.org/10.1016/0550-3213(71)90297-5} {\bibfield
  {journal} {\bibinfo  {journal} {Nucl. Phys. B}\ }\textbf {\bibinfo {volume}
  {33}},\ \bibinfo {pages} {436} (\bibinfo {year} {1971})}\BibitemShut
  {NoStop}%
\bibitem [{\citenamefont {Slavnov}(1972)}]{Slavnov:1972fg}%
  \BibitemOpen
  \bibfield  {author} {\bibinfo {author} {\bibfnamefont {A.}~\bibnamefont
  {Slavnov}},\ }\href {https://doi.org/10.1007/BF01090719} {\bibfield
  {journal} {\bibinfo  {journal} {Theor. Math. Phys.}\ }\textbf {\bibinfo
  {volume} {10}},\ \bibinfo {pages} {99} (\bibinfo {year} {1972})}\BibitemShut
  {NoStop}%
\bibitem [{\citenamefont {Aguilar}\ \emph
  {et~al.}(2023{\natexlab{d}})\citenamefont {Aguilar}, \citenamefont
  {Ferreira}, \citenamefont {Oliveira}, \citenamefont {Papavassiliou},\ and\
  \citenamefont {Santos}}]{Aguilar:2023mdv}%
  \BibitemOpen
  \bibfield  {author} {\bibinfo {author} {\bibfnamefont {A.~C.}\ \bibnamefont
  {Aguilar}}, \bibinfo {author} {\bibfnamefont {M.~N.}\ \bibnamefont
  {Ferreira}}, \bibinfo {author} {\bibfnamefont {B.~M.}\ \bibnamefont
  {Oliveira}}, \bibinfo {author} {\bibfnamefont {J.}~\bibnamefont
  {Papavassiliou}},\ and\ \bibinfo {author} {\bibfnamefont {L.~R.}\
  \bibnamefont {Santos}},\ }\href
  {https://doi.org/10.1140/epjc/s10052-023-12058-w} {\bibfield  {journal}
  {\bibinfo  {journal} {Eur. Phys. J. C}\ }\textbf {\bibinfo {volume} {83}},\
  \bibinfo {pages} {889} (\bibinfo {year} {2023}{\natexlab{d}})}\BibitemShut
  {NoStop}%
\bibitem [{\citenamefont {Ball}\ and\ \citenamefont
  {Chiu}(1980)}]{Ball:1980ax}%
  \BibitemOpen
  \bibfield  {author} {\bibinfo {author} {\bibfnamefont {J.~S.}\ \bibnamefont
  {Ball}}\ and\ \bibinfo {author} {\bibfnamefont {T.-W.}\ \bibnamefont
  {Chiu}},\ }\href {https://doi.org/10.1103/PhysRevD.22.2550} {\bibfield
  {journal} {\bibinfo  {journal} {Phys. Rev. D}\ }\textbf {\bibinfo {volume}
  {22}},\ \bibinfo {pages} {2550} (\bibinfo {year} {1980})},\ \bibinfo {note}
  {[Erratum: Phys. Rev. D 23, 3085 (1981)]}\BibitemShut {NoStop}%
\bibitem [{\citenamefont {Davydychev}\ \emph {et~al.}(1996)\citenamefont
  {Davydychev}, \citenamefont {Osland},\ and\ \citenamefont
  {Tarasov}}]{Davydychev:1996pb}%
  \BibitemOpen
  \bibfield  {author} {\bibinfo {author} {\bibfnamefont {A.~I.}\ \bibnamefont
  {Davydychev}}, \bibinfo {author} {\bibfnamefont {P.}~\bibnamefont {Osland}},\
  and\ \bibinfo {author} {\bibfnamefont {O.}~\bibnamefont {Tarasov}},\ }\href
  {https://doi.org/10.1103/PhysRevD.59.109901} {\bibfield  {journal} {\bibinfo
  {journal} {Phys. Rev. D}\ }\textbf {\bibinfo {volume} {54}},\ \bibinfo
  {pages} {4087} (\bibinfo {year} {1996})},\ \bibinfo {note} {[Erratum: Phys.
  Rev. D 59, 109901 (1999)]}\BibitemShut {NoStop}%
\bibitem [{\citenamefont {Gracey}\ \emph
  {et~al.}(2019{\natexlab{b}})\citenamefont {Gracey}, \citenamefont {Kißler},\
  and\ \citenamefont {Kreimer}}]{Gracey:2019mix}%
  \BibitemOpen
  \bibfield  {author} {\bibinfo {author} {\bibfnamefont {J.}~\bibnamefont
  {Gracey}}, \bibinfo {author} {\bibfnamefont {H.}~\bibnamefont {Kißler}},\
  and\ \bibinfo {author} {\bibfnamefont {D.}~\bibnamefont {Kreimer}},\ }\href
  {https://doi.org/10.1103/PhysRevD.100.085001} {\bibfield  {journal} {\bibinfo
   {journal} {Phys. Rev. D}\ }\textbf {\bibinfo {volume} {100}},\ \bibinfo
  {pages} {085001} (\bibinfo {year} {2019}{\natexlab{b}})}\BibitemShut
  {NoStop}%
\bibitem [{\citenamefont {Aguilar}\ \emph {et~al.}(2014)\citenamefont
  {Aguilar}, \citenamefont {Binosi},\ and\ \citenamefont
  {Papavassiliou}}]{Aguilar:2014tka}%
  \BibitemOpen
  \bibfield  {author} {\bibinfo {author} {\bibfnamefont {A.~C.}\ \bibnamefont
  {Aguilar}}, \bibinfo {author} {\bibfnamefont {D.}~\bibnamefont {Binosi}},\
  and\ \bibinfo {author} {\bibfnamefont {J.}~\bibnamefont {Papavassiliou}},\
  }\href {https://doi.org/10.1103/PhysRevD.89.085032} {\bibfield  {journal}
  {\bibinfo  {journal} {Phys. Rev.}\ }\textbf {\bibinfo {volume} {D89}},\
  \bibinfo {pages} {085032} (\bibinfo {year} {2014})}\BibitemShut {NoStop}%
\bibitem [{\citenamefont {Bjorken}\ and\ \citenamefont
  {Drell}(1965)}]{Bjorken:1965zz}%
  \BibitemOpen
  \bibfield  {author} {\bibinfo {author} {\bibfnamefont {J.~D.}\ \bibnamefont
  {Bjorken}}\ and\ \bibinfo {author} {\bibfnamefont {S.~D.}\ \bibnamefont
  {Drell}},\ }\href@noop {} {\emph {\bibinfo {title} {{Relativistic quantum
  fields}}}},\ International Series In Pure and Applied Physics\ (\bibinfo
  {publisher} {McGraw-Hill},\ \bibinfo {address} {New York},\ \bibinfo {year}
  {1965})\BibitemShut {NoStop}%
\bibitem [{\citenamefont {Cui}\ \emph {et~al.}(2020)\citenamefont {Cui},
  \citenamefont {Zhang}, \citenamefont {Binosi}, \citenamefont {de~Soto},
  \citenamefont {Mezrag}, \citenamefont {Papavassiliou}, \citenamefont
  {Roberts}, \citenamefont {Rodríguez-Quintero}, \citenamefont {Segovia},\
  and\ \citenamefont {Zafeiropoulos}}]{Cui:2019dwv}%
  \BibitemOpen
  \bibfield  {author} {\bibinfo {author} {\bibfnamefont {Z.-F.}\ \bibnamefont
  {Cui}}, \bibinfo {author} {\bibfnamefont {J.-L.}\ \bibnamefont {Zhang}},
  \bibinfo {author} {\bibfnamefont {D.}~\bibnamefont {Binosi}}, \bibinfo
  {author} {\bibfnamefont {F.}~\bibnamefont {de~Soto}}, \bibinfo {author}
  {\bibfnamefont {C.}~\bibnamefont {Mezrag}}, \bibinfo {author} {\bibfnamefont
  {J.}~\bibnamefont {Papavassiliou}}, \bibinfo {author} {\bibfnamefont {C.~D.}\
  \bibnamefont {Roberts}}, \bibinfo {author} {\bibfnamefont {J.}~\bibnamefont
  {Rodríguez-Quintero}}, \bibinfo {author} {\bibfnamefont {J.}~\bibnamefont
  {Segovia}},\ and\ \bibinfo {author} {\bibfnamefont {S.}~\bibnamefont
  {Zafeiropoulos}},\ }\href {https://doi.org/10.1088/1674-1137/44/8/083102}
  {\bibfield  {journal} {\bibinfo  {journal} {Chin. Phys. C}\ }\textbf
  {\bibinfo {volume} {44}},\ \bibinfo {pages} {083102} (\bibinfo {year}
  {2020})}\BibitemShut {NoStop}%
\bibitem [{\citenamefont {Eichmann}\ \emph {et~al.}(2014)\citenamefont
  {Eichmann}, \citenamefont {Williams}, \citenamefont {Alkofer},\ and\
  \citenamefont {Vujinovic}}]{Eichmann:2014xya}%
  \BibitemOpen
  \bibfield  {author} {\bibinfo {author} {\bibfnamefont {G.}~\bibnamefont
  {Eichmann}}, \bibinfo {author} {\bibfnamefont {R.}~\bibnamefont {Williams}},
  \bibinfo {author} {\bibfnamefont {R.}~\bibnamefont {Alkofer}},\ and\ \bibinfo
  {author} {\bibfnamefont {M.}~\bibnamefont {Vujinovic}},\ }\href
  {https://doi.org/10.1103/PhysRevD.89.105014} {\bibfield  {journal} {\bibinfo
  {journal} {Phys. Rev.}\ }\textbf {\bibinfo {volume} {D89}},\ \bibinfo {pages}
  {105014} (\bibinfo {year} {2014})}\BibitemShut {NoStop}%
\bibitem [{\citenamefont {Blum}\ \emph {et~al.}(2014)\citenamefont {Blum},
  \citenamefont {Huber}, \citenamefont {Mitter},\ and\ \citenamefont {von
  Smekal}}]{Blum:2014gna}%
  \BibitemOpen
  \bibfield  {author} {\bibinfo {author} {\bibfnamefont {A.}~\bibnamefont
  {Blum}}, \bibinfo {author} {\bibfnamefont {M.~Q.}\ \bibnamefont {Huber}},
  \bibinfo {author} {\bibfnamefont {M.}~\bibnamefont {Mitter}},\ and\ \bibinfo
  {author} {\bibfnamefont {L.}~\bibnamefont {von Smekal}},\ }\href
  {https://doi.org/10.1103/PhysRevD.89.061703} {\bibfield  {journal} {\bibinfo
  {journal} {Phys. Rev.}\ }\textbf {\bibinfo {volume} {D89}},\ \bibinfo {pages}
  {061703} (\bibinfo {year} {2014})}\BibitemShut {NoStop}%
\bibitem [{\citenamefont {Mitter}\ \emph {et~al.}(2015)\citenamefont {Mitter},
  \citenamefont {Pawlowski},\ and\ \citenamefont
  {Strodthoff}}]{Mitter:2014wpa}%
  \BibitemOpen
  \bibfield  {author} {\bibinfo {author} {\bibfnamefont {M.}~\bibnamefont
  {Mitter}}, \bibinfo {author} {\bibfnamefont {J.~M.}\ \bibnamefont
  {Pawlowski}},\ and\ \bibinfo {author} {\bibfnamefont {N.}~\bibnamefont
  {Strodthoff}},\ }\href {https://doi.org/10.1103/PhysRevD.91.054035}
  {\bibfield  {journal} {\bibinfo  {journal} {Phys. Rev.}\ }\textbf {\bibinfo
  {volume} {D91}},\ \bibinfo {pages} {054035} (\bibinfo {year}
  {2015})}\BibitemShut {NoStop}%
\bibitem [{\citenamefont {Pawlowski}\ \emph {et~al.}(2022)\citenamefont
  {Pawlowski}, \citenamefont {Schneider},\ and\ \citenamefont
  {Wink}}]{Pawlowski:2022oyq}%
  \BibitemOpen
  \bibfield  {author} {\bibinfo {author} {\bibfnamefont {J.~M.}\ \bibnamefont
  {Pawlowski}}, \bibinfo {author} {\bibfnamefont {C.~S.}\ \bibnamefont
  {Schneider}},\ and\ \bibinfo {author} {\bibfnamefont {N.}~\bibnamefont
  {Wink}},\ }\Eprint {https://arxiv.org/abs/2202.11123} {arXiv:2202.11123
  [hep-th]}  (\bibinfo {year} {2022})\BibitemShut {NoStop}%
\bibitem [{\citenamefont {Press}\ \emph {et~al.}(1992)\citenamefont {Press},
  \citenamefont {Teukolsky}, \citenamefont {Vetterling},\ and\ \citenamefont
  {Flannery}}]{Press:1992zz}%
  \BibitemOpen
  \bibfield  {author} {\bibinfo {author} {\bibfnamefont {W.~H.}\ \bibnamefont
  {Press}}, \bibinfo {author} {\bibfnamefont {S.~A.}\ \bibnamefont
  {Teukolsky}}, \bibinfo {author} {\bibfnamefont {W.~T.}\ \bibnamefont
  {Vetterling}},\ and\ \bibinfo {author} {\bibfnamefont {B.~P.}\ \bibnamefont
  {Flannery}},\ }\href@noop {} {\emph {\bibinfo {title} {{Numerical Recipes in
  FORTRAN: The Art of Scientific Computing}}}}\ (\bibinfo  {publisher}
  {{Cambridge University Press}},\ \bibinfo {year} {1992})\BibitemShut
  {NoStop}%
\bibitem [{\citenamefont {Altarelli}(1982)}]{Altarelli:1981ax}%
  \BibitemOpen
  \bibfield  {author} {\bibinfo {author} {\bibfnamefont {G.}~\bibnamefont
  {Altarelli}},\ }\href {https://doi.org/10.1016/0370-1573(82)90127-2}
  {\bibfield  {journal} {\bibinfo  {journal} {Phys. Rept.}\ }\textbf {\bibinfo
  {volume} {81}},\ \bibinfo {pages} {1} (\bibinfo {year} {1982})}\BibitemShut
  {NoStop}%
\bibitem [{\citenamefont {von Smekal}\ \emph {et~al.}(1998)\citenamefont {von
  Smekal}, \citenamefont {Hauck},\ and\ \citenamefont
  {Alkofer}}]{vonSmekal:1997ern}%
  \BibitemOpen
  \bibfield  {author} {\bibinfo {author} {\bibfnamefont {L.}~\bibnamefont {von
  Smekal}}, \bibinfo {author} {\bibfnamefont {A.}~\bibnamefont {Hauck}},\ and\
  \bibinfo {author} {\bibfnamefont {R.}~\bibnamefont {Alkofer}},\ }\href
  {https://doi.org/10.1006/aphy.1998.5806, 10.1006/aphy.1998.5864} {\bibfield
  {journal} {\bibinfo  {journal} {Annals Phys.}\ }\textbf {\bibinfo {volume}
  {267}},\ \bibinfo {pages} {1} (\bibinfo {year} {1998})},\ \bibinfo {note}
  {[Erratum: Annals Phys. 269, 182 (1998)]}\BibitemShut {NoStop}%
\bibitem [{\citenamefont {Fischer}\ \emph {et~al.}(2002)\citenamefont
  {Fischer}, \citenamefont {Alkofer},\ and\ \citenamefont
  {Reinhardt}}]{Fischer:2002eq}%
  \BibitemOpen
  \bibfield  {author} {\bibinfo {author} {\bibfnamefont {C.~S.}\ \bibnamefont
  {Fischer}}, \bibinfo {author} {\bibfnamefont {R.}~\bibnamefont {Alkofer}},\
  and\ \bibinfo {author} {\bibfnamefont {H.}~\bibnamefont {Reinhardt}},\ }\href
  {https://doi.org/10.1103/PhysRevD.65.094008} {\bibfield  {journal} {\bibinfo
  {journal} {Phys. Rev. D}\ }\textbf {\bibinfo {volume} {65}},\ \bibinfo
  {pages} {094008} (\bibinfo {year} {2002})}\BibitemShut {NoStop}%
\bibitem [{\citenamefont {Pennington}\ and\ \citenamefont
  {Wilson}(2011)}]{Pennington:2011xs}%
  \BibitemOpen
  \bibfield  {author} {\bibinfo {author} {\bibfnamefont {M.~R.}\ \bibnamefont
  {Pennington}}\ and\ \bibinfo {author} {\bibfnamefont {D.~J.}\ \bibnamefont
  {Wilson}},\ }\href {https://doi.org/10.1103/PhysRevD.84.094028,
  10.1103/PhysRevD.84.119901} {\bibfield  {journal} {\bibinfo  {journal} {Phys.
  Rev.}\ }\textbf {\bibinfo {volume} {D84}},\ \bibinfo {pages} {119901}
  (\bibinfo {year} {2011})}\BibitemShut {NoStop}%
\bibitem [{\citenamefont {Celmaster}\ and\ \citenamefont
  {Gonsalves}(1979{\natexlab{a}})}]{Celmaster:1979dm}%
  \BibitemOpen
  \bibfield  {author} {\bibinfo {author} {\bibfnamefont {W.}~\bibnamefont
  {Celmaster}}\ and\ \bibinfo {author} {\bibfnamefont {R.~J.}\ \bibnamefont
  {Gonsalves}},\ }\href {https://doi.org/10.1103/PhysRevLett.42.1435}
  {\bibfield  {journal} {\bibinfo  {journal} {Phys. Rev. Lett.}\ }\textbf
  {\bibinfo {volume} {42}},\ \bibinfo {pages} {1435} (\bibinfo {year}
  {1979}{\natexlab{a}})}\BibitemShut {NoStop}%
\bibitem [{\citenamefont {Celmaster}\ and\ \citenamefont
  {Gonsalves}(1979{\natexlab{b}})}]{Celmaster:1979km}%
  \BibitemOpen
  \bibfield  {author} {\bibinfo {author} {\bibfnamefont {W.}~\bibnamefont
  {Celmaster}}\ and\ \bibinfo {author} {\bibfnamefont {R.~J.}\ \bibnamefont
  {Gonsalves}},\ }\href {https://doi.org/10.1103/PhysRevD.20.1420} {\bibfield
  {journal} {\bibinfo  {journal} {Phys. Rev. D}\ }\textbf {\bibinfo {volume}
  {20}},\ \bibinfo {pages} {1420} (\bibinfo {year}
  {1979}{\natexlab{b}})}\BibitemShut {NoStop}%
\bibitem [{\citenamefont {Alles}\ \emph {et~al.}(1997)\citenamefont {Alles},
  \citenamefont {Henty}, \citenamefont {Panagopoulos}, \citenamefont
  {Parrinello}, \citenamefont {Pittori},\ and\ \citenamefont
  {Richards}}]{Alles:1996ka}%
  \BibitemOpen
  \bibfield  {author} {\bibinfo {author} {\bibfnamefont {B.}~\bibnamefont
  {Alles}}, \bibinfo {author} {\bibfnamefont {D.}~\bibnamefont {Henty}},
  \bibinfo {author} {\bibfnamefont {H.}~\bibnamefont {Panagopoulos}}, \bibinfo
  {author} {\bibfnamefont {C.}~\bibnamefont {Parrinello}}, \bibinfo {author}
  {\bibfnamefont {C.}~\bibnamefont {Pittori}},\ and\ \bibinfo {author}
  {\bibfnamefont {D.~G.}\ \bibnamefont {Richards}},\ }\href
  {https://doi.org/10.1016/S0550-3213(97)00483-5} {\bibfield  {journal}
  {\bibinfo  {journal} {Nucl. Phys.}\ }\textbf {\bibinfo {volume} {B502}},\
  \bibinfo {pages} {325} (\bibinfo {year} {1997})}\BibitemShut {NoStop}%
\bibitem [{\citenamefont {Boucaud}\ \emph {et~al.}(2009)\citenamefont
  {Boucaud}, \citenamefont {De~Soto}, \citenamefont {Leroy}, \citenamefont
  {Le~Yaouanc}, \citenamefont {Micheli} \emph {et~al.}}]{Boucaud:2008gn}%
  \BibitemOpen
  \bibfield  {author} {\bibinfo {author} {\bibfnamefont {P.}~\bibnamefont
  {Boucaud}}, \bibinfo {author} {\bibfnamefont {F.}~\bibnamefont {De~Soto}},
  \bibinfo {author} {\bibfnamefont {J.}~\bibnamefont {Leroy}}, \bibinfo
  {author} {\bibfnamefont {A.}~\bibnamefont {Le~Yaouanc}}, \bibinfo {author}
  {\bibfnamefont {J.}~\bibnamefont {Micheli}}, \emph {et~al.},\ }\href
  {https://doi.org/10.1103/PhysRevD.79.014508} {\bibfield  {journal} {\bibinfo
  {journal} {Phys. Rev.}\ }\textbf {\bibinfo {volume} {D79}},\ \bibinfo {pages}
  {014508} (\bibinfo {year} {2009})}\BibitemShut {NoStop}%
\bibitem [{\citenamefont {Huber}(2017)}]{Huber:2017txg}%
  \BibitemOpen
  \bibfield  {author} {\bibinfo {author} {\bibfnamefont {M.~Q.}\ \bibnamefont
  {Huber}},\ }\href {https://doi.org/10.1140/epjc/s10052-017-5310-y} {\bibfield
   {journal} {\bibinfo  {journal} {Eur. Phys. J.}\ }\textbf {\bibinfo {volume}
  {C77}},\ \bibinfo {pages} {733} (\bibinfo {year} {2017})}\BibitemShut
  {NoStop}%
\bibitem [{\citenamefont {Binosi}\ \emph {et~al.}(2014)\citenamefont {Binosi},
  \citenamefont {Iba\~nez},\ and\ \citenamefont
  {Papavassiliou}}]{Binosi:2014kka}%
  \BibitemOpen
  \bibfield  {author} {\bibinfo {author} {\bibfnamefont {D.}~\bibnamefont
  {Binosi}}, \bibinfo {author} {\bibfnamefont {D.}~\bibnamefont {Iba\~nez}},\
  and\ \bibinfo {author} {\bibfnamefont {J.}~\bibnamefont {Papavassiliou}},\
  }\href {https://doi.org/10.1007/JHEP09(2014)059} {\bibfield  {journal}
  {\bibinfo  {journal} {{JHEP}}\ }\textbf {\bibinfo {volume} {09}},\ \bibinfo
  {pages} {059} (\bibinfo {year} {2014})}\BibitemShut {NoStop}%
\bibitem [{\citenamefont {Cyrol}\ \emph {et~al.}(2015)\citenamefont {Cyrol},
  \citenamefont {Huber},\ and\ \citenamefont {von Smekal}}]{Cyrol:2014kca}%
  \BibitemOpen
  \bibfield  {author} {\bibinfo {author} {\bibfnamefont {A.~K.}\ \bibnamefont
  {Cyrol}}, \bibinfo {author} {\bibfnamefont {M.~Q.}\ \bibnamefont {Huber}},\
  and\ \bibinfo {author} {\bibfnamefont {L.}~\bibnamefont {von Smekal}},\
  }\href {https://doi.org/10.1140/epjc/s10052-015-3312-1} {\bibfield  {journal}
  {\bibinfo  {journal} {Eur. Phys. J.}\ }\textbf {\bibinfo {volume} {C75}},\
  \bibinfo {pages} {102} (\bibinfo {year} {2015})}\BibitemShut {NoStop}%
\bibitem [{\citenamefont {Pawlowski}\ \emph {et~al.}(2023)\citenamefont
  {Pawlowski}, \citenamefont {Schneider}, \citenamefont {Turnwald},
  \citenamefont {Urban},\ and\ \citenamefont {Wink}}]{Pawlowski:2022zhh}%
  \BibitemOpen
  \bibfield  {author} {\bibinfo {author} {\bibfnamefont {J.~M.}\ \bibnamefont
  {Pawlowski}}, \bibinfo {author} {\bibfnamefont {C.~S.}\ \bibnamefont
  {Schneider}}, \bibinfo {author} {\bibfnamefont {J.}~\bibnamefont {Turnwald}},
  \bibinfo {author} {\bibfnamefont {J.~M.}\ \bibnamefont {Urban}},\ and\
  \bibinfo {author} {\bibfnamefont {N.}~\bibnamefont {Wink}},\ }\href
  {https://doi.org/10.1103/PhysRevD.108.076018} {\bibfield  {journal} {\bibinfo
   {journal} {Phys. Rev. D}\ }\textbf {\bibinfo {volume} {108}},\ \bibinfo
  {pages} {076018} (\bibinfo {year} {2023})}\BibitemShut {NoStop}%
\bibitem [{\citenamefont {Cola\c{c}o}\ \emph {et~al.}(2024)\citenamefont
  {Cola\c{c}o}, \citenamefont {Oliveira},\ and\ \citenamefont
  {Silva}}]{Colaco:2024gmt}%
  \BibitemOpen
  \bibfield  {author} {\bibinfo {author} {\bibfnamefont {M.}~\bibnamefont
  {Cola\c{c}o}}, \bibinfo {author} {\bibfnamefont {O.}~\bibnamefont
  {Oliveira}},\ and\ \bibinfo {author} {\bibfnamefont {P.~J.}\ \bibnamefont
  {Silva}},\ }\href {https://doi.org/10.1103/PhysRevD.109.074502} {\bibfield
  {journal} {\bibinfo  {journal} {Phys. Rev. D}\ }\textbf {\bibinfo {volume}
  {109}},\ \bibinfo {pages} {074502} (\bibinfo {year} {2024})}\BibitemShut
  {NoStop}%
\bibitem [{\citenamefont {Aguilar}\ \emph {et~al.}(2024)\citenamefont
  {Aguilar}, \citenamefont {Ferreira}, \citenamefont {Papavassiliou},\ and\
  \citenamefont {Santos}}]{Aguilar:2024fen}%
  \BibitemOpen
  \bibfield  {author} {\bibinfo {author} {\bibfnamefont {A.~C.}\ \bibnamefont
  {Aguilar}}, \bibinfo {author} {\bibfnamefont {M.~N.}\ \bibnamefont
  {Ferreira}}, \bibinfo {author} {\bibfnamefont {J.}~\bibnamefont
  {Papavassiliou}},\ and\ \bibinfo {author} {\bibfnamefont {L.~R.}\
  \bibnamefont {Santos}},\ }\Eprint {https://arxiv.org/abs/2402.16071}
  {arXiv:2402.16071 [hep-ph]}  (\bibinfo {year} {2024})\BibitemShut {NoStop}%
\end{thebibliography}
\end{document}